\documentclass[hyper]{JHEP3}

\usepackage[dvips]{graphicx}
\usepackage{epsfig}
\usepackage{multicol}
\usepackage{bbm}
\usepackage{amsmath}
\usepackage{cite}
\usepackage{lscape}
\usepackage{multirow}
\usepackage{amscd} 
\usepackage{remreset}   

\makeatletter
\@removefromreset{paragraph}{section}
\makeatother

\newcommand{\beq}{\begin{equation}}
\newcommand{\eeq}{\end{equation}}
\newcommand{\bea}{\begin{eqnarray}}
\newcommand{\eea}{\end{eqnarray}}
\newcommand{\ud}{\,\mathrm{d}} 
\newcommand{\BAR}{\,\mathcal{C}_{\mathrm{B}}}

\newcommand{\shatmin}{\,\sqrt{\hat s}_{\mathrm{min}}}
\newcommand{\smin}{\,\sqrt{s}_{\mathrm{min}}}

\DeclareMathOperator{\sech}{sech}

\newcommand{\ellp}{\ell^{\scriptscriptstyle{+}}}
\newcommand{\ellm}{\ell^{\scriptscriptstyle{-}}}

\DeclareMathSymbol{\minus}
    {\mathord}{operators}{"2D}

\newcommand{\lsim}{
\mathrel{\hbox{\rlap{\hbox{\lower4pt\hbox{$\sim$}}}\hbox{$<$}}}}
\newcommand{\gsim}{
\mathrel{\hbox{\rlap{\hbox{\lower4pt\hbox{$\sim$}}}\hbox{$>$}}}}

\newcommand{\met}{\not \!\! E_T}

\allowdisplaybreaks[1]

\title{Spin effects in the antler event topology at hadron colliders }

\author{Lisa Edelh\"auser \\
Institute for Theoretical Particle Physics and Cosmology, RWTH Aachen University, Aachen,
Germany\\
E-mail: \email{ledelhaeuser@physik.rwth-aachen.de}}

\author{Konstantin T.~Matchev \\ 
        Physics Department, University of Florida,
        Gainesville, FL 32611, USA \\
        E-mail: \email{matchev@phys.ufl.edu}
        }

\author{Myeonghun Park\\
        TH Division, CERN, Geneva, Switzerland\\
        E-mail: \email{Myeonghun.Park@cern.ch}
        }

\abstract{We investigate spin correlation effects in the ``antler" event topology 
$pp\to A\to B_1B_2\to (\ell^- C_1)(\ell^+ C_2)$ at the LHC.
We study the shapes of several kinematic variables, including 
the relative pseudorapidity, relative azimuthal angle and the energies of the two leptons, 
as well as several mass variables $M_{\ell\ell}$, $M_{\mathrm{eff}}$,
$\smin$, $M_{T2}$, $M_{CT}$ and $M_{CTx}$.
We focus on the two kinematic extremes of $\sqrt{s}$ --- threshold and infinity ---
and derive analytical expressions for the differential distributions of several variables,
most notably the $\cos{\theta_{\ellm\ellp}^\ast}$ variable proposed by Barr in hep-ph/0511115.
For all possible spin assignments of particles A, B and C, we derive 
the $\cos{\theta_{\ellm\ellp}^\ast}$ differential distribution at threshold, 
including the effects of spin correlations. Our analytical results help
identify the problematic cases for spin discrimination.
}


\preprint{
         May 9, 2012
          }

\begin{document}

\section{Introduction and motivation}
 
Once a signal of new physics is discovered at the Large Hadron Collider (LHC) at CERN, 
it must be interpreted in terms of the production of new particles with definite properties:
mass, spin, charge, coupling strength, chirality, etc. Typically, one studies the properties 
of new, short-lived resonances by examining their decay products which are seen in the detector.
For example, the electric charge of the parent particle (a short-lived resonance) is easily obtained by adding up the electric 
charges of its daughters (particles from a decaying process), the mass is similarly determined by forming the combined invariant mass
of the daughter particles, etc.

Notoriously, this procedure breaks down when some of the daughter particles are very weakly interacting 
particles and are not visible in the detector (the prototypical examples being the neutrinos
of the Standard Model (SM) or hypothetical dark matter WIMPs). Mass and spin measurements then
become very challenging, especially at hadron colliders where the partonic center-of-mass (CM) energy
and total longitudinal momentum in the event are a priori unknown. This has prompted a long line of research 
into developing new kinematic methods for determining the masses (see, e.g. \cite{Barr:2010zj,Barr:2011xt} 
and references therein) and spins \cite{Barr:2004ze,Battaglia:2005zf,Smillie:2005ar,Datta:2005zs,Datta:2005vx,Barr:2005dz,Meade:2006dw,Alves:2006df,Athanasiou:2006ef,
Wang:2006hk,Smillie:2006cd,Kilic:2007zk,Alves:2007xt,Csaki:2007xm,Buckley:2007th,Wang:2008sw,Buckley:2008pp,Burns:2008cp,Cho:2008tj,Gedalia:2009ym,
Boudjema:2009fz,Ehrenfeld:2009rt,Edelhauser:2010gb,Horton:2010bg,Cheng:2010yy,Buckley:2010jv,Chen:2010ek,Nojiri:2011qn,MoortgatPick:2011ix,Edelhauser:2011vj,Melia:2011cu,Chen:2011cya}
of new semi-invisibly decaying parent particles.

In this paper we tackle the more challenging of those two issues, namely, the determination of the intrinsic spin of the new particles.
Broadly speaking, most methods proposed so far in the literature fall into the following categories:
\begin{itemize}
\item Methods applicable to the case of lepton colliders, where the CM energy and longitudinal momentum of the 
initial state are known \cite{Battaglia:2005zf,Buckley:2007th}. Here we shall be interested in
hadron colliders like the LHC, where those methods would not be directly applicable.
\item Methods attempting either exact \cite{Cheng:2010yy} or approximate \cite{Cho:2008tj} reconstruction of the
kinematics of the invisible particles on an event by event basis. Unfortunately, in order to be able to solve 
for the missing momenta, one needs a relatively long decay chain, with a sufficiently large number of 
intermediate resonances \cite{Burns:2008va}. In contrast, in this paper we shall consider the case of 
the shortest decay chain possible (a chain with a single two-body decay), for which there are not enough kinematic constraints 
for reconstructing the missing momenta, so that one has to resort to likelihood methods \cite{Horton:2010bg,Chen:2010ek}.
\item Methods which study the shapes of the invariant mass distributions of the visible particles 
originating from {\em one} given decay chain \cite{Barr:2004ze,Smillie:2005ar,Datta:2005zs,Athanasiou:2006ef,Wang:2006hk}.
These methods again require a relatively long decay chain, which would produce at least two visible SM particles,
and are not applicable to the short decay chain which will be considered here.
\item Methods which simultaneously utilize the measured visible particles from both decay chains\footnote{In a typical 
new physics model, the longevity of the dark matter particle is enforced by an exact symmetry, the 
simplest option being $Z_2$ parity, under which all SM particles are even, while the new particles (superpartners, KK-partners, etc.) 
are odd. A consequence 
of such $Z_2$ parity is that the new particles are pair-produced, and each event has two cascade decays. }.
Here one studies the distributions of suitably defined quantities, e.g.~angular variables \cite{Barr:2005dz}, 
asymmetry-like variables \cite{Chen:2010ek,Melia:2011cu}, or global event variables like $M_{\mathrm{eff}}$ or $\met$
\cite{Nojiri:2011qn}.
\end{itemize}

In this paper we shall focus on the most challenging case for spin determination: that of a single stage, two-body decay chain,
in which a new parent particle $B$ is produced and then decays to a single visible SM particle $c$
and an invisible daughter particle $C$:
\beq
B \to c + C.
\eeq
Motivated by the dark matter problem, we shall further assume that the new particles $C$ and $B$ are $Z_2$-parity odd, so that 
the LHC produces the parent particles in pairs:
\beq
pp\to B\bar{B}\to c \bar{c}  C  \bar{C}.
\label{BBbaratLHC}
\eeq
In general, the process (\ref{BBbaratLHC}) can be due to $s$, $t$ and $u$ channel diagrams. 
Which of those diagrams are present and/or dominant is a very model-dependent issue, which
also depends on the nature of the visible particle $c$. For example, if $c$ carries color (i.e. it is a quark or a gluon),
one generically expects all three types of diagrams to be present. On the other hand, if $c$
is a SM lepton, which carries lepton number but no color, then it is reasonable to expect (\ref{BBbaratLHC}) 
to be mediated by $s$-channel diagrams only\footnote{In principle, $t$ and $u$ channel diagrams cannot be
ruled out completely --- for example, the dark matter particle $C$ itself may also carry lepton number
(although the sneutrino option in supersymmetry is disfavored by experiment), or the $t$ and $u$ channel
diagrams could be mediated by exotic particles carrying both color and lepton number (e.g. leptoquarks),
or the process (\ref{BBbaratLHC}) may involve lepton number violating interactions.}. 
 In what follows, for simplicity we shall assume that the SM particle $c$ is a lepton $\ell$,
 and that the process (\ref{BBbaratLHC}) is due to the exchange of a single $Z_2$-parity even 
 particle $A$ in an $s$-channel. The resulting event topology is shown in Fig.~\ref{fig:top}
 and has been dubbed ``the antler event topology'' \cite{Han:2009ss}, due to its resemblance to the
 headgear of the common wildlife species in Wisconsin.
\FIGURE[t]{
\centerline{ 
\includegraphics[height=6cm]{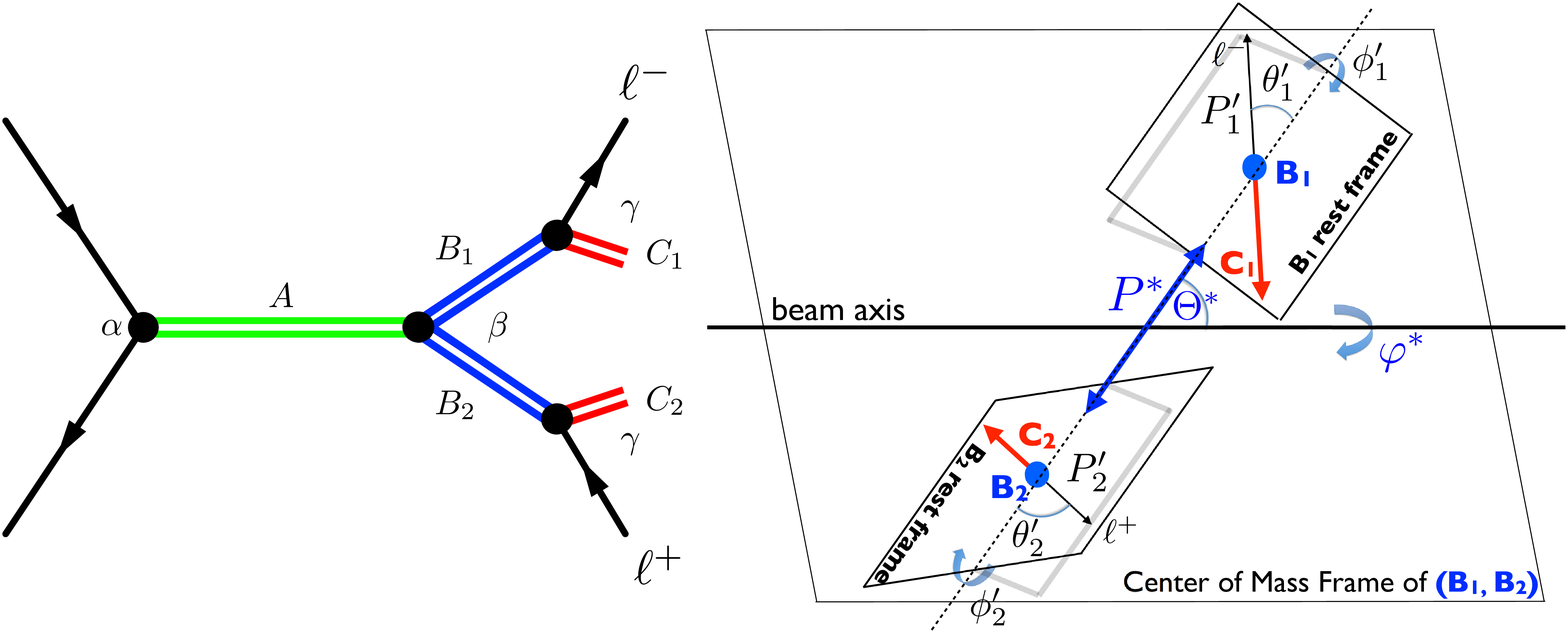} 
}
\caption{Left: the antler event topology under consideration in this paper. $B_i$ and $C_i$ are new $Z_2$-odd
particles, whose spins (as well as the chirality $\beta$ and $\gamma$ of their couplings) are a priori unknown.
The intermediate resonance $A$ is a $Z_2$-even particle which may or may not be on-shell, and furthermore,
its spin (and the chirality $\alpha$ of its coupling to the initial state partons) is also 
unknown. Right: pictorial definition of the kinematic variables used in this paper. Starred quantities refer to the
CM frame of the $B_1B_2$ system while primed variables refer to the individual rest frames of the parent particles $B_i$. }
\label{fig:top}
}
In general, the $s$-channel particle $A$ in Fig.~\ref{fig:top} can be a SM gauge boson (a photon or a $Z$-boson), a SM Higgs 
boson, or some other new bosonic particle. Depending on the relative masses of particles $A$ and $B$, the 
$s$-channel resonance $A$ could be on-shell (if $M_A>2M_B$) or off-shell (if $M_A<2M_B$). In the following 
we shall discuss both of those cases. 

In summary, as illustrated in Fig.~\ref{fig:top}, the process under study here will be
\beq
pp\to A \to B_1 B_2 \to \ell^+\ell^-  C_1  C_2.
\label{theantler}
\eeq
An important benefit of considering the leptonic (i.e.~$c=\ell$) version of the antler topology 
is that it does not suffer from large QCD backgrounds (as would be the case if $c$ were a jet).
Another advantage is that it allows us to safely focus on the $s$-channel topology only, where we only 
need to introduce the very minimum number of unknown parameters (see Section~\ref{sec:antler} below)
and tackle the problem of spin measurements in a rather model independent way.

In this paper, we shall analyze the kinematics of the antler event topology (\ref{theantler}) with several goals in mind:
\begin{itemize}
\item First, we shall perform a comparative study of the several kinematic variables already discussed in the literature,
supplementing them with a few variables of our own (see Sec.~\ref{sec:variables}). We shall be mostly interested in the question to what extent 
the shapes of the different kinematic distributions are sensitive to spin effects. The main objective of this initial study 
(presented below in Section~\ref{sec:comparison}) will be to identify the kinematic variables which retain the largest amount of
spin information, and are thus most suitable for measuring the spins of particles $A$, $B$ and $C$.
\item Next, we shall address the question how to infer the presence of spin effects in the distribution of any given kinematic variable.
Clearly, in order to firmly establish the presence of spin effects, one must first know the expected shape of the distribution
{\em in the absence} of any spin correlations, i.e. in the pure ``phase space approximation". 
Then, any deviation from the expected ``phase space" distribution will indicate spin effects, although 
a priori it will not be immediately clear whether those are due to non-trivial intrinsic spins of the particles $\{A,B,C\}$,
non-trivial coupling chiralities, or some combination of the two \cite{Burns:2008cp}. 
In Section~\ref{sec:phasespace} we therefore provide the expected ``phase space" distributions for 
several kinematic variables, including the 
angular variable $\cos{\theta_{\ellm\ellp}^\ast}$ proposed in \cite{Barr:2005dz} (see Sec.~\ref{sec:Barr} below for its exact definition).
We shall pay special attention to the dependence of the kinematic variables on the total CM energy $\sqrt{s}$ in the event,
and in Section~\ref{sec:BarrPhase} we shall derive a new formula for the shape of the $\cos{\theta_{\ellm\ellp}^\ast}$ 
distribution at threshold ($\sqrt{s}=2M_B$). 
\item Having observed a deviation from the expected ``pure phase space" kinematic shape, one can be sure that non-trivial 
spin effects are in play. But how much of the effect is due to the spins and how much is due to the 
chirality structure of the couplings? This question is tackled next in Section~\ref{sec:Chiralities}.
In order to explicitly separate the two effects, we derive analytically the $\cos{\theta_{\ellm\ellp}^\ast}$ distributions 
at threshold for various spin assignments and arbitrary chiralities. 
These results provide useful intuition and guidance for the study which comes next --- the measurement of the spins of particles $A$, $B$ and $C$.
\item Section~\ref{sec:discrimination} contrasts the shapes of the $\cos{\theta_{\ellm\ellp}^\ast}$ distributions obtained for 
various spin and chirality assignments. The ultimate goal there is to identify which spin combinations 
for the particles $\{A,B,C\}$ can be distinguished (at least in principle)
in such a model-independent way (i.e.~without any additional assumptions about the chirality of the fermion couplings)
and which can be confused with each other. We shall specify the dangerous ``twin" spin scenarios with 
similar $\cos{\theta_{\ellm\ellp}^\ast}$ distributions and discuss the possibility to discriminate them through the distributions of 
other, complementary kinematic variables.
\end{itemize}

Before taking up these four goals, we first introduce our setup and notation in Section~\ref{sec:antler}. 
 
\section{Setup, notation and conventions}
\label{sec:antler}

In this paper we study the process (\ref{theantler}) 
depicted in Fig.~\ref{fig:top}. The two particles $C_1$ and $C_2$ are invisible in the detector, while 
the two massless leptons $\ellp$ and $\ellm$ are visible. 
The double lines in the antler event diagram of Fig.~\ref{fig:top} denote particles whose spin is unknown and needs to be determined. 
The masses of the particles $A$, $B_i$ and $C_i$ are also initially unknown, 
but can eventually be determined from purely kinematic methods \cite{Gripaios:2007is,Barr:2007hy,Burns:2008va,Barr:2009jv,Han:2009ss,Matchev:2009fh,Matchev:2009ad,Konar:2009wn,Konar:2009qr,Cohen:2010wv,Cho:2010vz,Park:2011uz,Cheng:2011ya},
in spite of the challenge presented by the short decay chain.
In what follows, we shall therefore assume that the masses $M_A$, $M_B$ and $M_C$ of the new particles 
have already been measured and we concentrate on the issue of spin determination alone.

In order to perform a model independent spin measurement, we will need to separate 
pure kinematic effects (see Section~\ref{sec:kinematics}) from spin and chirality 
variables (see Sections~\ref{sec:spin} and \ref{sec:chirality}).

\subsection{Kinematics}
\label{sec:kinematics}

In what follows, we shall find it convenient to describe the kinematics of the process
(\ref{theantler}) in terms of variables which refer to three different reference frames
(two of those are illustrated in Fig.~\ref{fig:top}):
\begin{enumerate}
\item{\bf The LAB frame.} This is the frame of the experiment, in which the visible particles are measured.
Variables related to the LAB frame will carry no special designations, e.g. the 4-momenta of the two 
visible particles (the leptons) will simply be
\bea
P^\mu_{\ell^-}&\equiv& P_1^\mu = \left( E_1,\vec{P}_{1T}, P_{1z}\right) = \left( E_1, P_1, \Theta_1, \varphi_1 \right) ;  \label{eq:Plep1}\\
P^\mu_{\ell^+}&\equiv& P_2^\mu = \left( E_2,\vec{P}_{2T}, P_{2z}\right) = \left( E_2, P_2, \Theta_2, \varphi_2 \right),  \label{eq:Plep2}
\eea
where $\Theta_i$ and $\varphi_i$ are the usual spherical angular coordinates.
Note that the angles $\Theta_i$ are measured from the beam axis ($z$-axis)
and $\varphi_i$ are measured in the fixed plane (common to all events) transverse to the beam.
Since the leptons are assumed to be massless, we also have $P_i=E_i$.
The LAB variables are {\em not} shown in Fig.~\ref{fig:top}.
\item{\bf CMBB: the CM frame of the parent pair $B_1B_2$.}  This is the relevant frame 
for describing the $2\to2$ process $pp\to B_1B_2$ and the corresponding kinematic variables 
will be denoted with an asterisk (Fig.~\ref{fig:top}). The final state of
the $B_1B_2$ pair is described by three degrees of freedom, which can be taken to be:
the CM energy $\sqrt{s^\ast}$ and the spherical angular coordinates of particle $B_1$, which are  
$\Theta^\ast$ (still measured from the beam axis) and $\varphi^\ast$. 
Since $\sqrt{s^\ast}$ and $\varphi^\ast$ are invariant under longitudinal boosts,
they are the same in the LAB frame:
\bea
\sqrt{s^\ast}&=&\sqrt{s},  \\
\varphi^\ast&=&\varphi,
\eea
so that in what follows for simplicity we shall omit the asterisks on $\sqrt{s}$ and $\varphi$.
Instead of $\sqrt{s}$, one could equivalently consider the (magnitude of the) momentum $P^\ast$
of the $B_i$ parent particle
\beq
P^\ast = \frac{\sqrt{s}}{2}\left(1-\frac{4 M_B^2}{s}\right)^{\frac{1}{2}}\, ,
\label{eq:PB}
\eeq
or the corresponding boost factor $\eta^\ast$
\beq
\eta^\ast=\cosh^{\minus 1}\left({\frac{\sqrt{s}}{2 M_B}}\right)=\sinh^{\minus 1}\left(\frac{P^\ast}{M_B}\right) \, .
\label{eq:eta}
\eeq
The main variables $P^\ast$, $\Theta^\ast$ and $\varphi^\ast$ of the CMBB frame are depicted in Fig.~\ref{fig:top} in blue.
\item{\bf CMB1 (CMB2): the CM frame of an individual parent $B_i$.}  The kinematics of each final state lepton 
is most easily described in the rest frame of its parent particle, where variables will be 
denoted by a ``prime'' superscript  (Fig.~\ref{fig:top}) and will also carry a subscript ``1" or ``2"
to indicate the CM frame of $B_1$ CMB1 or the CM frame of $B_2$ CMB2.
The three degrees of freedom describing the kinematics of each lepton can be taken to be: 
the magnitude of the momentum $P_i'$, the polar angle $\theta_i'$ and the azimuthal angle $\phi_i' $.
Note that, as shown in Fig.~\ref{fig:top}, the polar angles $\theta_i'$ are measured with respect to the direction of the parent particle
$P_i$ and not the beam axis, as was the case for the angle $\Theta$. Similarly, the azimuthal angles
$\phi_i'$ are measured differently from the azimuthal angle $\varphi$ in the CMBB frame.
The momentum of each lepton in the corresponding parent CM frame (CMB1 or CMB2) is a constant which
only depends on the mass spectrum:
\beq
P_\ell ' \equiv P_1' = P_2' = \frac{M_B}{2}\left(1-\frac{M_C^2}{M_B^2}\right)\, .
\label{eq:Piprime}
\eeq
In place of the polar angle $\theta_i'$ one could alternatively use the 
corresponding pseudorapidity
\beq
\eta_i'=   -\ln\left[\tanh\left(\frac{\theta_i'}{2}\right)\right]\, .
\label{eq:etaprime}
\eeq
The frame CMB1 (CMB2) is obtained from the frame CMBB by boosting with a Lorentz factor $\gamma=\cosh\eta^\ast$
along the direction of $B_1$ ($B_2$).
\end{enumerate}
 
\subsection{Spin assignments}
\label{sec:spin}

Since the particles $A$, $B$ and $C$ are not directly measured, their spins are 
a priori unknown, and each can be a spin zero scalar (S), 
a spin $\frac{1}{2}$ fermion (F), a spin 1 vector particle (V), etc. 
Following \cite{Athanasiou:2006ef,Burns:2008cp}, we shall consider all
allowed spin assignments
\beq
(\mathrm{ABC}) \in \biggl\{(\mathrm{SSF}),(\mathrm{SVF}),(\mathrm{VSF}),(\mathrm{VVF}),(\mathrm{SFS}),(\mathrm{VFS}),(\mathrm{SFV}),(\mathrm{VFV})\biggr\}
\label{eq:3spins}
\eeq
in the antler event topology of Fig.~\ref{fig:top} with spins up to one, i.e.~consistent with 
renormalizable interactions.
The Feynman diagrams for these 8 cases are listed in
Table~\ref{tab:models}, together with a representative example from a 
popular new physics model.

For the underlying $2\rightarrow 2$ process $pp\to BB$, there are in principle
6 possibilities for the spins of the $(\mathrm{AB})$ pair:
\beq
(\mathrm{AB})\in \biggl\{\mathrm{(SS),(SV),(VS),(VV),(SF),(VF)}\biggr\}
\label{eq:2spins}
\eeq
which will be discussed in Sections~\ref{sec:infiniteS} and \ref{sec:offshell}.
 
\TABLE[t]{
\begin{tabular}{|c|c|c|}
\hline
Spin assignments		 & Antler diagram         &  Examples       \\ \hline
\raisebox{0.6cm}{SSF}	&	\includegraphics[height=1.5cm]{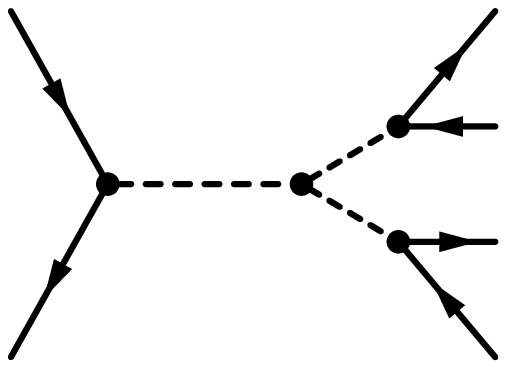} & \raisebox{0.6cm}{$H_2^0 \to H_1^+ H_1^- \to \ell^+\ell^- \nu_1 \bar{\nu}_1$}
\\	\hline
\raisebox{0.6cm}{SVF}		&  \includegraphics[height=1.5cm]{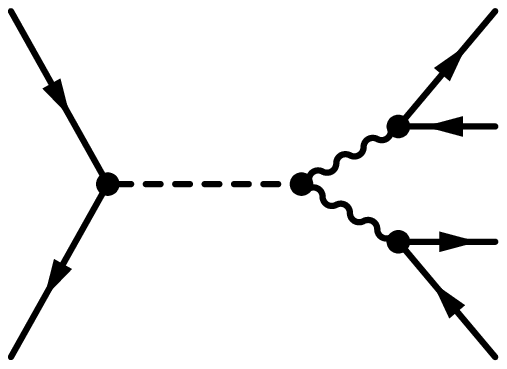}&\raisebox{0.6cm}{ $H^0\to W^+W^-\to  \ell^+\ell^- \nu \bar{\nu}$	}
\\  \hline
\raisebox{0.6cm}{SFS}		&  \includegraphics[height=1.5cm]{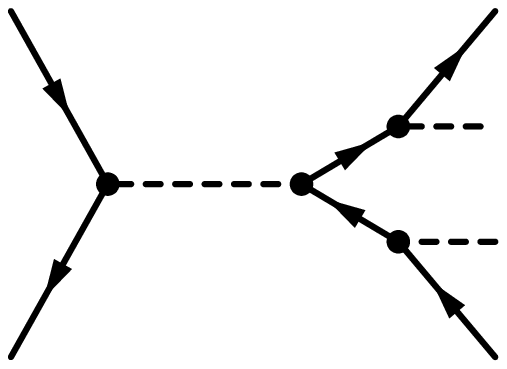}&\raisebox{0.6cm}{ $H_2\to \ell^+_1 \ell^-_1 \to \ell^+\ell^-B_{H1} B_{H1}$	\cite{Chang:2011vt} }
\\  \hline
\raisebox{0.6cm}{SFV}		&  \includegraphics[height=1.5cm]{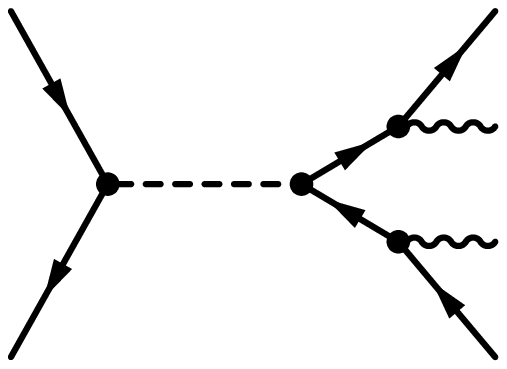}&\raisebox{0.6cm}{ $H_2\to \ell^+_1 \ell^-_1 \to \ell^+\ell^-B_1 B_1$ 	\cite{Chang:2011vt} }
\\  \hline
\hline
\raisebox{0.6cm}{VSF}		&  \includegraphics[height=1.5cm]{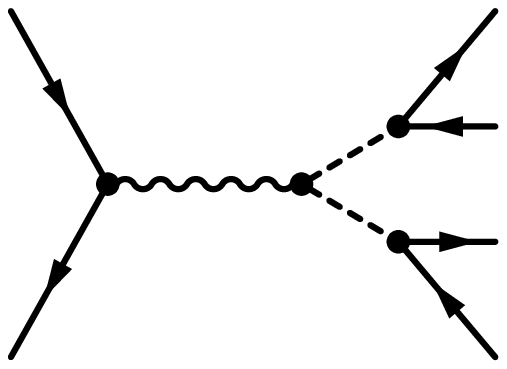}&\raisebox{0.6cm}{$Z'\to \tilde\ell^+\tilde\ell^- \to \ell^+\ell^-\tilde\chi^0_1\tilde\chi^0_1$ \cite{Baumgart:2006pa,Han:2009ss}	}
\\  \hline
\raisebox{0.6cm}{VVF}	         &  \includegraphics[height=1.5cm]{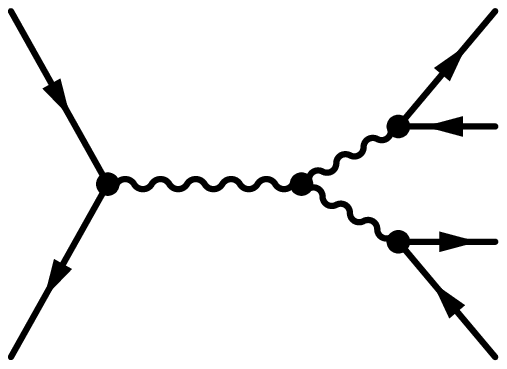}& \raisebox{0.6cm}{ $Z'\to W^+ W^- \to \ell^+\ell^- \nu \bar{\nu}$ \cite{Eboli:2011bq}  }			
                                          \\ \hline
\raisebox{0.6cm}{VFS}		&  \includegraphics[height=1.5cm]{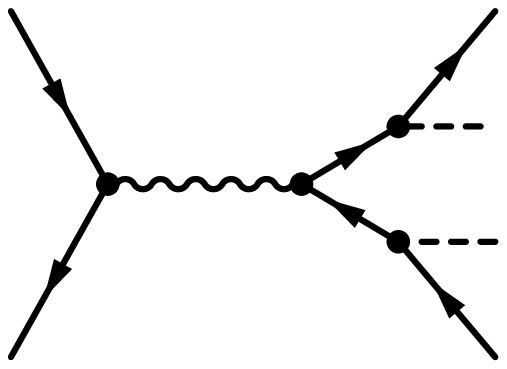}& \raisebox{0.6cm}{  $Z_2\to \ell^+_1 \ell^-_1 \to \ell^+\ell^-B_{H1} B_{H1}$   \cite{Burdman:2006gy,Dobrescu:2007xf} }
	\\ \hline
\raisebox{0.6cm}{VFV}	         &  \includegraphics[height=1.5cm]{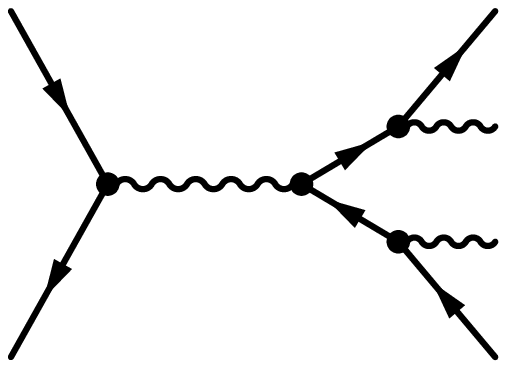}&  \raisebox{0.6cm}{$Z_2\to \ell^+_1 \ell^-_1 \to \ell^+\ell^-B_1 B_1$   \cite{Cheng:2002ab,Datta:2005zs}  } 			
\\ \hline
\end{tabular}
\caption{The 8 different spin assignments for the unknown particles $A$, $B$ and $C$ in the antler topology of Fig.~\ref{fig:top}.
The last column lists a few typical examples from popular theoretical models, where supersymmetric particles are denoted by a tilde,
while a subscript $n$ indicates a level $n$ Kaluza-Klein mode in theories with Universal Extra Dimensions. }
\label{tab:models}
}

\subsection{Coupling assignments}
\label{sec:chirality}

As we shall see below, the shapes of the kinematic distributions are affected not just by the spins of the intermediate particles, 
but also by the chiralities of their couplings. Since the observed particles (leptons) are fermions, the corresponding vertices of the
antler diagram may contain chiral projectors. In the spirit of Refs.~\cite{Burns:2008cp,Chen:2011cya}, in each vertex of the
antler diagram, we allow for an arbitrary linear combination of right-handed and left-handed couplings. 
For example, when $C$ is a fermion and $B$ is a vector boson, the $B\ell C$ coupling is given by
\beq
\mathcal{L}_{int} = \bar{\psi}_l\gamma_\mu\left(c_LP_L+c_RP_R\right)\psi_C A_B^\mu,
\eeq
with arbitrary coefficients $c_L$ and $c_R$.  As indicated in Fig.~\ref{fig:top}, 
the relative chirality of that vertex will then be parameterized in terms of a single parameter
\beq
\gamma=\left( \frac{c_L^2-c_R^2}{c_L^2+c_R^2} \right) \, ,
\label{gammadef}
\eeq
so that $\gamma=0$ corresponds to a vector-like coupling, while $\gamma=\pm 1$ 
is a purely chiral coupling.

One can extend this technology to the other two vertices as well \cite{Burns:2008cp}
and define the relative chirality of the $AB_1B_2$ vertex as
\beq
\beta=\left( \frac{b_L^2-b_R^2}{b_L^2+b_R^2} \right) \, ,
\label{betadef}
\eeq
and the relative chirality of the $q\bar{q}A$ vertex as
\beq
\alpha=\left( \frac{a_L^2-a_R^2}{a_L^2+a_R^2} \right) \, .
\label{alphadef}
\eeq
Our notation is chosen so that the chiral coefficients 
$a_L$, $a_R$ and the relative chirality $\alpha$ refer to the production vertex $q\bar{q}A$ of particle $A$; 
$b_L$, $b_R$ and the relative chirality\footnote{Of course, the definition of $\beta$ does not apply when $A$ and $B$ are both bosons.} 
$\beta $ refer to the production vertex $AB_1B_2$ of particle $B$, while
$c_L$, $c_R$ and $\gamma$ refer to the production vertex $B\ell C$ of particle $C$.
 
\TABULAR[t]{cccc}{
\hline\hline
Pos. &Vertex	&	Lagrangian term &	Vertex	\\ \hline
$\alpha:$&	\includegraphics[width=1cm]{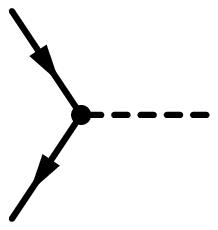}	&	$ a \bar{\psi}_q \left(P_L+e^{i\delta_a}P_R\right)\psi_q$	&	$a\left(P_L+e^{i\delta_a}P_R\right) $		\\ 
\raisebox{1.5ex}{\includegraphics[width=1cm]{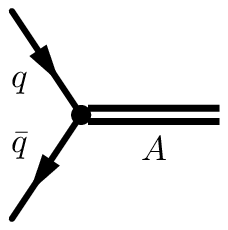}}	&	\includegraphics[width=1cm]{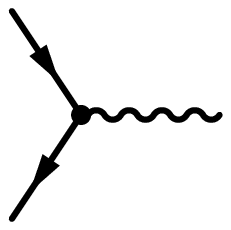}	& 	$  \bar{\psi}_q\gamma_\mu\left(a_L P_L+a_R P_R\right)\psi_q A_A^\mu$	&	$ \gamma_\mu\left(a_LP_L+a_RP_R\right)$		\\ \hline
$\beta	:$&	\includegraphics[width=1cm]{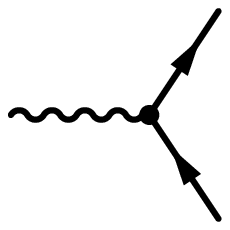}	&	$\bar{\psi}_B\gamma_\mu\left(b_L P_L+b_R P_R\right)\psi_B A_A^\mu $	&	$ \gamma_\mu\left(b_L P_L+b_R P_R\right)$		\\ 
	&	\includegraphics[width=1cm]{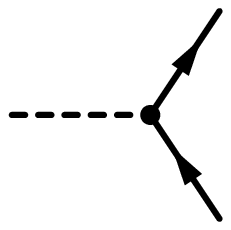}	&$ b\, \bar{\psi}_B \left(P_L+e^{i\delta_b}P_R\right)\psi_B \phi_A$	&	$b\left(P_L+e^{i\delta_b}P_R\right) $		\\ 
	&	\includegraphics[width=1cm]{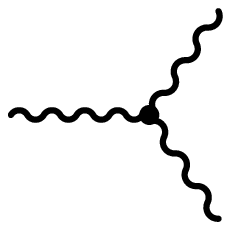}	&	$ F_{\mu\nu} F^{\mu\nu}$
		&	$b\,\left((p-k)^\rho g^{\mu\nu}+(q-p)^\nu g^{\rho\mu}+(k-q)^\mu g^{\rho\nu}\right) $		\\ 
	\raisebox{3.5ex}{\includegraphics[width=1cm]{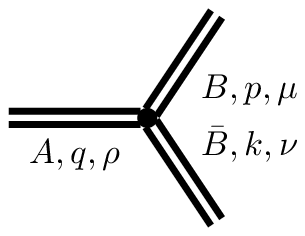}}&	\includegraphics[width=1cm]{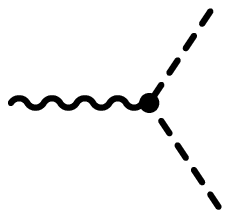}	&	$(D_\mu\phi_B)^\dagger(D^\mu \phi_B)  $	&	$b\,(p-k)^\rho $		\\ 
	&	\includegraphics[width=1cm]{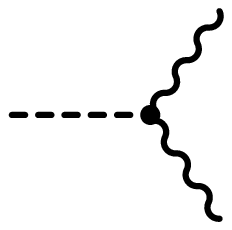}	&	$ b\, A^\mu_B A_{\mu,B}\phi_A$	&	$b\, g^{\mu\nu} $		\\ 
	&	\includegraphics[width=1cm]{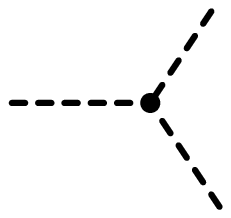}	&	$b\, \phi_B\phi_B^\dagger \phi_A$	&	$b $		\\ \hline
$\gamma:$&	\includegraphics[width=1cm]{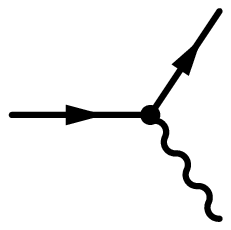}	&	$ \bar{\psi}_l\gamma_\mu\left(c_LP_L+c_RP_R\right)\psi_B A_C^\mu+h.c.$	&	$ \gamma_\mu\left(c_LP_L+c_RP_R\right) $		\\ 
	&	\includegraphics[width=1cm]{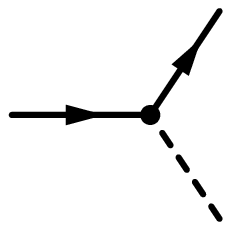}	&	$\bar{\psi}_l\left(c_LP_L+c_RP_R\right)\psi_B \phi_C+h.c. $	&	$ \left(c_LP_L+c_RP_R\right)$		\\ 
\raisebox{2.5ex}{\includegraphics[width=1cm]{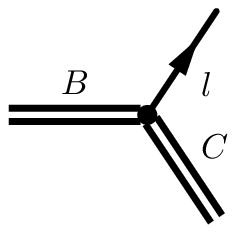}}	&	\includegraphics[width=1cm]{fig/decay3}	&	$ \bar{\psi}_l\gamma_\mu\left(c_LP_L+c_RP_R\right)\psi_C A_B^\mu+h.c.$	&	$ \gamma_\mu\left(c_LP_L+c_RP_R\right)$		\\ 
	&	\includegraphics[width=1cm]{fig/decay4}	&	$ \bar{\psi}_l\left(c_LP_L+c_RP_R\right)\psi_C \phi_B+h.c.$	&	$\left(c_LP_L+c_RP_R\right) $		\\ \hline\hline
}{Summary of the Lagrangian terms and the corresponding Feynman rules for the vertices in the antler diagram. 
\label{tab:nomenclature}}

Table~\ref{tab:nomenclature} lists the explicit form of the coupling in each vertex for all possible spin assignments for $A$, $B$ and $C$
which we are considering.
 
 \section{Kinematic variables for the antler topology}
\label{sec:variables}

\TABULAR[ht]{|c||c|c||c|c||}{
\hline
Particle & \multicolumn{2}{|c||}{SUSY-like} &  \multicolumn{2}{c||}{UED-like} \\ \hline\hline
$A$      & $\gamma^\ast/Z^\ast$    & -        & $\gamma^\ast/Z^\ast$ & - \\ \hline
$B$     & $\tilde{\ell}_R$     & $176.62$      & $\ell_1 $ & $505.45$ \\ \hline
$C$     & $\tilde{\chi}_1^0$ & $158.18$     & $\gamma_{1}$ & $500.89$ \\ \hline
}{\label{table:mass} Mass spectrum in GeV for the two study points considered in the text. 
The SUSY-like mass spectrum corresponds to the CMS LM6 study point \cite{Ball:2007zza}, while
the UED-like spectrum comes from the Minimal UED model with 
$R^{-1}=500$ GeV, $\lambda R=20$ and $m_h=120$ GeV \cite{Battaglia:2005zf,Smillie:2005ar,Datta:2005zs}.}
At a hadron collider like the LHC, the energy and momentum of the initial state partons 
vary from one event to another. The lack of knowledge about the kinematics of the 
initial state prevents us from computing the momenta of the missing particles $C_i$ 
in the simple antler topology.
(At a linear collider, where the energy and momentum of the initial state are known, 
the missing momenta in the antler topology can be reconstructed only up to a two-fold 
ambiguity\cite{Buckley:2007th}, even assuming precise knowledge of the masses of the intermediate particles).
Thus we are limited to variables constructed out of the measured 4-momenta $P_{i}^\mu$ 
(defined in eqs.~(\ref{eq:Plep1}-\ref{eq:Plep2})) of the visible particles $\ell^{\pm}$ in the LAB frame.
In this section, we review a number of such variables which have been discussed in the literature.
In order to gauge their sensitivity to spin effects in the antler topology, we shall investigate 
their distributions for two conventional study points, whose mass spectra are listed in Table~\ref{table:mass}. 
For each of these two mass spectra, we shall consider two spin configurations:
$\mathrm{VSF}$ for SUSY and $\mathrm{VFV}$ for UED, postponing the comprehensive analysis including
the remaining spin combinations until Sections~\ref{sec:Chiralities} and \ref{sec:discrimination}. 
The resulting four cases will be contrasted in Fig.~\ref{fig:var}, where we shall plot the unit normalized distributions for
all variables introduced in this section.

\subsection{The relative pseudorapidity variable $\cos{\theta_{\ellm\ellp}^\ast}$}
\label{sec:Barr}

In \cite{Barr:2005dz} A.~Barr introduced an angular variable $\theta_{\ellm\ellp}^\ast$
which is related to the relative pseudorapidity of the visible particles in the LAB frame as
\bea
\BAR \equiv \cos{\theta_{\ellm\ellp}^\ast} &=& \tanh{\left(\frac{\Delta{\eta}_{\ellm\ellp}}{2}\right)}, \label{eq:barangle}  \\
\Delta\eta_{\ellm\ellp} &\equiv& \eta_{\ellm}-\eta_{\ellp} = \eta_1-\eta_2.
\eea
For brevity, from now on we shall use $\BAR$ to denote $\cos{\theta_{\ellm\ellp}^\ast}$. 
The pseudorapidities of the two visible particles in the LAB frame are defined as
\beq
\eta_i \equiv \frac{1}{2} \ln{\left(\frac{P_i+P_{i\, z}}{P_i-P_{i\, z}}\right)}
\label{eq:rapidity}
\eeq
in terms of the measured momenta $P_i$ and $P_{iz}$ from eqs.~(\ref{eq:Plep1}-\ref{eq:Plep2}).

\subsection{Azimuthal angular difference $\Delta \varphi$}
\label{sec:dphi}

The azimuthal angular difference of the visible particles
\beq
\Delta \varphi \equiv  \cos^{\minus 1}\left( \frac{\vec{P}_{1T}\cdot \vec{P}_{2T}}{P_{1T}P_{2T}} \right)  \eeq
has also been shown to be sensitive to spin effects \cite{Buckley:2010jv,Chen:2010ek} and we shall 
include it among our set of observables here as well.

\subsection{Invariant mass $M_{\ellp \ellm}$}
\label{sec:mll}

One can also consider the invariant mass of the two visible particles in the antler topology
\beq
M_{\ellp \ellm} \equiv  \sqrt{\left( P_1+P_2\right)^{\mu} \left(P_1+P_2\right)_{\mu}}.
\label{Mlldef}
\eeq
The shape of the $M_{\ellp \ellm}$ distribution was studied in detail in \cite{Han:2009ss}
for the case of resonant production ($M_A>2M_B$). It was shown that
the $M_{\ellp \ellm}$ distribution exhibits a point (called a ``cusp"), 
where it is not continuously differentiable. If the mass spectrum is such that 
\beq
M_A \le (2 M_B) \cosh\left(\frac{1}{2}\right) \approx \left(2.26\right) M_B, 
\eeq
then the cusp appears at the very peak of the distribution. On the other hand, when
$M_A>2.26 M_B$, the peak is smooth and the cusp appears somewhere below the peak \cite{Han:2009ss}. 

In what follows, we shall prefer to work with dimensionless variables, so we need to rescale 
our mass variables by a suitable mass constant. Since the mass spectrum of the antler topology
is assumed to be already known, the momentum $P_\ell'$ of each lepton in its mother's rest frame
(\ref{eq:Piprime}) is also known and we can use it for normalization. We therefore define the
rescaled invariant mass\footnote{Rescaled quantities will be denoted with a hat.} as
\beq
\hat M_{\ellp \ellm} = \frac{M_{\ellp \ellm}}{2 P_\ell'}\, .
\label{Mllrescaled}
\eeq

\subsection{Contransverse mass $M_{CT}$}

The contransverse mass variable 
\beq
M_{CT}=\sqrt{2\left(P_{1T} P_{2T}+\vec{P}_{1T} \cdot \vec{P}_{2T}\right)}
\label{MCTdef}
\eeq
proposed in \cite{Tovey:2008ui} is also suitable for the antler topology. 
It is invariant under longitudinal boosts from the LAB frame to the CMBB frame,
and under ``back to back" boosts of the parent particles.
The $M_{CT}$ distribution has an upper kinematic endpoint
\beq
M_{CT}^{(max)} = 2P_\ell' = M_B\left(1-\frac{M_C^2}{M_B^2}\right)\, ,
\label{MCTmax}
\eeq
which can be used for mass measurements \cite{Tovey:2008ui}.
Here we shall focus on the shape of the distribution of the corresponding
rescaled $M_{CT}$ variable
\beq
\hat M_{CT}=  \frac{M_{CT}}{2 P_\ell'}\, .
\label{MCThat}
\eeq
Eqs.~(\ref{MCTmax}) and (\ref{MCThat}) imply that the endpoint of the ${\hat M}_{CT}$ distribution 
is always at $\hat M_{CT}=1$ and this will be evident in the plots below.

\subsection{Doubly projected contransverse mass $M_{CTx}$}

The doubly projected contransverse mass variable $M_{CT_\perp}$ was proposed in order to remove
the upstream boost effect from initial state radiation (ISR) \cite{Matchev:2009ad}. 
$M_{CT_\perp}$ is the one-dimensional analogue of (\ref{MCTdef}) which uses the
transverse momenta components which are orthogonal (hence the subscript ``$\perp$") 
to the upstream $\vec{P}_T$ which can be caused by ISR or previous decays. 
Here we shall not consider ISR effects, so 
we are free to choose any one of the two transverse axes to define the ``$\perp$" components, e.g.
\beq
M_{CTx} = \sqrt{2 \left(|P_{1x}P_{2x}|+ P_{1x} P_{2x}\right)}\, .
\eeq
The corresponding rescaled variable is
\beq
\hat M_{CTx} = \frac{M_{CTx} }{2 P_\ell'}
\eeq
and its endpoint is also at $\hat M_{CTx}=1$. For our purposes, $\hat M_{CTx}$ 
has an important advantage --- the analytical formula for its differential distribution 
(in the absence of spin correlations) is already known \cite{Matchev:2009ad},
providing an important benchmark for spin studies (see Sec.~\ref{sec:mctx} below).

\subsection{Effective mass $M_{\mathrm{eff}}$} 

The effective mass variable $M_{\mathrm{eff}}$ \cite{Tovey:2000wk} has been widely utilized in SUSY searches. 
When applied to the antler topology, it reads
\beq
M_{\mathrm{eff}} =  P_{1T} + P_{2T}+\met \, .
\eeq
Here we use the rescaled variable
\beq
\hat M_{\mathrm{eff}} = \frac{M_{\mathrm{eff}}}{ 4 P_\ell'}\, .
\eeq
Many SUSY analyses use $M_{\mathrm{eff}}$ as one of their selection cuts, 
since it is correlated with the mass scale of the particles produced in the event.

\subsection{The $\smin$ variable} 

The $\smin$ variable was advertized in \cite{Konar:2008ei,Konar:2010ma} as a better estimator of the 
mass scale of the hard scattering, since it explicitly accounts for the masses of the invisible particles, 
and is defined in a theoretically rigorous way. 
$\smin$ is the minimum value of $\sqrt{s}$ which is required in order to account for
the observed visible particles and the measured $\met$ in the event. In the case of the antler
event topology, the general formula for $\smin$ from \cite{Konar:2008ei} reduces to
\beq
\smin = \sqrt{M_{\ellp \ellm}^2
+|\vec{P}_{1T}+\vec{P}_{2T}|^2}+\sqrt{4M_C^2+\met^2}\, ,
\eeq
already accounting for the fact that the combined mass of all invisible particles present in the event 
(in this case the two $C$'s) is $2M_C$.
It was noted in \cite{Konar:2008ei} that for typical SUSY events the $\smin$ distribution peaks near
the mass threshold ($2M_B$) for production of the two parents. This is why we choose to 
rescale the $\smin$ variable by $2M_B$ and define
\beq
\shatmin = \frac{\smin}{2 M_B}\, ,
\eeq
which we expect to peak near $\shatmin=1$, as confirmed in Fig.~\ref{fig:var}.

We anticipate that the $\smin$ variable will be useful for spin studies, particularly in the
off-shell scenario when $M_A<2M_B$. In that case, the two $B$ particles are produced near threshold,
and the threshold suppression of the $\sqrt{s}$ distribution will be sensitive to the spins of the $A$ and $B$ particles
(see Sec.~\ref{sec:offshell} below). Since $\smin$ is designed to correlate with the 
actual $\sqrt{s}$ in the event, any spin effects present in the $\sqrt{s}$ distribution will 
be inherited to some extent by the $\smin$ distribution as well.

\subsection{Lepton energy $E_\ell$} 

The lepton energy $E_i$ is a classic variable for kinematic studies at a linear collider. Here we revisit this variable
for the case of a hadron collider, rescaling as usual
\beq
\hat E_i = \frac{E_{i}}{ P_\ell'}\, .
\label{Ehat}
\eeq
In what follows we shall not distinguish the two lepton energies $E_1$ and $E_2$ 
and will always show them together on the same plot as $E_\ell$.

\subsection{The Cambridge $M_{T2}$ variable}  

The Cambridge $M_{T2}$ variable was originally proposed \cite{Lester:1999tx} for precisely 
the antler topology case of Fig.~\ref{fig:top}. Since then, it has been extensively applied
for measuring the mass spectrum of the new particles $A$, $B$ and $C$. 
In the case of the antler topology, $M_{T2}$ is given by \cite{Cho:2007dh,Konar:2009wn}
\bea
M_{T2}(M_C) &=& \sqrt{A_T} + \sqrt{A_T+M_C^2},   \label{MT2def} \\
 A_T     &=& \frac{1}{2}\left(P_{1T}P_{2T}+\vec{P}_{1T}\cdot \vec{P}_{2T}\right).   \label{ATdef}
\eea
Just like $\smin$, $M_{T2}$ requires an input mass for the missing particle, which
in eq.~(\ref{MT2def}) has been taken to be the correct value $M_C$.
In that case, the $M_{T2}$ endpoint is equal to the parent mass $M_B$: 
\beq
M_{T2}^{(max)}(M_C) \equiv \max \left\{M_{T2}(M_C)\right\} = M_B\, . 
\eeq
This motivates us to rescale the $M_{T2}$ variable as
\beq
\hat M_{T2} = \frac{M_{T2}(M_C)-M_C}{M_B-M_C}\, ,
\eeq
so that the distribution of $\hat M_{T2}$ ranges from $0$ to $1$.

\section{Comparison of different variables}
\label{sec:comparison}

Having defined the nine kinematic variables of interest in the previous section, 
we shall now compare their sensitivity to spin effects in the antler topology.
In order to get some preliminary idea about this, in Fig.~\ref{fig:var} we
show the distributions of our nine variables in the four cases, obtained by pairing
the two mass spectra from Table~\ref{table:mass} (SUSY-like or UED-like) with the corresponding
spin assignments (VSF for the MSSM and VFV for MUED).
If a particular distribution is affected by spins, one would expect 
a visible difference between the two dotted lines, showing the two different 
spin scenarios for the same SUSY-like mass spectrum. Similarly, 
one would also expect a difference between the two solid lines, which 
show the two different spin scenarios for a UED-like mass spectrum.

\FIGURE[t]{
\centerline{
\epsfig{file=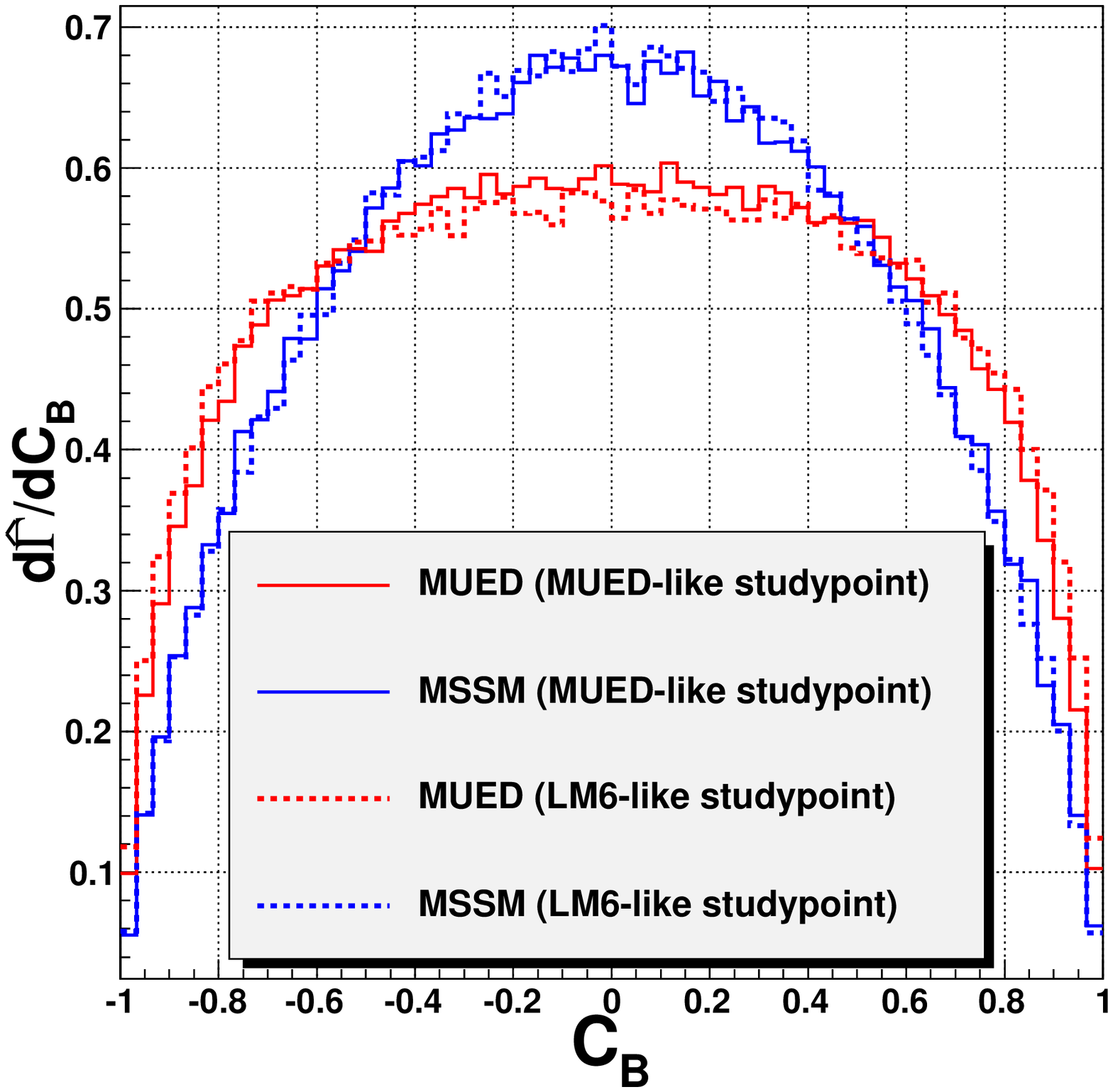,width=5.cm} 
\epsfig{file=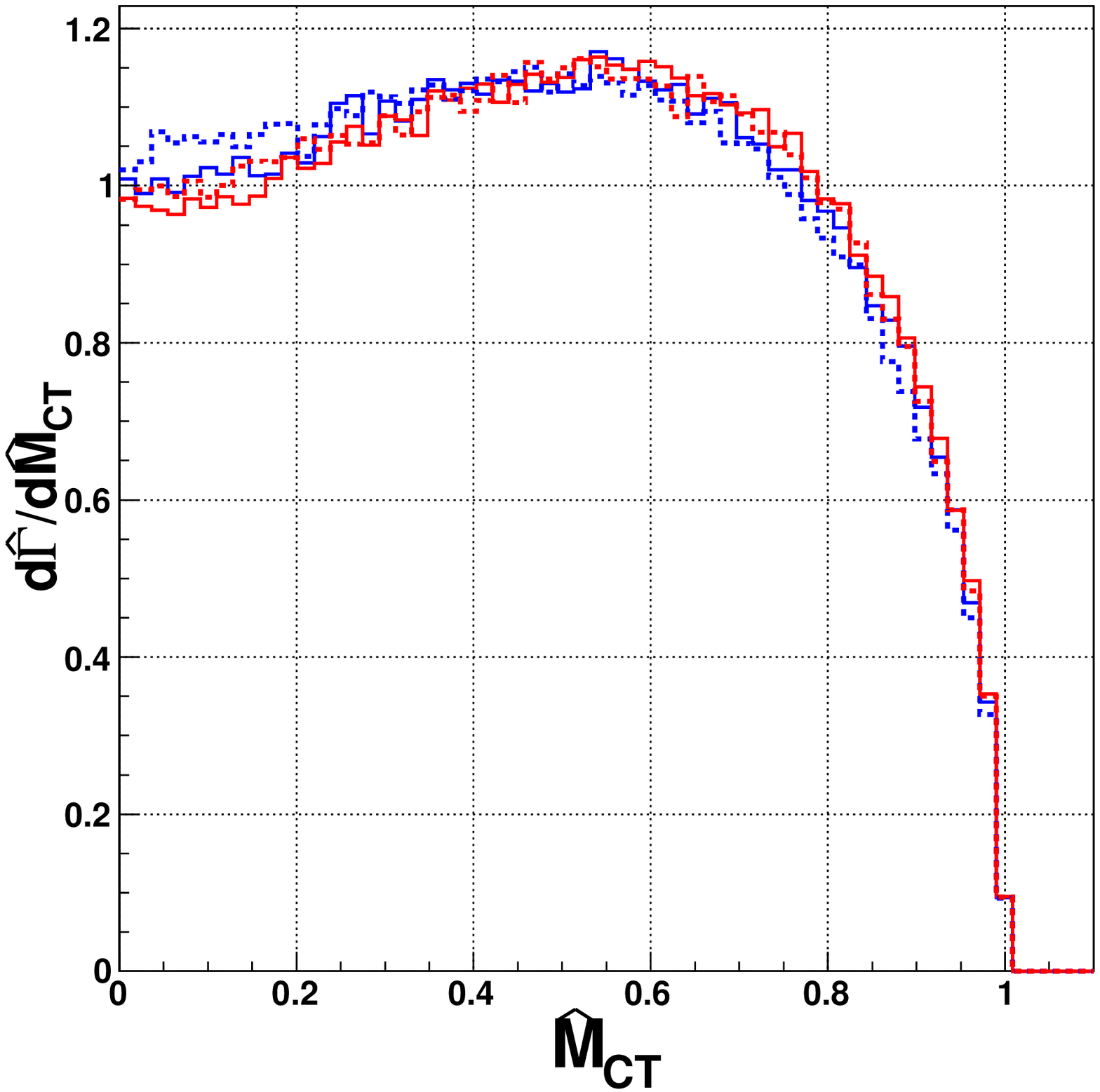,width=5.cm}
\epsfig{file=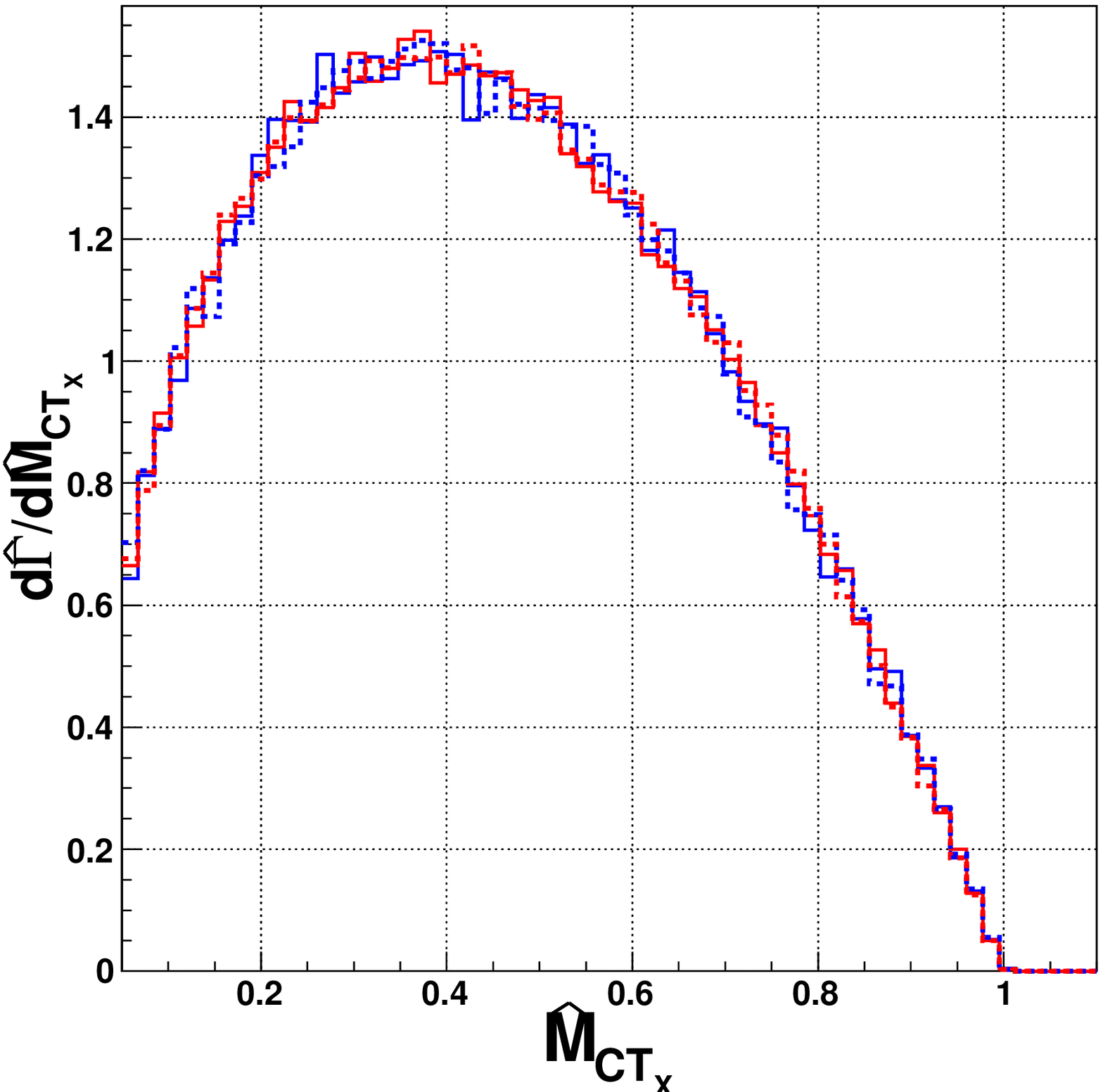,width=5cm}}\\
\centerline{
\epsfig{file=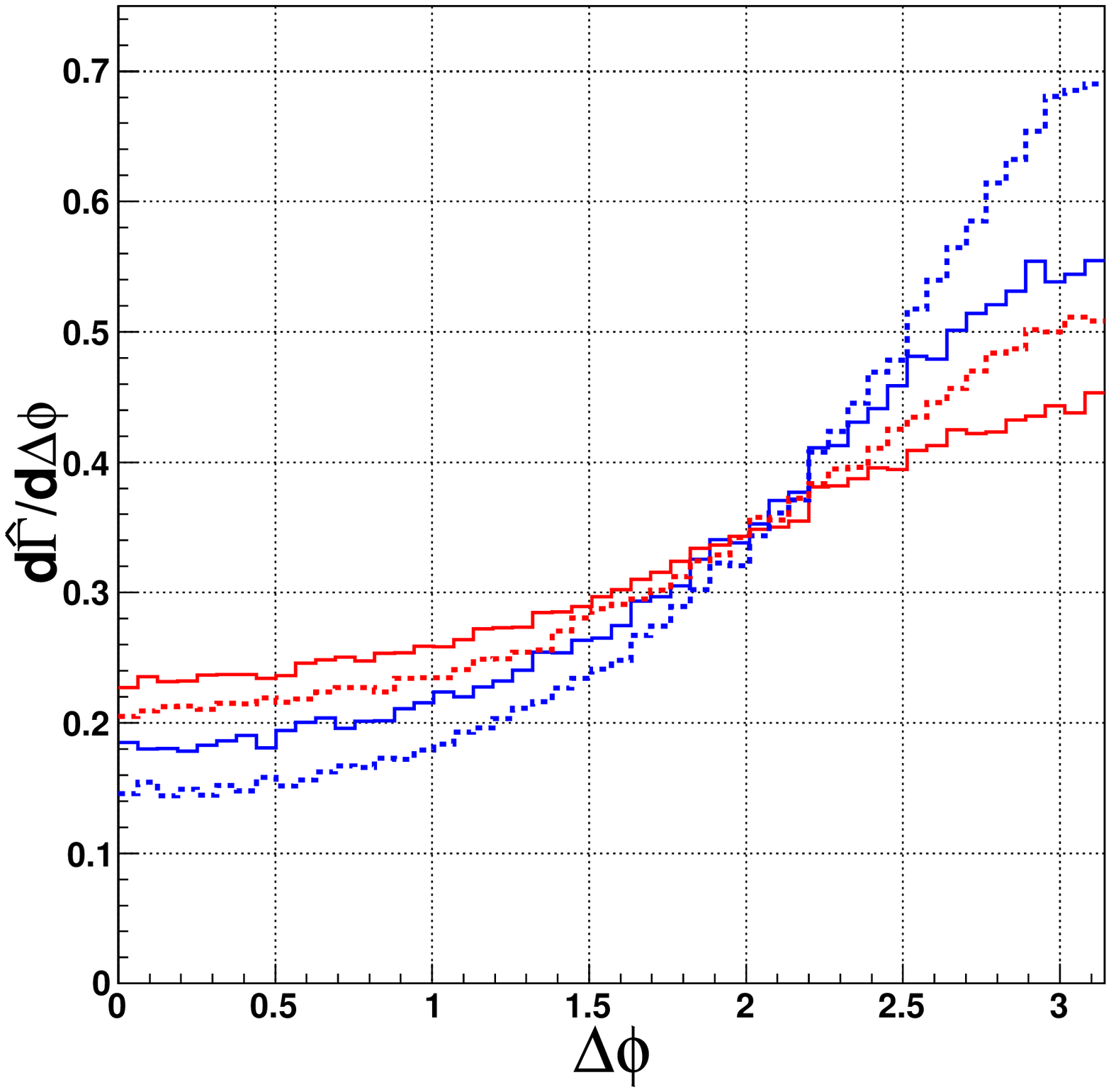,width=5.cm}
\epsfig{file=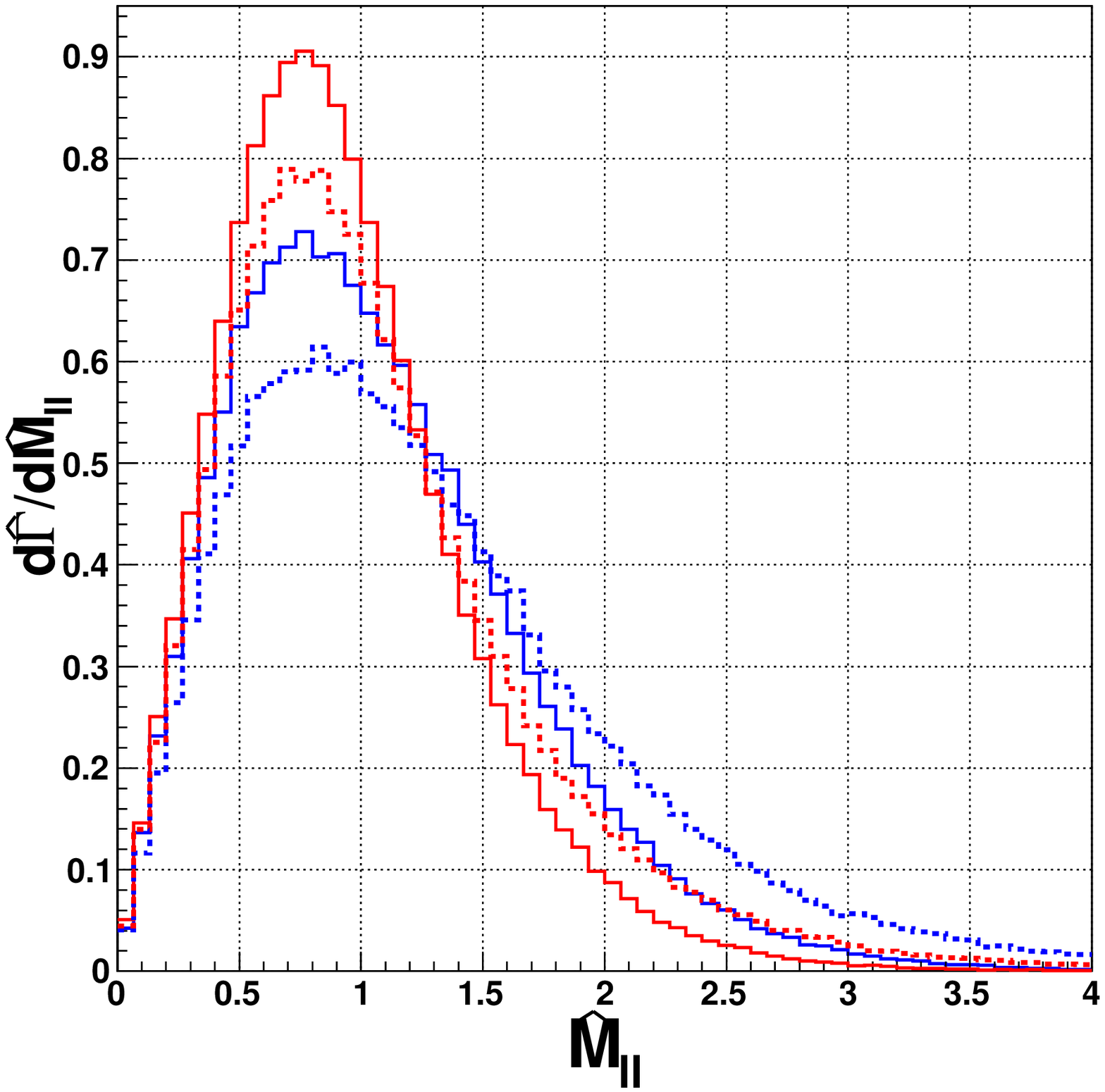,width=5.cm}
\epsfig{file=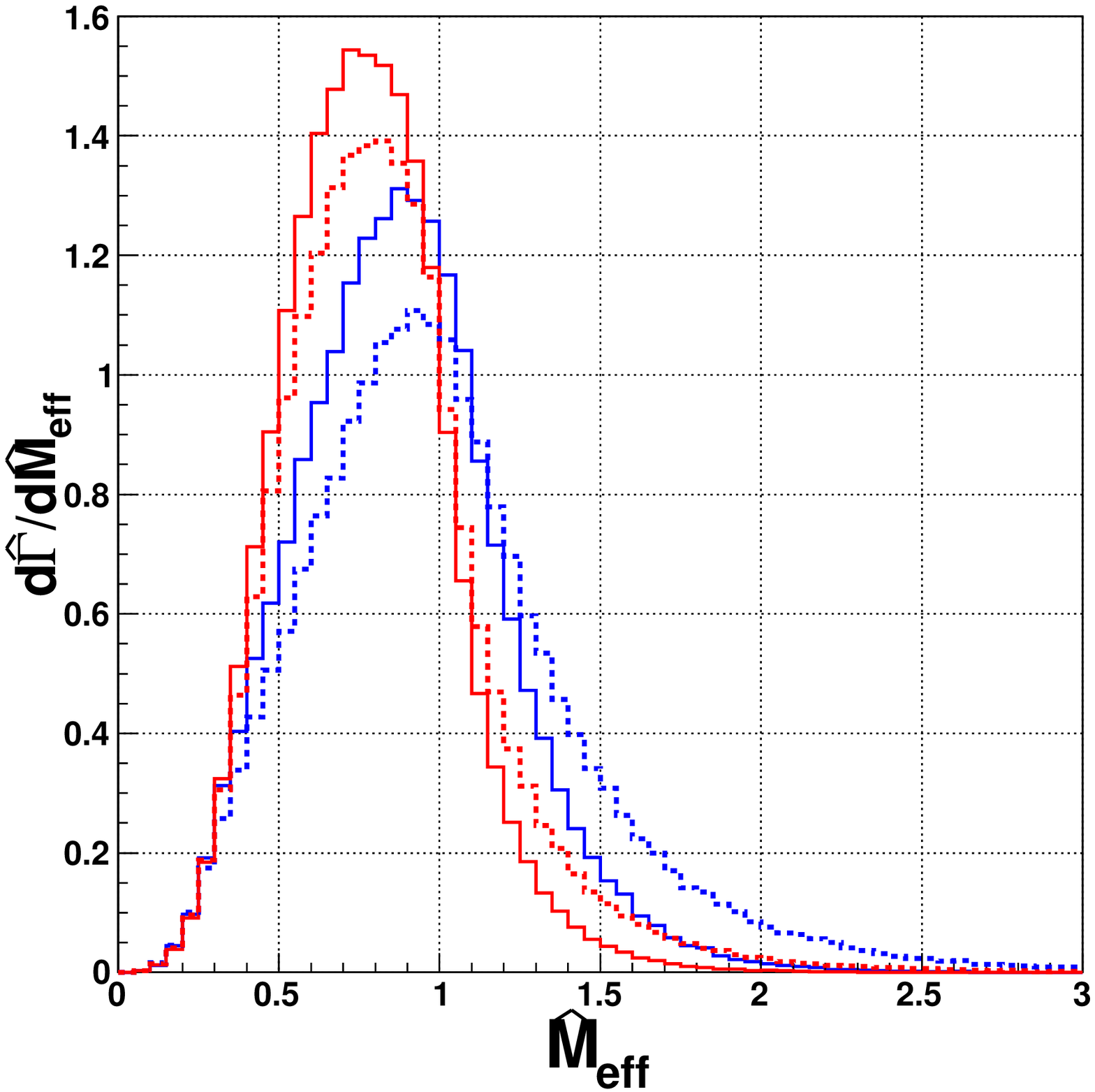,width=5.cm}}\\
\centerline{
\epsfig{file=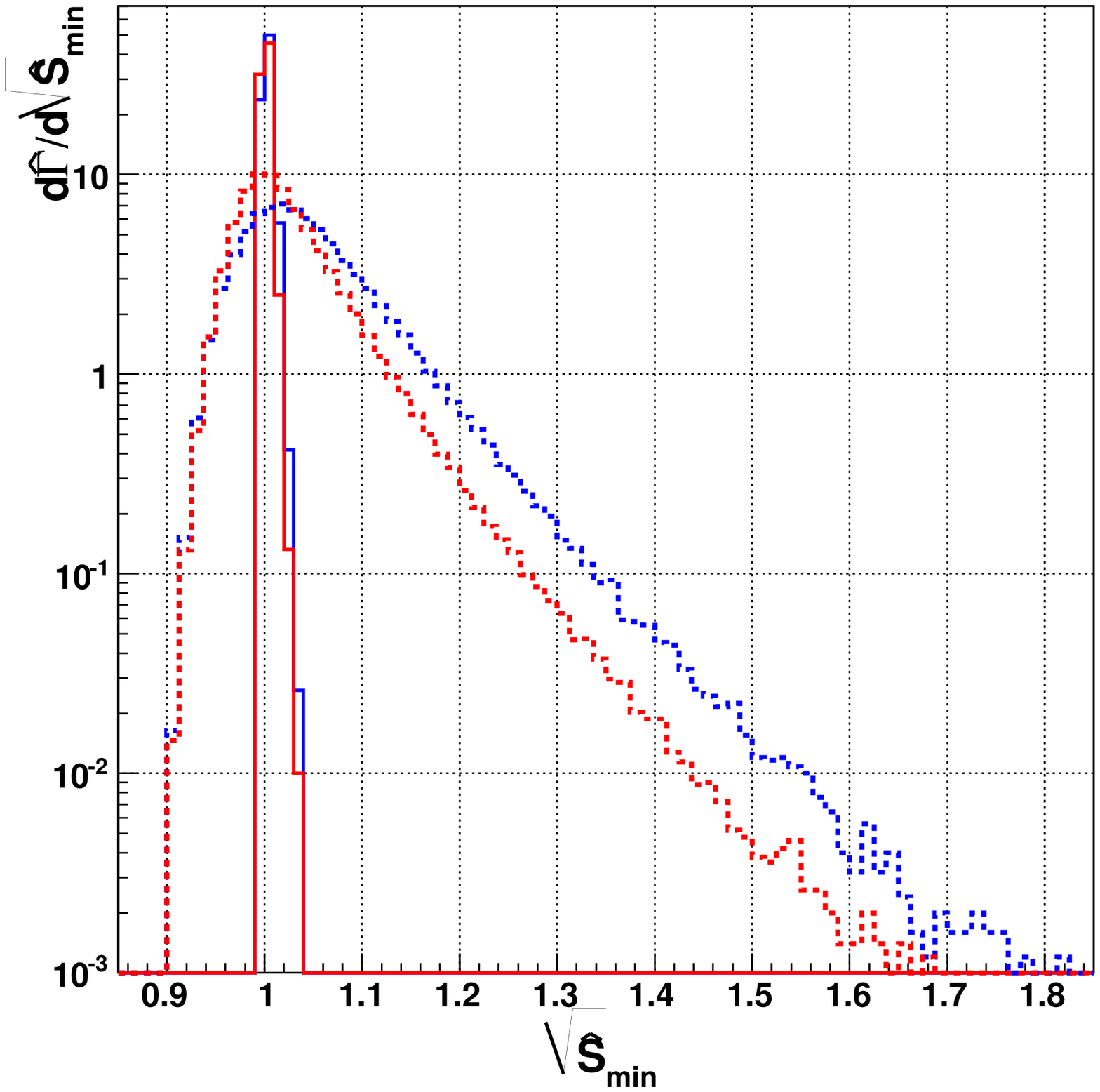,width=5.cm}
\epsfig{file=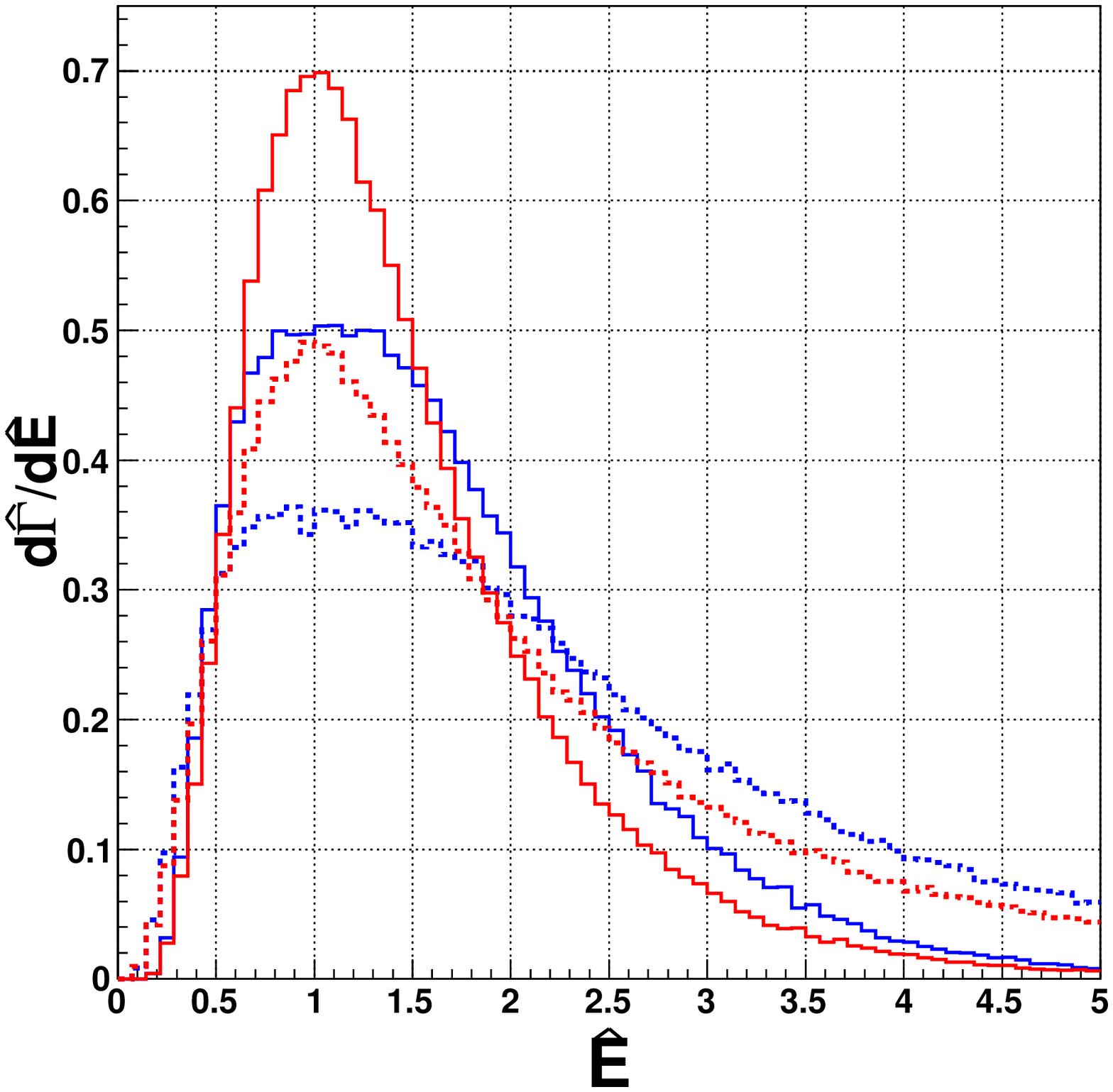,width=5.cm}
\epsfig{file=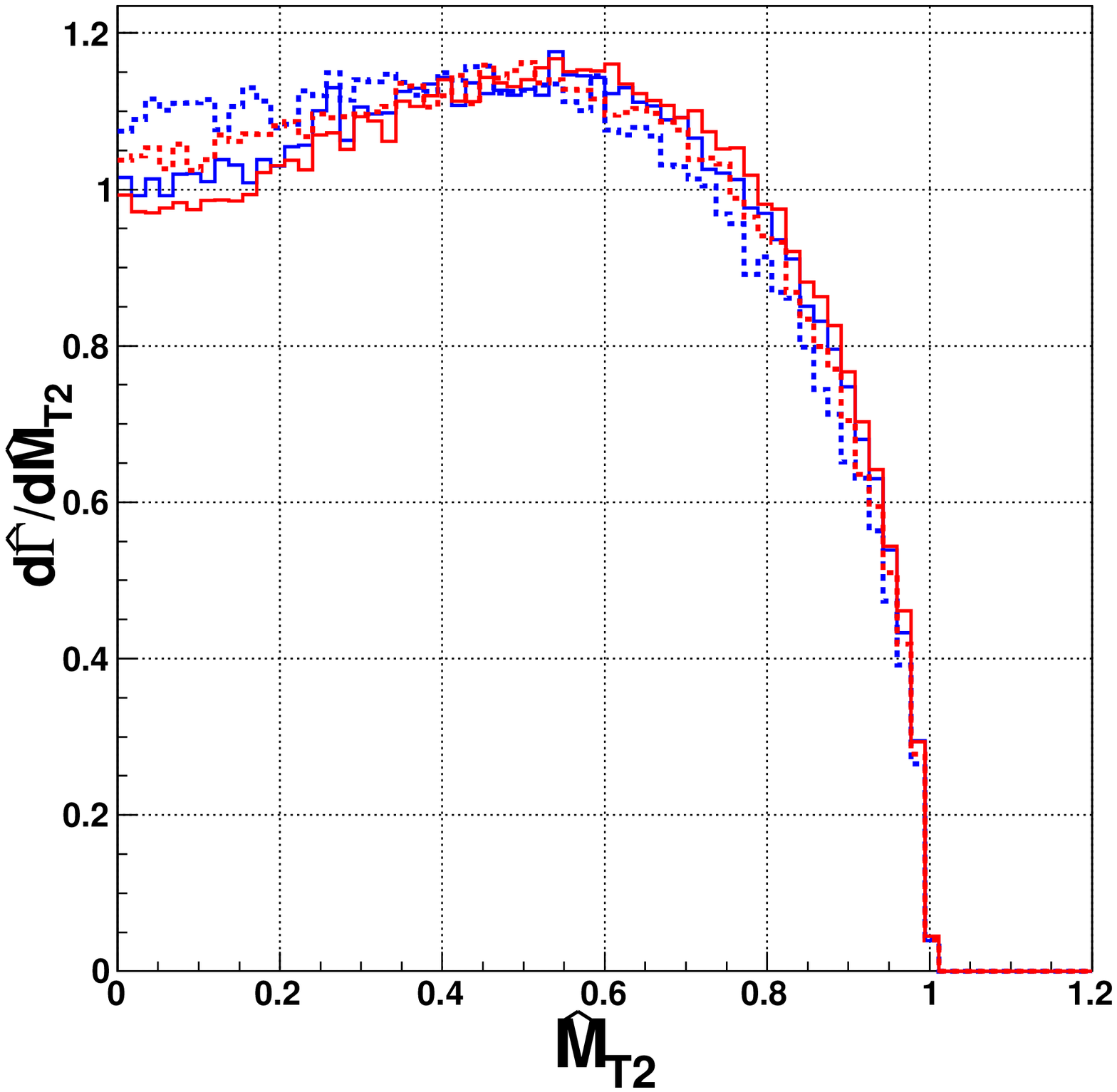,width=5.cm}
}
\caption{Unit-normalized distributions of the nine variables introduced in Section~\ref{sec:variables} at a 7 TeV LHC.
Each panel shows 4 cases, depending on the mass spectrum from Table~\ref{table:mass}: SUSY-like (dotted lines) or UED-like (solid lines);
and the spin configuration: VSF for the MSSM (blue lines) and VFV for MUED (red lines).  }
\label{fig:var}
}

Fig.~\ref{fig:var} reveals that the nine variables from Section \ref{sec:variables}
exhibit varying degrees of sensitivity to spins. The distributions of some variables, like 
$M_{CT}$, $M_{CTx}$ and $M_{T2}$, show very little variation and appear to be
relatively insensitive to changes in the spins and masses. The case of $M_{CTx}$
is particularly striking --- all four $M_{CTx}$ distributions are virtually identical.
This begs the question whether the $M_{CTx}$ distribution has any dependence on the
spins at all. This issue will be tackled in Section~\ref{sec:mctx} below.

At the same time, Fig.~\ref{fig:var} also demonstrates that the remaining 6 variables,
$\BAR$,  $\Delta \varphi$, $M_{\ellp\ellm}$, $M_{\mathrm{eff}}$, $\smin$
and $E_\ell$, all have a certain degree of sensitivity to spins, and are generally 
promising for spin studies. However, at this point one should not read too much into Fig.~\ref{fig:var},
since the observed variations in the kinematic shapes can be attributed to several 
different factors, not all of which are related to spins:
\begin{itemize}
\item First, we see that in Fig.~\ref{fig:var} there is a noticeable difference between 
lines of the same color, i.e.~when the same spin scenario is shown for two different mass spectra. 
This means that the shapes of the kinematic distributions are affected by the mass
spectrum (which in turn determines the available phase space). This is not necessarily a problem for spin 
studies per se, since the mass spectrum will be known in advance. However,
this does present a problem in the sense that any conclusions that one might draw as to which variables are
most sensitive to spins, will necessarily be subject to the choice of mass spectrum.
In other words, when it comes to spin determinations, it may very well be that for one 
particular mass spectrum a certain kinematic variable performs best, but for a different 
spectrum another variable is the winner. This is why in what follows we shall take
great care in illustrating our results throughout the full mass parameter space, and not
just a couple of study points as in Fig.~\ref{fig:var} (see Sec.~\ref{sec:detailed} below).
\item One should also keep in mind that the $\sqrt{s}$ distribution also varies from one model
point to another, and this will impact the distributions of kinematic variables which carry
a dependence on the underlying $\sqrt{s}$ in the event. Fig.~\ref{fig:shatstudy} shows the
$\sqrt{s}$ distributions for the four cases discussed in Fig.~\ref{fig:var}. As expected, the 
SUSY spin scenario (VSF) leads to a significantly harder $\sqrt{s}$ distribution (blue lines in Fig.~\ref{fig:shatstudy}),
due to the $p$-wave threshold suppression for scalar $B_1B_2$ production.
Now returning to Fig.~\ref{fig:var}, it should be rather intuitive that $\BAR$,  $\Delta \varphi$, 
$M_{\ellp\ellm}$, $M_{\mathrm{eff}}$, $\smin$ and $E_\ell$ are all sensitive to 
the value of $\sqrt{s}$, so it is not immediately clear how much of the
differences between the red and blue lines in their distributions were 
simply due to the different $\sqrt{s}$ distributions in Fig.~\ref{fig:shatstudy} as opposed to spins. 
Conversely, the variables $M_{CT}$ and $M_{CTx}$ are less sensitive to $\sqrt{s}$
because of the back-to-back boost invariance. Therefore they will not be subject to
the $\sqrt{s}$ effects in Fig.~\ref{fig:shatstudy}, and will have much smaller variations,
as indeed observed in Fig.~\ref{fig:var}.
\item Of course, the shapes of the kinematic distributions do carry information
about spins. As already explained in Sec.~\ref{sec:spin}, this information can be extracted
by comparing the predictions from the eight different spin scenarios in Table~\ref{tab:models} to the observed data.
\item As emphasized in \cite{Burns:2008cp}, spin correlations appear not only because
of spins, but also due to the chirality of the fermion couplings. In reality, one is measuring 
the combined effects from spins and chirality, and the blueprint for disentangling these two 
effects can be found in \cite{Burns:2008cp,Boudjema:2009fz,Ehrenfeld:2009rt,Chen:2010ek}.
We shall investigate the chirality effect in Section~\ref{sec:Chiralities}.
\end{itemize}

\FIGURE[t]{
\centerline{
\epsfig{file=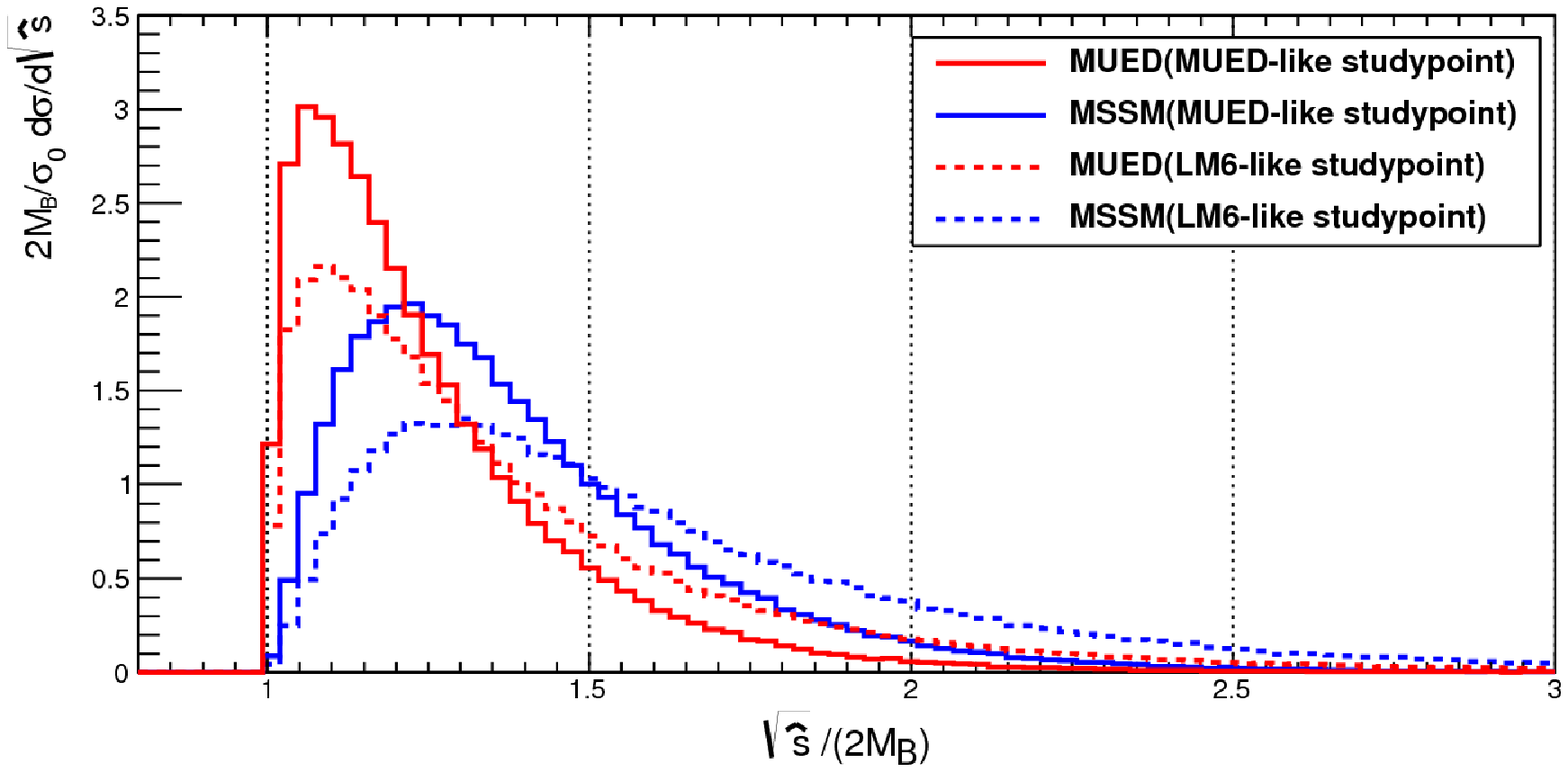,width=14cm}
}\\
\caption{Unit-normalized $\sqrt{s}$ distributions for the four cases presented in Fig.~\ref{fig:var}.}
\label{fig:shatstudy}
}

\subsection{Kinematic dependence on $\sqrt{s}$ and the mass spectrum}
\label{sec:detailed}

We shall now discuss in some more detail to what extent our conclusions from
the simple exercise in Fig.~\ref{fig:var} persist as we vary the mass spectrum 
and the energy $\sqrt{s}$ in the event. In Fig.~\ref{fig:ALLILC} we compare the
VSF spin scenario of the MSSM to the VFV spin scenario in MUED, 
as a function of $\sqrt{s}$ (plotted on the $x$-axis) and 
the mass splitting between $B$ and $C$, parameterized through the mass ratio
$M_C/M_B$ (plotted on the $y$-axis), for a fixed $M_B=500$ GeV. 
Each panel shows a temperature plot of the $\chi^2$ measure for 
one of the nine kinematic variables from Fig.~\ref{fig:var}.
Larger values of $\chi^2$ (warmer colors in Fig.~\ref{fig:ALLILC}) indicate larger differences in the predicted
shapes between the SUSY and MUED scenarios, and smaller values
of $\chi^2$ indicate rather similar shapes, where the two scenarios might look the same. 

As already mentioned, the observed differences in shapes in general could be due to
spin correlations, but could also be explained by different kinematics, as 
suggested by Fig.~\ref{fig:shatstudy}. In order to fairly assess the performance 
of the different kinematic variables with respect to pure spin effects, 
one should compare the $\chi^2$  values on the different panels in Fig.~\ref{fig:ALLILC} 
for a common value of $\sqrt{s}$, thus taking kinematics out of the equation. 
Then, the variable with the larger $\chi^2$ for a given value of 
$\sqrt{s}$ would  be more sensitive to spin correlations, and vice versa.

\FIGURE[ht!]{
\centerline{
\epsfig{file=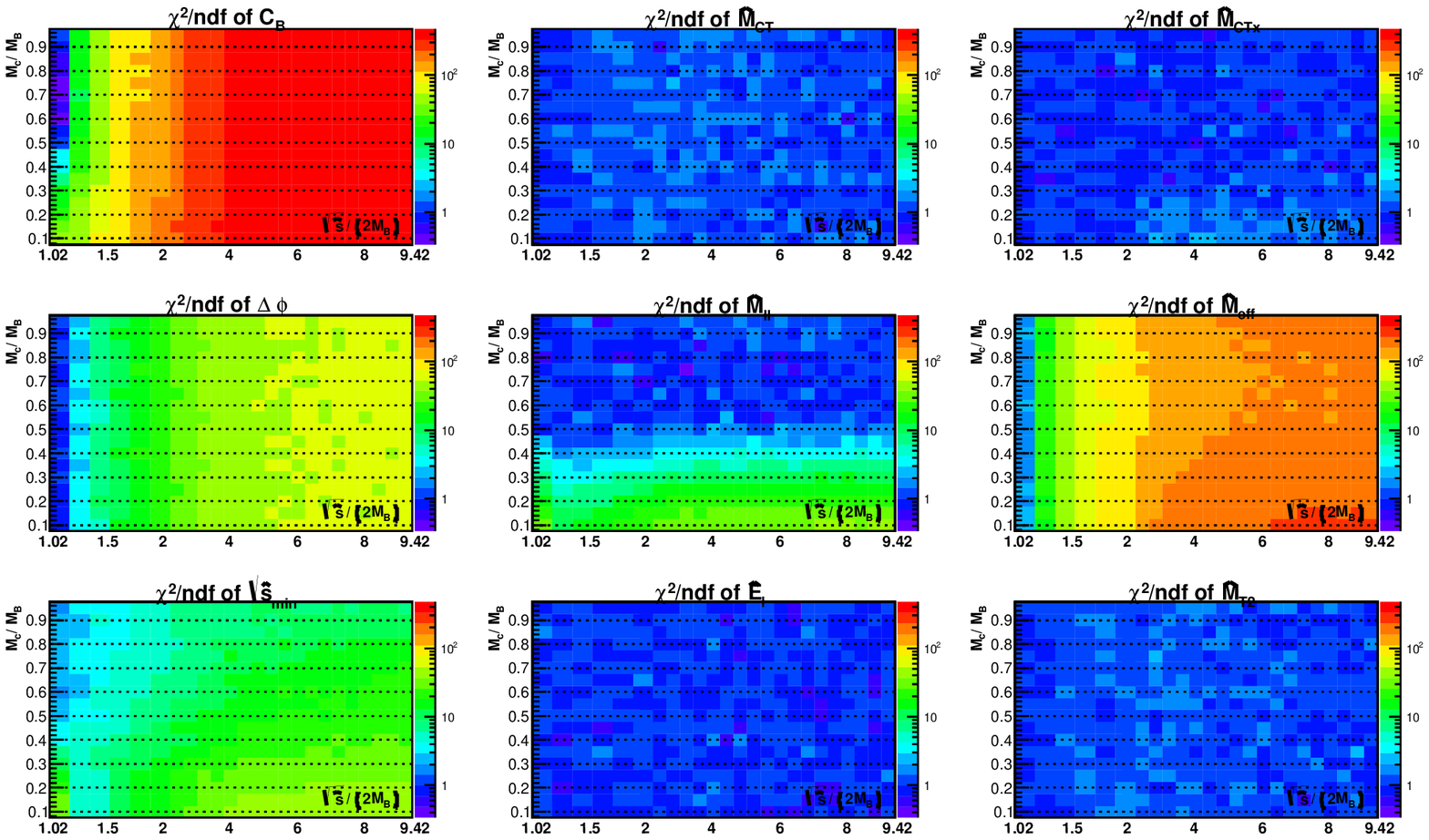,width=16.5cm}
}\\
\caption{$\chi^2$-comparisons of the two spin scenarios (VSF in the MSSM versus VFV in MUED),
for each of the nine observables from Fig.~\ref{fig:var}. We fix the mass of $B$ as $M_B=500$ GeV,
and then vary the CM energy $\sqrt{s}$ and the mass $M_C$ of $C$.}
\label{fig:ALLILC}
}

There are several lessons to be learned from Fig.~\ref{fig:ALLILC}. First, not surprisingly, 
different variables exhibit different degrees of sensitivity to spin correlations. 
In this sense, $M_{CT}$, $M_{CTx}$, $M_{T2}$ and $E_\ell$ appear to be the worst performers.
For the first three variables, this might have been expected, based on the result from Fig.~\ref{fig:var}.
The surprising member of this group is the lepton energy $E_\ell$: its distributions
now appear identical in the two spin scenarios, in spite of the much more promising differences 
seen previously in Fig.~\ref{fig:var}. This suggests that the differences 
in the $E_\ell$ distributions in Fig.~\ref{fig:var} can be attributed solely to the
different $\sqrt{s}$ kinematics seen in Fig.~\ref{fig:shatstudy} and not to true spin correlations.

Fig.~\ref{fig:ALLILC} also reveals that the contrast between the two spin scenarios depends quite 
significantly on the energy $\sqrt{s}$, especially in the case of $\BAR$, $\Delta\varphi$,
$M_{\mathrm{eff}}$ and $\smin$. In general, as the value of $\sqrt{s}$ increases, differences 
become more pronounced. For the largest values of $\sqrt{s}$ seen on the plots
(several times above threshold) the most discriminating variable appears to be $\BAR$,
however such conclusion would be rather premature. As Fig.~\ref{fig:shatstudy} shows,
at hadron colliders particles are produced near threshold, with typical values of $\sqrt{s}$
only 10-20\% above the $2M_B$ threshold. Unfortunately, Fig.~\ref{fig:ALLILC} shows that 
for those low values of $\sqrt{s}$, the discriminating power of the $\BAR$ variable is 
significantly reduced, and it behaves similarly to the other alternatives like 
$M_{\mathrm{eff}}$ and $\Delta\varphi$. 

\FIGURE[t]{
\centerline{
\epsfig{file=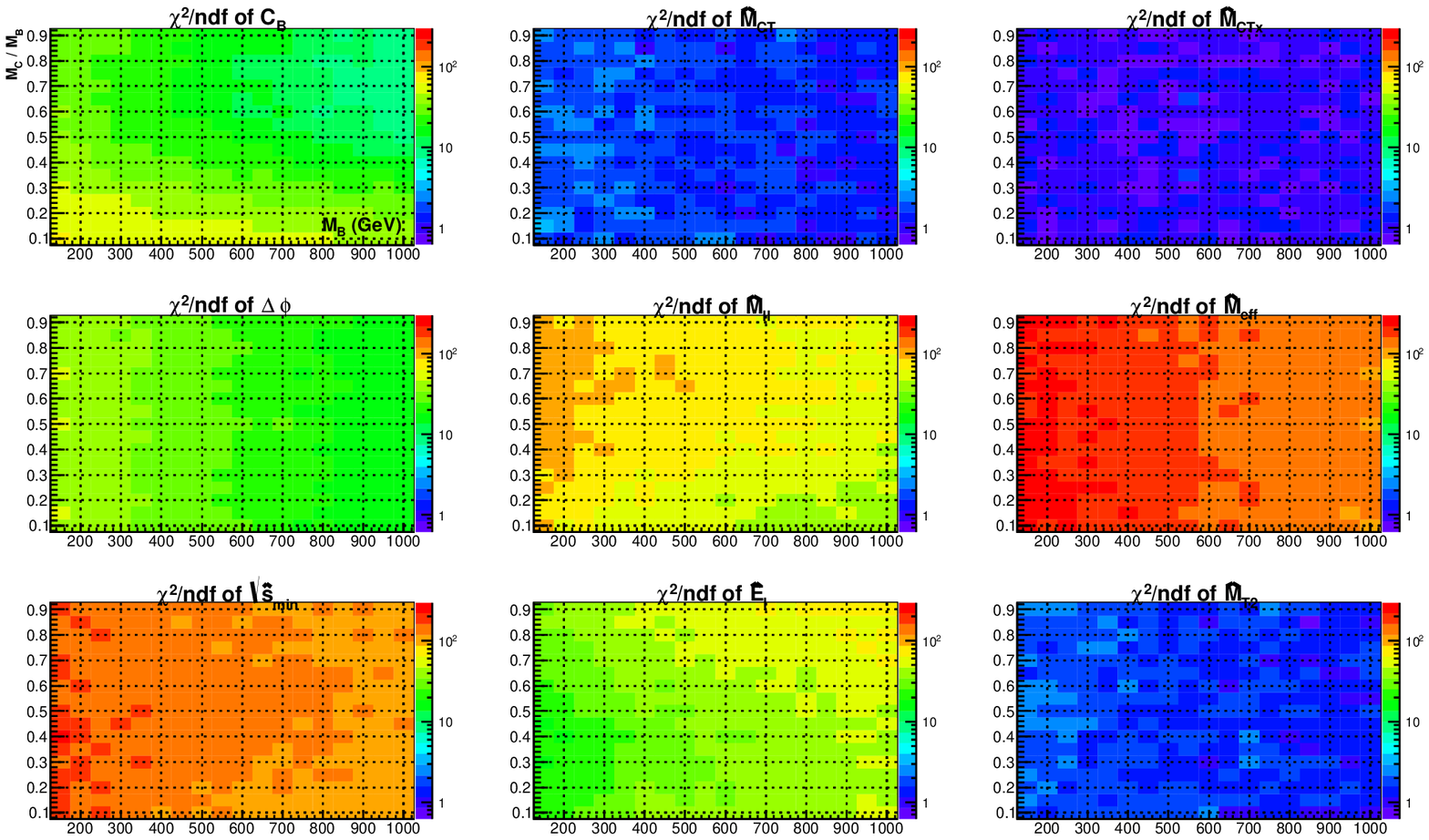,width=16.5cm}
}\\
\caption{The same as Fig.~\ref{fig:ALLILC}, but instead of a linear collider at 
a fixed CM energy $\sqrt{s}$, results are shown for the 8 TeV LHC. 
Here we vary both $M_B$ and $M_C$ and plot in the $(M_B, \frac{M_C}{M_B})$ plane.}
\label{fig:ALLLHC}
}

In order to get a sense of the full picture at the LHC (currently running at 8 TeV CM energy),
one needs to re-weight the results from Fig.~\ref{fig:ALLILC} by the proper $\sqrt{s}$
probability distributions analogous to Fig.~\ref{fig:shatstudy}. The result is shown in Fig.~\ref{fig:ALLLHC},
where this time we vary both $M_B$ and $M_C$ and plot in the $(M_B, \frac{M_C}{M_B})$ plane.
Interestingly, Fig.~\ref{fig:ALLLHC} shows that in practice the variables whose distributions are most
likely to show a noticeable difference between SUSY and MUED are $M_{\mathrm{eff}}$, $\smin$
and $M_{\ellp\ellm}$. All three of these variables are sensitive to the $\sqrt{s}$ in the event, and are 
therefore capable of discriminating the different shapes of the $\sqrt{s}$ distributions in
Fig.~\ref{fig:shatstudy}.

\section{Expected shapes in the absence of spin correlations}
\label{sec:phasespace}

In this section, we shall be interested in analytical predictions of the shapes of some of the observables
from the previous two sections. Consider a generic kinematic variable $V$. In general, it is a function of the
underlying event kinematics described in Sec.~\ref{sec:kinematics}:
\bea
V &=& V(\sqrt{s},\Theta^\ast, \varphi; \theta'_1, \phi'_1; \theta'_2, \phi'_2), \label{Vdef1}\\ [1mm]
    &=& V(\eta^\ast,\Theta^\ast, \varphi; \eta'_1, \phi'_1; \eta'_2, \phi'_2). \label{Vdef2}
\eea
Once this function is known, the pure ``phase space" distribution of the variable $V$ for a given fixed $\sqrt{s}$ 
can be obtained by integrating over $d\Omega_1'$ for the $B_1\to \ell^{-} C_1$ decay in the CMB1 frame,
over $d\Omega_2'$ for the $B_2\to \ell^+ C_2$ decay in the CMB2 frame,
and over $d\Omega$ for the initial $pp\to B_1B_2$ scattering in the CMBB frame:
\beq
\left(\frac{\ud N}{\ud V} \right)(V,{\sqrt{s}})\propto
\int d\Omega \int d\Omega_1' \int d\Omega_2'\, \delta\left(V-V(\sqrt{s},\Theta^\ast, \varphi; \theta'_1, \phi'_1; \theta'_2, \phi'_2)\right).
\label{dNdVs}
\eeq
Finally, at a hadron collider one also needs to convolute the fixed $\sqrt{s}$ distribution (\ref{dNdVs}) with 
the parton luminosity $L(\sqrt{s})$ (recall Fig.~\ref{fig:shatstudy}), so that the experimentally observed distribution will be given by
\bea
\frac{\ud N}{\ud V} &\propto&
\int d\sqrt{s}\, L(\sqrt{s}) \left(\frac{\ud N}{\ud V} \right)(V,{\sqrt{s}}) \label{dNdVintegrate} \\ [2mm]
&=& \int d\sqrt{s}\, L(\sqrt{s})\int d\Omega \int d\Omega_1' \int d\Omega_2'\, \delta\left(V-V(\sqrt{s},\Theta^\ast, \varphi; \theta'_1, \phi'_1; \theta'_2, \phi'_2)\right).
\label{dNdV}
\eea
Unfortunately, even if we concentrate on the fixed $\sqrt{s}$ case of (\ref{dNdVs}),
the 6 integrations in it are typically quite involved, and exact analytical expressions for 
the resulting distribution $dN/dV(V,\sqrt{s})$ are known only in very few special cases (reviewed below).
To make matters worse, the fixed $\sqrt{s}$ distribution of (\ref{dNdVs}) still needs to be integrated 
numerically over the parton luminosities as in (\ref{dNdVintegrate}).
Given all those difficulties, analytical studies of the kinematic shapes (\ref{dNdV})
(or even the simpler case of (\ref{dNdVs})) appear to be quite challenging, especially
for arbitrary values of $\sqrt{s}$.

In this and the following section, therefore, we shall limit ourselves to a much more manageable task:
instead of dealing with a general $\sqrt{s}$, we shall derive analytical formulas in some interesting and relevant 
$\sqrt{s}$ limits. There are two special values of $\sqrt{s}$, namely the endpoints of its definition interval:
the threshold value $\sqrt{s}_{th}=2M_B$ and the infinite energy limit $\sqrt{s}\to \infty$.
Taking either one of those limits leads to simplifications in the defining functions
(\ref{Vdef1},\ref{Vdef2}), and the integrations become easier, although success is not always guaranteed.
We shall now discuss selected kinematic observables for general $\sqrt{s}$ (where possible),
and in the two limits of $\sqrt{s}\to \sqrt{s}_{th}$ and $\sqrt{s}\to \infty$. 
We have already learned from Fig.~\ref{fig:shatstudy} that out of those two limiting scenarios, the
threshold case is much more relevant at the LHC, since the mother particles $B$ are much more likely 
to be produced near threshold as opposed to very large boosts.

\subsection{The relative pseudorapidity variable $\cos{\theta_{\ellm\ellp}^\ast}$}
\label{sec:BarrPhase}

We begin with the $\BAR$ variable from Section~\ref{sec:Barr}
\beq
\BAR = \tanh{\left(\frac{\eta_1-\eta_2}{2}\right)}.
\label{eq:Barragain}
\eeq
The lepton pseudorapidities in the LAB frame $\eta_1$ and $\eta_2$ (which were defined in (\ref{eq:rapidity}))
can be conveniently expressed in terms of the respective pseudorapidities (\ref{eq:etaprime}) in 
the CMB1 and CMB2 frames as follows
\beq
\eta_i= 
\beta_z+\frac{1}{2} \ln{\left[
\frac{\cosh{\left(\eta_i'-(-1)^i \eta^*\right)}+\cos{\Theta^*} \sinh{\left(\eta_i'-(-1)^i \eta^*\right)}+\cos{\phi_i'} \sin{\Theta^*}}
{\cosh{\left(\eta_i'-(-1)^i \eta^*\right)}-\cos{\Theta^*} \sinh{\left(\eta_i'-(-1)^i \eta^*\right)}-\cos{\phi_i'} \sin{\Theta^*}}
\right]} \, .
\label{eq:etai}
\eeq
Substituting (\ref{eq:Barragain}) and (\ref{eq:etai}) into (\ref{dNdVs}), one can in principle obtain the 
$\BAR$ distribution for a given fixed $\sqrt{s}$. Unfortunately, we have not been able to obtain closed form expressions for
general $\sqrt{s}$, and shall consider the two limiting cases instead.
 
In the infinite energy limit $\sqrt{s} \rightarrow \infty$, eq. (\ref{eq:eta}) implies that $\eta^* \rightarrow \infty$ 
and then (\ref{eq:etai}) reduces to
\bea
\eta_1&& \xrightarrow[\eta^* \rightarrow \infty]{} \beta_z+\ln\left[\cot{\left(\frac{\Theta^*}{2}\right)}\right] \, , \nonumber \\
\eta_2&& \xrightarrow[\eta^* \rightarrow \infty] {} \beta_z+\ln\left[\tan{\left(\frac{\Theta^*}{2}\right)}\right]\, .
\eea
Then (\ref{eq:Barragain}) simplifies to
\beq
\BAR  \xrightarrow[\eta^\ast \rightarrow \infty]{}  \cos{\Theta^*}\, .
\label{BAReqTheta}
\eeq
Therefore, in the infinite energy limit, the $\BAR$ distribution reduces to the 
$\cos\Theta^*$ distribution:
\beq
\frac{\ud N}{\ud \BAR} \xrightarrow[\eta^\ast \rightarrow \infty]{}  \frac{\ud N}{\ud \cos\Theta^\ast}.
\label{dBAReqdTheta}
\eeq
This fact served as the original motivation for introducing the 
$\BAR$ variable in the first place \cite{Barr:2005dz}. It is well known (see Section~\ref{sec:infiniteS} below)
that the $\cos\Theta^*$ distribution is directly probing the spins of the particles $A$ and $B$,
and so $\BAR$ will inherit some spin sensitivity through (\ref{dBAReqdTheta}).

\FIGURE[t]{
\centerline{
\epsfig{file=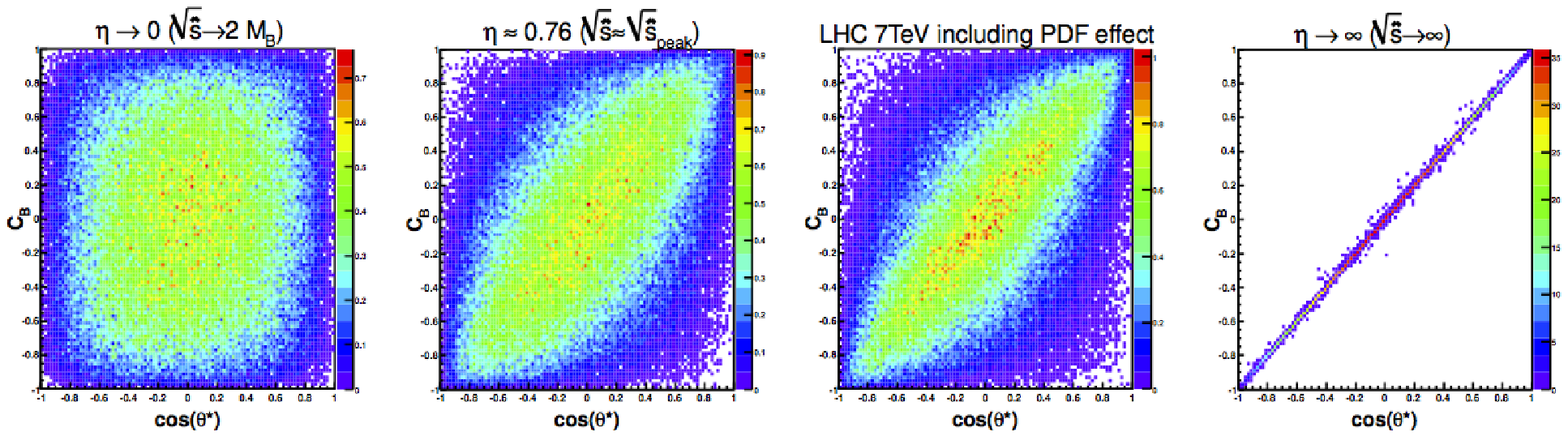,width=15.5cm } }\\
\centerline{
\epsfig{file=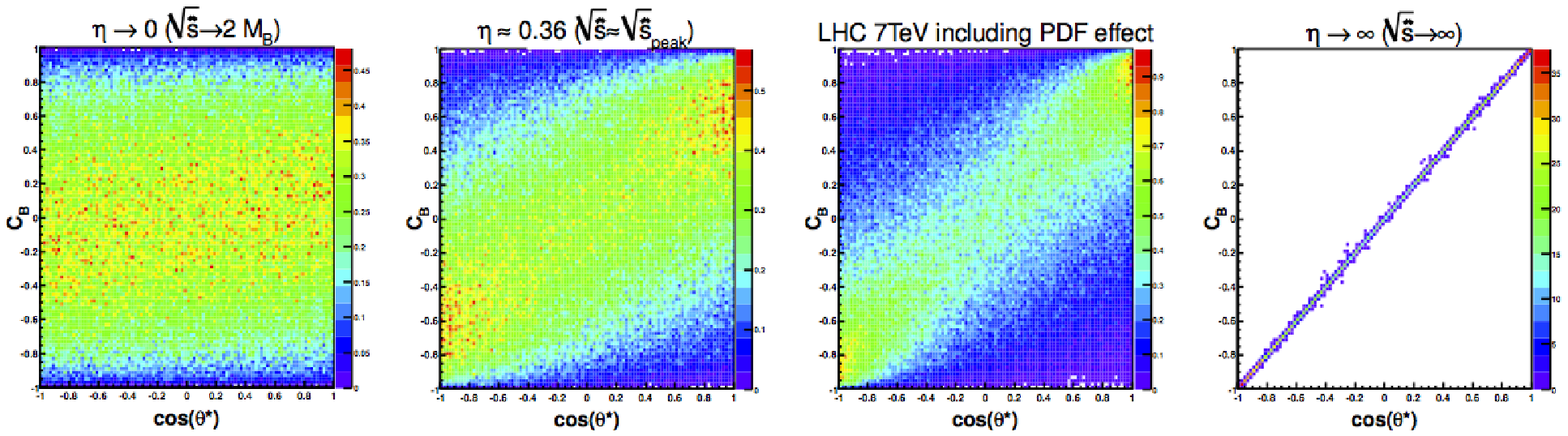,width=15.5cm}
}
\caption{Correlation between $\BAR$ and $\cos{\Theta^*}$, for different boost factors $\eta^\ast$.
The upper (lower) row of plots is for the MSSM (MUED) study point from Table~\ref{table:mass}. }
\label{fig:corr}
}

The correlation (\ref{BAReqTheta}) is pictorially illustrated in Fig.~\ref{fig:corr},
which shows scatter plots of $\BAR$ versus $\cos\Theta^*$ for several different boost factors $\eta^\ast$
for the two study points from Table~\ref{table:mass}.
We see that (\ref{BAReqTheta}) holds only as long as $\sqrt{s}$ is sufficiently large.
As $\sqrt{s}$ gets closer to threshold, the correlation is lost and so is the sensitivity of the
$\BAR$ variable to the spin effects encoded in the $\cos\Theta^*$ distribution
(this effect was already evident in Fig.~\ref{fig:ALLILC}).

Given that most events at a hadron collider are produced at threshold, 
we now turn to the opposite limit of $\sqrt{s}\sim \sqrt{s}_{th}$ or equivalently,
$\eta^\ast \to 0$. In that case, the angle $\Theta^\ast$ becomes arbitrary 
and can be set to zero, so that (\ref{eq:Barragain}) and (\ref{eq:etai}) give
\beq
\BAR  \xrightarrow[
\eta^\ast \rightarrow 0, 
\Theta^\ast \rightarrow 0
]{} \tanh \left(\frac{\eta_1'-\eta_2'}{2}\right) \, .
\label{BAReqtanh}
\eeq
With this simple result, one can now perform the integrations in (\ref{dNdVs})
and obtain the unit-normalized $\BAR$ distribution at threshold as
\beq
\frac{\ud N}{\ud \BAR} \xrightarrow[\eta^\ast \rightarrow 0]{} 
\frac{1- \BAR^2}{4 \BAR^3}
\left\{ -2 \BAR+\left(1+ \BAR^2\right)
 \ln\left[\frac{1+\BAR}{1- \BAR}\right]\right\}.
\label{dNdCBth}
\eeq
This formula is one of the main new results in this paper. It provides the benchmark shape of the
$\BAR$ distribution in the vicinity of the relevant energy regime. Eq.~(\ref{dNdCBth}) will be central 
to our understanding of the effects of spin correlations later on in Section~\ref{sec:Chiralities}. 
Of course, (\ref{dNdCBth}) is nothing but a crude approximation to the true $\BAR$ distribution
which is obtained by integrating over all energies $\sqrt{s}$ as in (\ref{dNdV}).
Nevertheless, for practical purposes (\ref{dNdCBth}) appears to be a more relevant limit 
than (\ref{dBAReqdTheta}), since the typical values of $\sqrt{s}$ are closer to 
threshold than to $\infty$ --- see Fig.~\ref{fig:shatstudy}.

\subsection{Azimuthal angular difference $\Delta \varphi$}

The azimuthal angular difference $\Delta \varphi$ defined in Sec.~\ref{sec:dphi} is given by
\beq
\cos{\Delta\varphi}= \frac{\cos{\Delta\phi'} + D_0}{D_1D_2},
\eeq
where
\bea
D_0 &\equiv& \sin^2{\Theta^\ast} \left(\cos{\phi_1'}\cos{\phi_2'} -\sinh{\bar  \eta_1'}\sinh{\bar \eta_2'}\right) \nonumber \\
        &-&\frac{1}{2}\sin{2\Theta^\ast} \left(\cos{\phi_2'} \sinh{ \bar\eta_1'}-\cos{\phi_1'}\sinh{ \bar\eta_2'}\right), \\ [2mm]
D_1 &\equiv& \left\{1+\sin^2{\Theta^\ast} \left(\sinh^2{\bar  \eta_1'}- \cos^2{\phi_1'}\right)
-\sin{2\Theta^\ast} \cos{\phi_1'}\sinh{\bar \eta_1'}\right\}^{1/2},  \\ [2mm]
D_2 &\equiv& \left\{1+\sin^2{\Theta^\ast}  \left(\sinh^2{\bar  \eta_2'}- \cos^2{\phi_2'}\right)
+\sin{2\Theta^\ast} \cos{\phi_2'}\sinh{\bar \eta_2'}\right\}^{1/2},
\eea
and 
\beq
\bar \eta_1' \equiv  \eta^\ast + \eta_1', \qquad \bar \eta_2' \equiv  \eta^\ast - \eta_2'. 
\eeq
Now in the infinite energy limit we obtain
\beq
 \Delta\varphi  \xrightarrow[\eta^\ast \rightarrow \infty]{} \pi \, 
\label{dphiinfty}
\eeq
and the universal result
\beq
\frac{\ud N}{\ud  \Delta\varphi} \xrightarrow[\eta^\ast \rightarrow \infty]{}  \delta(\Delta\varphi-\pi).
\label{dNdphiinfty}
\eeq
In the other $\sqrt{s}$ extreme, at threshold, we obtain
\beq
 \Delta\varphi  \xrightarrow[\eta^\ast \rightarrow 0]{} \Delta \phi' \, 
\label{dphith}
\eeq
and a flat distribution
\beq
\frac{\ud N}{\ud  \Delta\varphi} \xrightarrow[\eta^\ast \rightarrow 0]{}  \frac{1}{\pi} = {\rm const}.
\label{dNdphith}
\eeq
We see that as $\sqrt{s}$ varies between its two extremes, 
the $\Delta\varphi$ distribution changes from being completely flat to 
a perfect delta-function peak at $\pi$. Therefore, the flatness
of the $\Delta\varphi$ distribution is a measure of the typical $\sqrt{s}$
in the event, which in turn is affected by spins as shown in Fig.~\ref{fig:shatstudy}.
Therefore one might expect that the $\Delta\varphi$ distribution
would also be sensitive to spin correlations through $\sqrt{s}$ effects,
and this is indeed what we saw previously in Figs.~\ref{fig:var}, \ref{fig:ALLILC} and \ref{fig:ALLLHC}.

\subsection{Invariant mass $M_{\ellp \ellm}$}

On our notation, the formula for the rescaled invariant mass variable (\ref{Mllrescaled}) is
\beq
\hat M_{\ellp \ellm}
=\sqrt{\frac{ \cosh{\left(\bar\eta_1'+\bar\eta_2'\right)}
-\cos{\left(\phi_1'-\phi_2'\right)} }{2\cosh{\eta_1'}\cosh{\eta_2'}}}.
\eeq
The integrations in (\ref{dNdVs}) can be performed for any fixed $\sqrt{s}>2M_B$ and the
result for the unit-normalized $\hat M_{\ellp \ellm}$ distribution is  \cite{Han:2009ss}
\beq
\frac{\ud N}{\ud \hat {M}_{\ellp \ellm}} =
\left\{
\begin{array}{ll}
\frac{4\eta^*}{\sinh{2 \eta^*}}\hat {M}_{\ellp\ellm},    & {\rm for\ }  \hat {M}_{\ellp \ellm} \le \hat {M}_{\ellp\ellm}^{\textrm{cusp}},    \\ [2mm]
\frac{2         }{\sinh{2 \eta^*}}  \hat {M}_{\ellp\ellm} \log\left(\frac{ \hat {M}_{\ellp \ellm}^{\textrm{max}}}{ \hat {M}_{\ellp \ellm}}\right),
  & {\rm for\ }  \hat {M}_{\ellp\ellm}^{\textrm{cusp}} < \hat {M}_{\ellp \ellm} \le \hat {M}_{\ellp \ellm}^{\textrm{max}}, 
 \end{array}
\right.
\label{Mlldist}
\eeq
where
\beq
\hat M_{\ellp \ellm}^{\textrm{cusp}}=e^{-\eta^*}, \quad
\hat M_{\ellp \ellm}^{\textrm{max}}= e^{\eta^*}.
\eeq
For events with a given fixed value of $\sqrt{s}$, the distribution (\ref{Mlldist}) exhibits 
a ``cusp'', i.e. a non-differentiable point, at $\hat {M}_{\ellp \ellm} = \hat {M}_{\ellp\ellm}^{\textrm{cusp}}$ \cite{Han:2009ss}.
However, since the events at hadron colliders occur at a variety of values of $\sqrt{s}$
(see Fig.~\ref{fig:shatstudy}), the cusp gets smeared, since its location depends on the boost factor
$\eta^\ast$ and thus on $\sqrt{s}$.

\subsection{Doubly projected contransverse mass $M_{CTx}$}
\label{sec:mctx}

\FIGURE[ht!]{
\centerline{
\epsfig{file=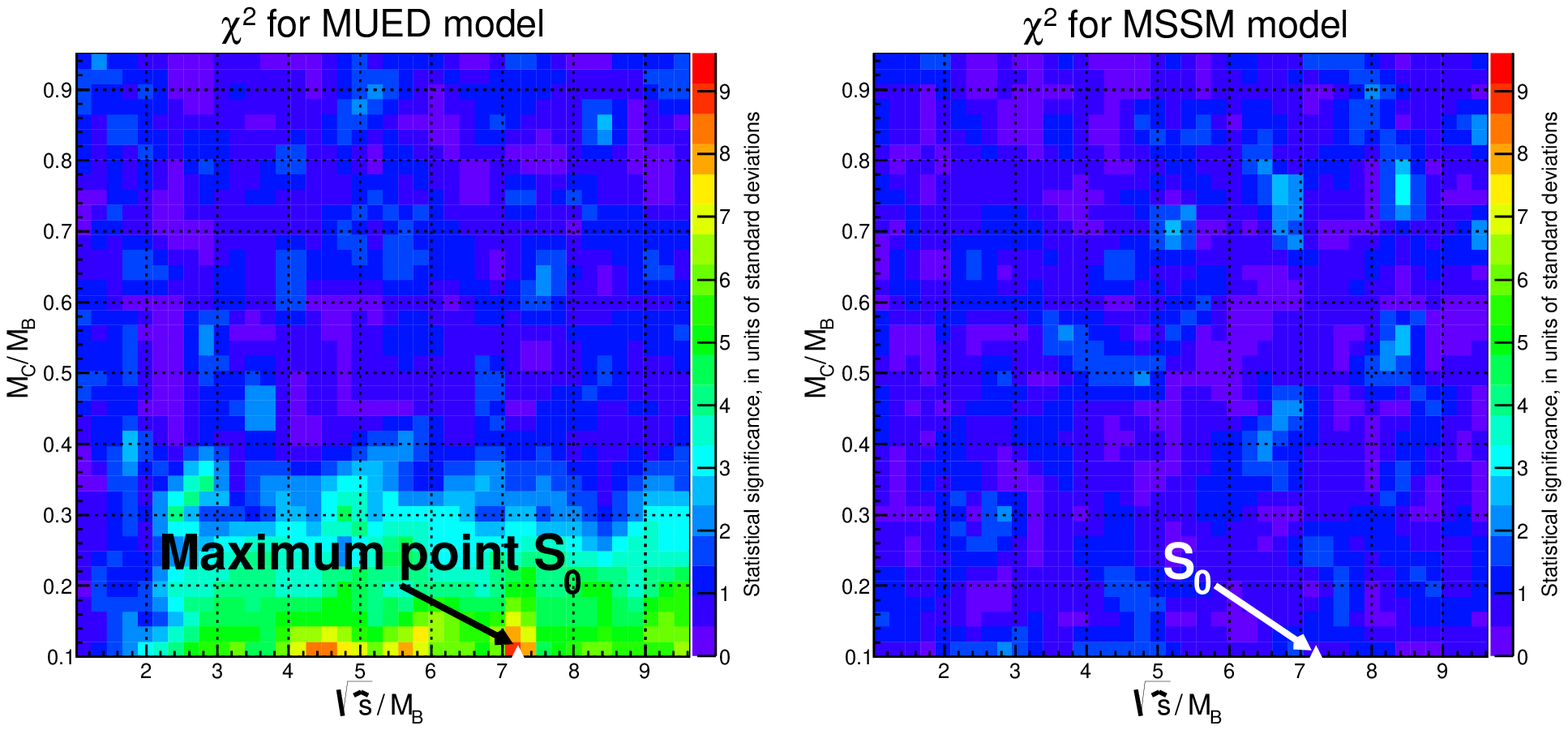,width=13.cm}
}\\
\centerline{
\epsfig{file=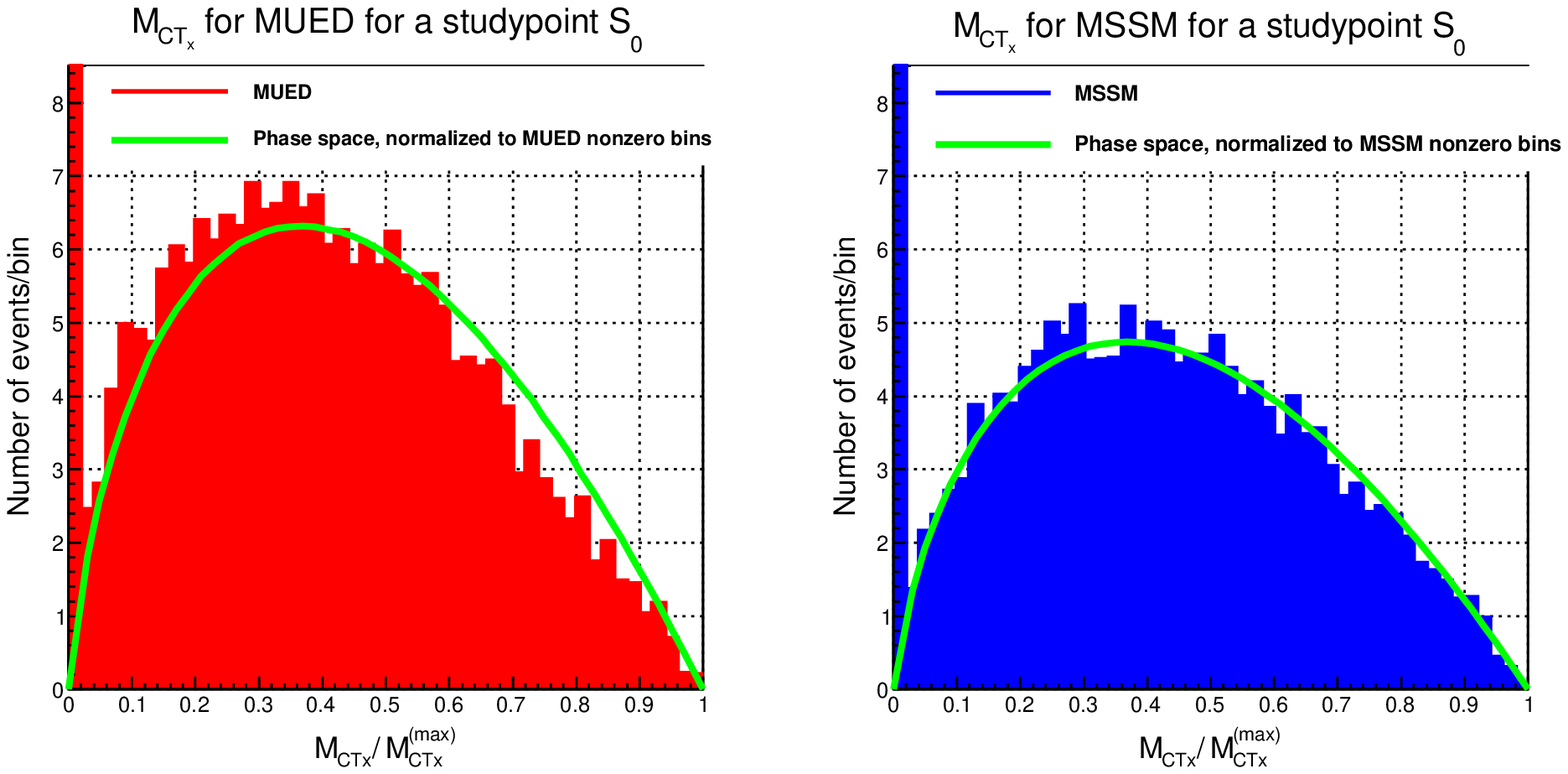,width=13.cm}
}\\
\caption{Top: $\chi^2$ tests of the predicted $M_{CTx}$ shape (\ref{MCTxshape}) against simulated data 
in MUED (left) and SUSY (right) for a fixed $M_B=500$ GeV and varying $\sqrt{s}$ and $M_C/M_B$.
Bottom: the $M_{CTx}$ distributions for the study point $S_0$ with the largest $\chi^2$.}
\label{fig:mctxCHI}
}

In the absence of spin correlations, the analytical formula for the 
$M_{CTx}$ distribution (restricted to events with non-vanishing $M_{CTx}$) 
is given by \cite{Matchev:2009ad}
\beq
\frac{\ud N}{\ud M_{CTx}} = -4 M_{CTx} \ln\left(\frac{M_{CTx}}{M_{CTx}^{(max)}}\right)\, .
\label{MCTxshape}
\eeq 
Interestingly, this formula does not carry any explicit dependence on $\sqrt{s}$,
which can be traced back to the invariance of the $M_{CTx}$ variable under
back-to-back transverse boosts of the parents $B_1$ and $B_2$ \cite{Tovey:2008ui,Matchev:2009ad}.
Even more peculiar is the observation
made in Figs.~\ref{fig:var}, \ref{fig:ALLILC} and \ref{fig:ALLLHC}: 
that the shape of the $M_{CTx}$ distribution appears to be pretty much
independent of the spins and the mass spectrum in the model.
One may therefore wonder whether the distribution (\ref{MCTxshape})
is indeed a universal function which is completely independent of the spin 
and the kinematics. 

In order to address this question, in Fig.~\ref{fig:mctxCHI} we test how well
the formula (\ref{MCTxshape}) fits the $M_{CTx}$ distributions
in the MUED and SUSY spin scenarios, over a wide range of mass spectra.
The upper two panels perform a $\chi^2$ test of the predicted shape (\ref{MCTxshape})
against the $M_{CTx}$ distribution obtained in our simulations for MUED (left panel) 
and SUSY (right panel). The test is repeatedly performed for various fixed values
of $\sqrt{s}$ (plotted on the $x$-axis) and mass hierarchies $M_C/M_B$ (plotted on the $y$-axis),
for a fixed $M_B=500$ GeV. We see that in the MSSM, the $M_{CTx}$ shape is very well predicted
by (\ref{MCTxshape}) in all cases. This could have been expected because of the scalar 
nature of the intermediate particles $B$, which causes the two leptons to be uncorrelated.
The formula (\ref{MCTxshape}) works very well in MUED as well, and only for $M_C<<M_B$ 
one starts to see deviations. In order to illustrate the size of the deviations, in the lower two panels
of Fig.~\ref{fig:mctxCHI} we plot the $M_{CTx}$ distributions for the point $S_0$ with the largest 
$\chi^2$ in the MUED case: $M_C=0.1M_B$ and $\sqrt{s}=14.4M_B$.  We see that even in the worst
case scenario of $S_0$, the two distributions are in reasonable agreement. 
We conclude that (\ref{MCTxshape}) is not a universal function and the shape of the
$M_{CTx}$ variable in principle does get affected by spin correlations, although the
effects are very minor.

\subsection{Lepton energy $E_{\ell}$}

The distribution of the lepton energy $E_\ell$ has the famous box-like shape for any given fixed $\sqrt{s}$:
\beq
\frac{\ud N}{\ud \hat E_{\ell}} = 
\left\{
\begin{array}{ll}
\frac{1}{\hat E_\ell^{(max)} -\hat E_\ell^{(min)}}, & \quad {\rm if\ }\hat E_\ell^{(min)} \le \hat E_\ell \le \hat E_\ell^{(max)},  \\
0,   & \quad {\rm otherwise},
\end{array}
\right.
\label{dNdE}
\eeq
where $\hat E_\ell$ is the rescaled energy variable from (\ref{Ehat}).
The two endpoints of the distribution (\ref{dNdE}) are given by
\beq
\hat E_\ell^{(min)} = e^{-\eta^\ast}\, ,\qquad 
\hat E_\ell^{(max)} = e^{\eta^\ast}. 
\label{Eendpoints}
\eeq
In the threshold limit of $\sqrt{s}\to 2M_B$ or $\eta^\ast\to 0$, (\ref{dNdE}) becomes simply
\beq
\frac{\ud N}{\ud \hat E_{\ell}} \xrightarrow[\eta^\ast \rightarrow 0]{}  \delta(\hat E_{\ell}-1).
\eeq
The observation that the two lepton energies are equal and constant at threshold 
was the main inspiration for the razor class of variables \cite{Rogan:2010kb}.
At hadron colliders, where $\sqrt{s}$ is varying from one event to another, 
the distribution (\ref{dNdE}) needs to be convoluted as in (\ref{dNdV})
and the two endpoints (\ref{Eendpoints}) become smeared, as can be seen 
in Fig.~\ref{fig:var}.

The distributions of the remaining four variables from Sec.~\ref{sec:variables} 
($M_{CT}$, $M_{\mathrm{eff}}$, $\smin$ and $M_{T2}$)
do not have compact analytical expressions and we shall not discuss them here.

\section{Spin correlations in the $\BAR$ distribution}
\label{sec:Chiralities}

In this section, we shall focus on the $\BAR$ variable and investigate 
how spin correlations affect the shape of its distribution. 
In Section~\ref{sec:BarrPhase} we saw that even in the pure phase space approximation 
(with no spin effects), we managed to obtain analytical formulas in closed form
only for the two limiting cases of $\sqrt{s}$ at threshold and at infinity.
Correspondingly, we shall now study the effect of spin correlations 
on $\BAR$ in those two limits as well.

\subsection{Spin effects at threshold}
\label{sec:dNdBARth}

We begin with the threshold limit $\sqrt{s} \sim \sqrt{s}_{th} \left(=2 M_B\right)$, where the $\BAR$ distribution is
given by the previously derived formula (\ref{dNdCBth}). Then we ask, how will 
the presence of spin correlations modify the phase space result (\ref{dNdCBth}).
For this purpose,  in Appendix~\ref{sec:derivation} we rederive the unit-normalized $\BAR$ distributions at threshold
for each of the 8 spin scenarios in Table~\ref{tab:models}, and for arbitrary
values of the relative chirality parameters $\alpha$, $\beta$ and $\gamma$
defined in (\ref{gammadef}-\ref{alphadef}). The result is
\begin{eqnarray}
\frac{\ud  N}{\ud \BAR } \Bigr\rvert^{\mathbf{SSF}}_{\mathbf{TH}}&&= \mathrm{J}_{\mathrm{PS}}  , 
\label{eq:ssfth}\\[2mm]
\frac{\ud  N}{\ud \BAR }  \Bigr\rvert^{\mathbf{SFS}}_{\mathbf{TH}}&&= \mathrm{J}_{\mathrm{PS}} + 
\frac{\gamma^2}{4} \cdot \mathrm{J}_{\mathrm{F}}   \, ,  
\label{eq:sfsth}\\[2mm]
\frac{\ud  N}{\ud \BAR }  \Bigr\rvert^{\mathbf{SFV}}_{\mathbf{TH}}&&= \mathrm{J}_{\mathrm{PS}} + 
\frac{\gamma^2}{4} \left(\frac{1-2 y}{1+2 y}\right)^2 \cdot
\mathrm{J}_{\mathrm{F}}   \, ,  
\label{eq:sfvth}\\[2mm]
\frac{\ud  N}{\ud \BAR }  \Bigr\rvert^{\mathbf{SVF}} _{\mathbf{TH}}&&= \frac{3(1+2y)}{(2+y)^2} 
\left\{ \mathrm{J}_{\mathrm{PS}} +
\frac{(1-y)^2}{4 (1+2 y)} \cdot \mathrm{J}_{\mathrm{V}} 
-\frac{\gamma^2}{2 (1+2y)} \cdot \mathrm{J}_{\mathrm{F}}  \right\} \, , 
\label{eq:svfth} \\[2mm]
\frac{\ud  N}{\ud \BAR } \Bigr\rvert^{\mathbf{VSF}}_{\mathbf{TH}}&&= \mathrm{J}_{\mathrm{PS}} \, ,  
\label{eq:vsfth}\\[2mm]
\frac{\ud  N}{\ud \BAR }  \Bigr\rvert^{\mathbf{VFS}}_{\mathbf{TH}}&&= \mathrm{J}_{\mathrm{PS}}  -
\frac{\gamma^2}{4} \cdot \mathrm{J}_{\mathrm{F}} \, ,  
\label{eq:vfsth}\\[2mm]
\frac{\ud  N}{\ud \BAR }  \Bigr\rvert^{\mathbf{VFV}}_{\mathbf{TH}}&&= \mathrm{J}_{\mathrm{PS}}  -
\frac{\gamma^2}{4} \left(\frac{1-2 y}{1+2 y}\right)^2 \cdot
\mathrm{J}_{\mathrm{F}}  \, , 
 \label{eq:vfvth}\\[2mm]
\frac{\ud  N}{\ud \BAR }  \Bigr\rvert^{\mathbf{VVF}} _{\mathbf{TH}}&&= \frac{9(1+y)}{2(2+y)^2} 
\left\{ \mathrm{J}_{\mathrm{PS}} -
\frac{(1-y)^2}{4 (1+ y)} \cdot \mathrm{J}_{\mathrm{V1}} 
-\frac{y(1-y)}{8 (1+y)} \cdot \mathrm{J}_{\mathrm{V2}} \right\} \, ,  
\label{eq:vvfth}
\end{eqnarray}
where 
\beq
y \equiv \frac{M_C^2}{M_B^2}
\eeq
and the basis functions ${\mathrm J}_i$ are defined as follows,
\bea
&&\mathrm{J}_{\mathrm{PS}}=
\frac{1- \BAR^2}{4\BAR^3}
\left\{ -2 \BAR+\left(1+\BAR^2\right)
 \ln\left[\frac{1+ \BAR}{1- \BAR}\right]\right\}  \, , 
 \label{eq:PHTH}
\\[2mm]
&&\mathrm{J}_{\mathrm{F}}= \frac{1-\BAR^2}{4 \BAR^5}
\left\{6\BAR+4 \BAR^3+6 \BAR^5-(3+\BAR^2+\BAR^4+3 \BAR^6)\ln\left(\frac{1+\BAR}{1-\BAR}\right) \right\}\, ,   
\label{eq:JF}
\\[2mm]
&&\mathrm{J}_{\mathrm{V}}= \frac{1-\BAR^2}{48 \BAR^7} \cdot
\bigg\{ 
-2 \BAR\left(45+22 \BAR^4+45 \BAR^8\right)  \nonumber \\
&&\qquad \quad+ 3\left(15-5\BAR^2+6\BAR^4+6\BAR^6-5 \BAR^8+15\BAR^{10}\right)\ln\left(\frac{1+\BAR}{1-\BAR}\right)\bigg \}\, ,  \label{JVdef} \\
 &&\mathrm{J}_{\mathrm{V1}}= \frac{1-\BAR^2}{24 \BAR^7} \cdot
\bigg\{ 
-2 \BAR\left(15+8 \BAR^2+10\BAR^4+8\BAR^6+15\BAR^8 \right) \nonumber \\[2mm]
&&\qquad \quad+ 3\left(5+\BAR^2+2\BAR^4+2\BAR^6+\BAR^8+5\BAR^{10}\right)\ln\left(\frac{1+\BAR}{1-\BAR}\right)\bigg \}\, , \label{JV1def} \\[2mm]
 &&\mathrm{J}_{\mathrm{V2}}= \frac{1}{3 \BAR^5} 
\left\{
 -2 \BAR\left(3+\BAR^2-\BAR^4-3\BAR^6 \right) 
 + 3\left(1-\BAR^8\right)\ln\left(\frac{1+\BAR}{1-\BAR}\right)\right \}\, .
 \label{eq:JV2}
\eea
The basis functions (\ref{eq:PHTH}-\ref{eq:JV2}) are pictorially illustrated in Fig.~\ref{fig:factorJ}.  The function $\mathrm{J}_{\mathrm{PS}}$
is simply the phase space limit (\ref{dNdCBth}) already derived in Sec.~\ref{sec:BarrPhase}, while the 
remaining 4 basis functions parameterize the different possible distortions of 
the pure phase space shape, for the various spin scenarios.

\FIGURE[ht]{
\centerline{ 
\includegraphics[width=10cm]{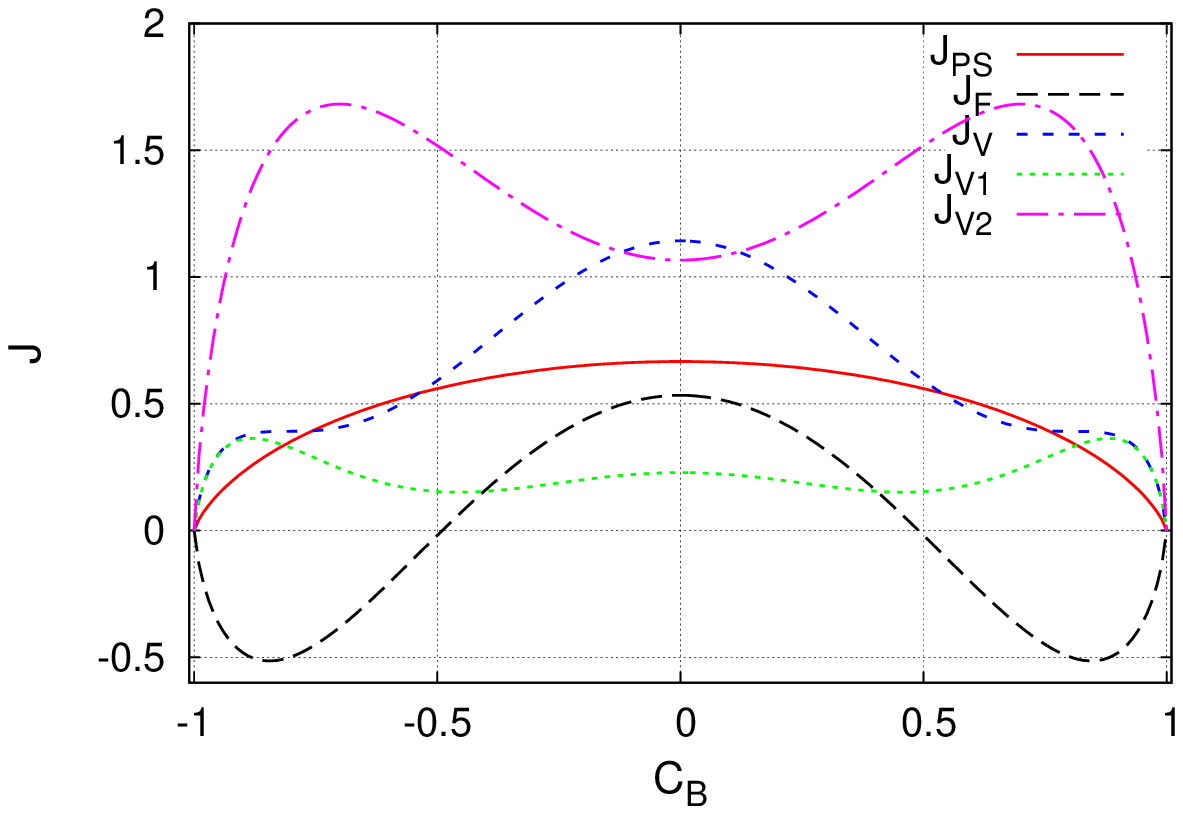}
}\\
\caption{The basis functions ${\mathrm J}(\BAR)$ defined in (\ref{eq:PHTH}-\ref{eq:JV2}). The solid red line is the pure phase space distribution 
(\ref{eq:PHTH}) or equivalently (\ref{dNdCBth}).}
\label{fig:factorJ}
}

Formulas (\ref{eq:ssfth}-\ref{eq:vvfth}) represent one of our main results.
Using (\ref{eq:ssfth}-\ref{eq:vvfth}), one can understand how the shape of the $\BAR$ distribution 
near threshold changes as a function of the mass spectrum (through the dependence on the $y$ parameter)
and as a function of the relative chirality parameter $\gamma$ defined in (\ref{gammadef}).
(The parameters $\alpha$ and $\beta$ decouple and do not enter the 
threshold distribution formulas.) For example, (\ref{eq:ssfth}) and (\ref{eq:vsfth})
show that in the cases where the intermediate particles $B$ are scalars, 
any spin correlations between the two leptons are wiped out and 
one obtains the pure phase space shape (\ref{eq:PHTH}), which
peaks at $\BAR = 0$ and vanishes at $\BAR=\pm 1$ (see the solid red line in Fig.~\ref{fig:factorJ}).
Furthermore, the predictions (\ref{eq:ssfth}) and (\ref{eq:vsfth}) are not sensitive to
the mass spectrum at all, since the mass ratio $y$ does not enter those formulas.

\FIGURE[t]{\centering
\includegraphics[width=0.45\textwidth]{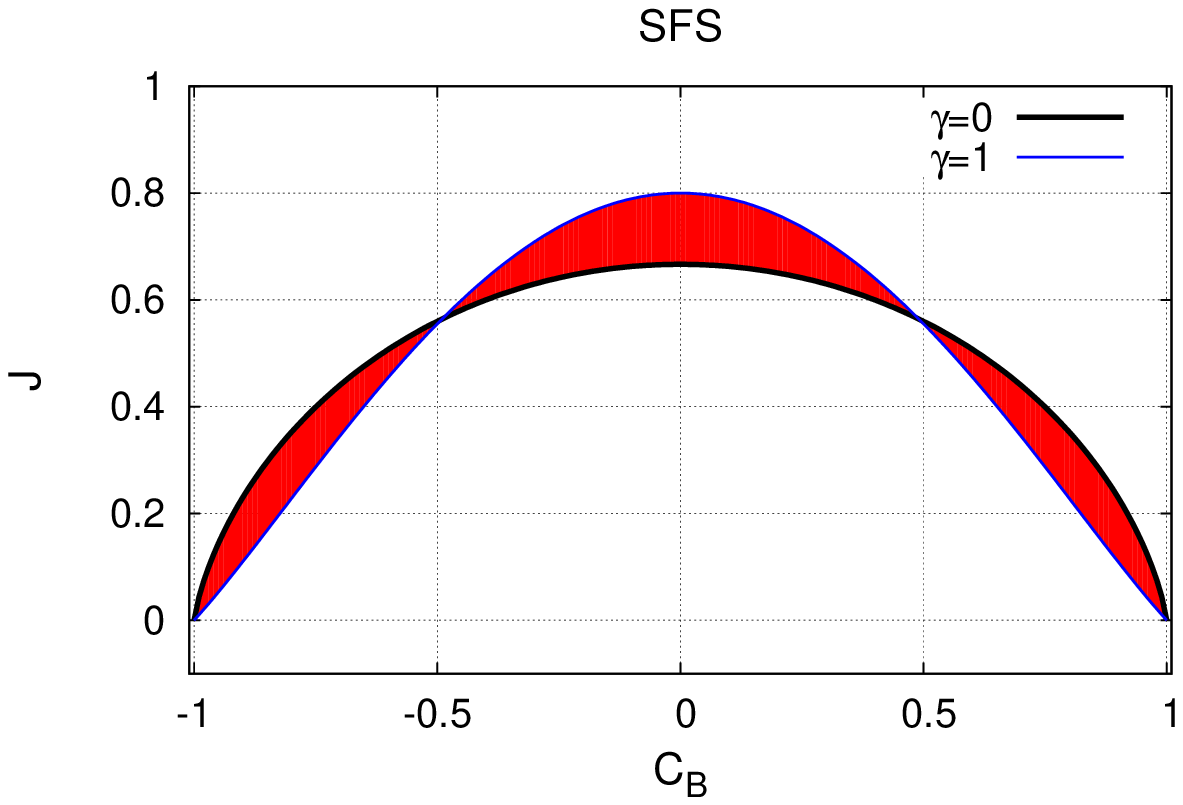}
\includegraphics[width=0.45\textwidth]{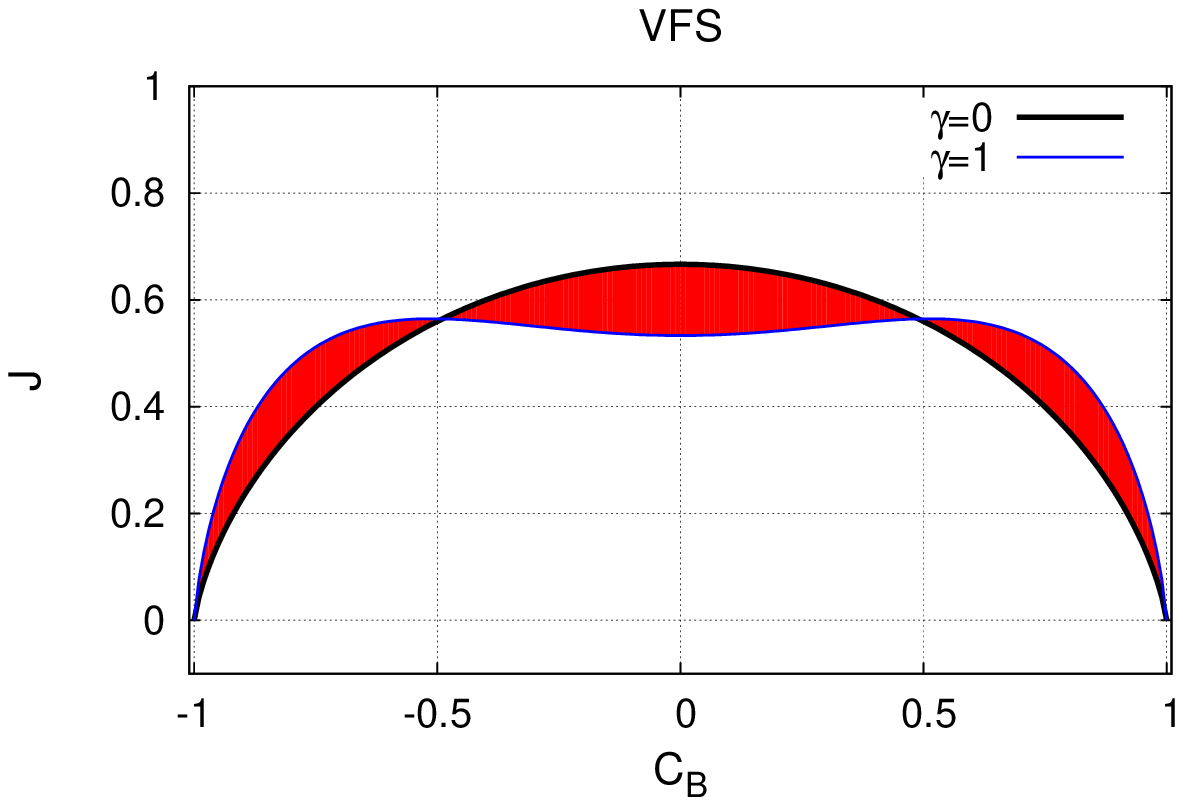}
\includegraphics[width=0.45\textwidth]{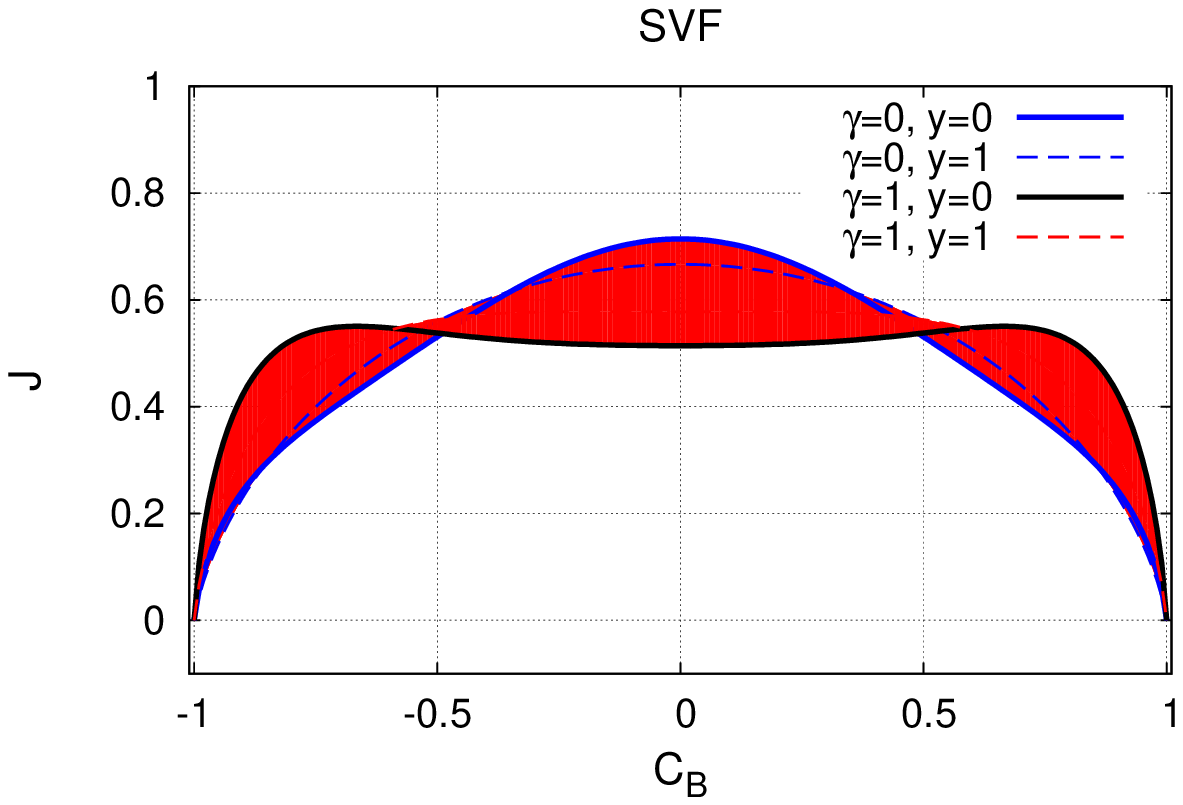}
\includegraphics[width=0.45\textwidth]{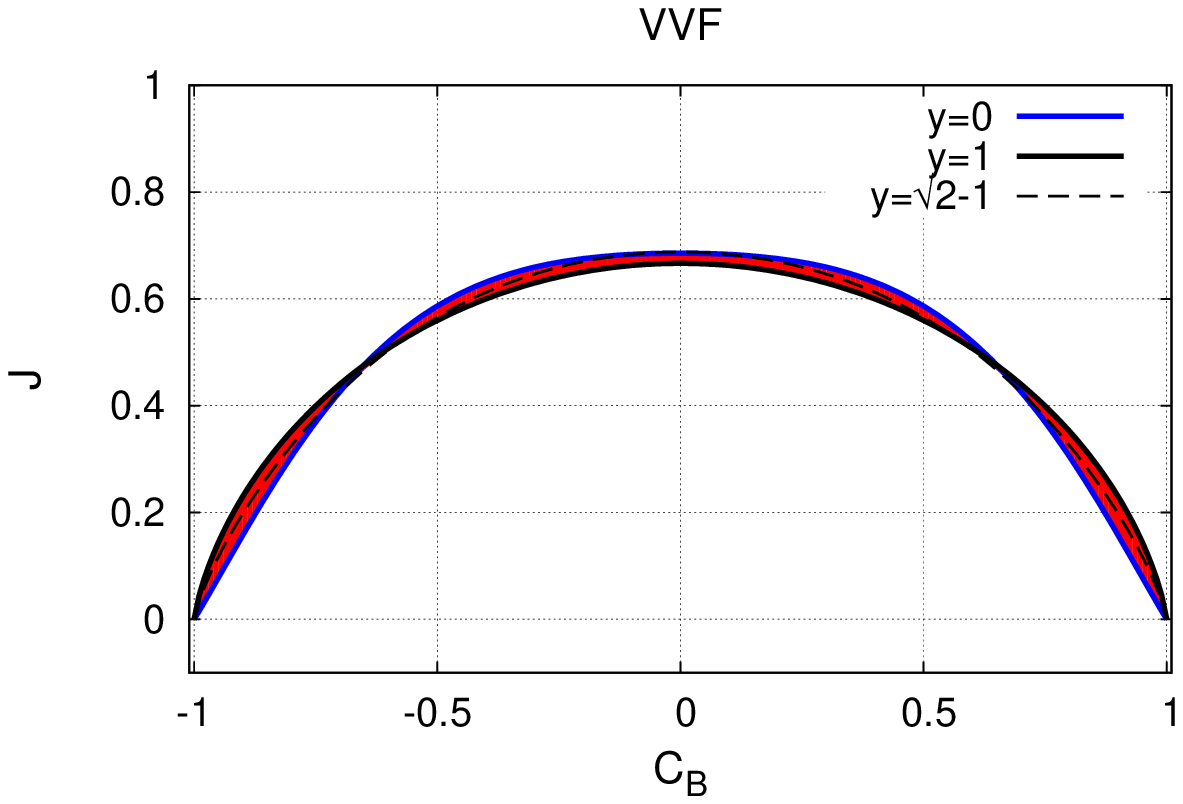}
\caption{Spin correlation effects on the $\BAR$ distribution at threshold, for various spin scenarios. 
In each case, we start with the pure phase space distribution
at $\gamma=0$ and/or $y=1$, and show the amount of shape distortion 
by varying $\gamma$ and $y$ over their full range. }
\label{fig:variation}
}

Now let us proceed to the SFS and VFS scenarios of (\ref{eq:sfsth}) and (\ref{eq:vfsth}), correspondingly.
Here we find that the spin correlations induce a second term involving the 
$\mathrm{J}_\mathrm{F}$ function, which has a maximum at $\BAR=0$ and two 
local minima at $\BAR=\pm 0.845$ (see the  black dashed line in Fig.~\ref{fig:factorJ}).
The coefficient of the $\mathrm{J}_\mathrm{F}$ term is positive in (\ref{eq:sfsth}) and 
negative in (\ref{eq:vfsth}). This means that in the SFS spin scenario, the addition of spin
correlations will make the central peak of the pure phase space distribution $\mathrm{J}_{\mathrm{PS}}$
even more pronounced. This is illustrated in the upper left panel of Fig.~\ref{fig:variation},
where we start with the pure phase space distribution at $\gamma=0$ (the black solid line)
and gradually increase the amount of spin correlations until we get purely chiral couplings 
at $\gamma=1$. As expected, the presence of spin correlations in the SFS scenario 
makes the central peak in the $\BAR$ distribution steeper. In the VFS scenario, on the other hand, 
spin correlations have just the opposite effect, since the $\mathrm{J}_\mathrm{F}$ term 
in (\ref{eq:vfsth}) comes with a minus sign. Now, the central peak will be suppressed,
and events will shift towards the two endpoints instead. The resulting shape distortion is 
illustrated in the upper right panel of Fig.~\ref{fig:variation}, where again we show the full range
from $\gamma=0$ (pure phase space) to $\gamma=1$ (purely chiral couplings).
We see that the presence of chiral couplings causes the VFS distribution to be much flatter than in
the pure phase space limit.

The SFS and VFS results shown in the upper two panels of Fig.~\ref{fig:variation}
are also independent of the mass spectrum, as the mass parameter $y$ does not
enter eqs.~(\ref{eq:sfsth}) and (\ref{eq:vfsth}). In contrast, the remaining four
spin cases to be discussed now will exhibit a dependence on the mass spectrum through $y$.
It is probably easiest to start with the SFV and VFV spin configurations, whose
$\BAR$ distributions are given by (\ref{eq:sfvth}) and (\ref{eq:vfvth}), respectively.
We find very similar behavior to the two cases of SFS and VFS just discussed --- the only 
difference is that the coefficient of the $\mathrm{J}_\mathrm{F}$ function is now
additionally suppressed by a factor of $(1-2y)^2/(1+2y)^2$, which incorporates 
the mass spectrum dependence.
Because of this suppression, one would generally expect smaller variations
than what we saw in the upper two panels of Fig.~\ref{fig:variation}.
There even exists a very special case with $y=1/2$ (i.e.~$M_B=\sqrt{2}M_C$)
when the $\BAR$ distribution reduces to the pure phase space prediction (\ref{eq:PHTH}).
For that particular mass spectrum, the SFV and VFV distributions at threshold would be 
identical to the SSF and VSF distributions, and spin discrimination would 
have to be done in a regime away from threshold.

Let us now move on to the SVF scenario of (\ref{eq:svfth}). This is perhaps the most
complicated case, since in addition to the $\mathrm{J}_\mathrm{PS}$ baseline shape
it has two extra terms --- one with the already familiar function 
$\mathrm{J}_\mathrm{F}$, and a second term involving the function $\mathrm{J}_\mathrm{V}$
(shown in Fig.~\ref{fig:factorJ} with the blue dashed line). The function $\mathrm{J}_\mathrm{V}$
peaks at $\BAR=0$ and has broad shoulders near the endpoints of the $\BAR$ interval.
The coefficient of $\mathrm{J}_\mathrm{V}$ in (\ref{eq:svfth}) is always positive, 
which means that the $\mathrm{J}_\mathrm{V}$ term will always tend to enhance the central peak.
At the same time, the coefficient of the $\mathrm{J}_\mathrm{F}$ term in (\ref{eq:svfth})
is always negative, which would tend to suppress the central peak, as already 
seen in the upper right panel of Fig.~\ref{fig:variation}. Thus the interplay of the two terms 
will lead to partial cancellations, which are illustrated in the lower left panel of Fig.~\ref{fig:variation}.
The dashed blue line represents the pure phase space limit in eq.~(\ref{eq:svfth}),
which is recovered when $\gamma=0$ and $y=1$. If we keep $\gamma=0$, the 
$\mathrm{J}_\mathrm{F}$ term is turned off, and by varying $y$, we can see the
impact of the $\mathrm{J}_\mathrm{V}$ term in (\ref{eq:svfth}), which is maximal at $y=0$
(the solid blue line). Conversely, if we keep $y=1$, the $\mathrm{J}_\mathrm{V}$ term is
turned off, and the variations in $\gamma$ reveal the effect of the $\mathrm{J}_\mathrm{F}$ term,
which is maximal at $\gamma=1$ (the red dashed line). The case when both terms are maximal
($\gamma=1$ and $y=0$) is given by the solid black line.

Finally, we discuss the VVF case of eq.~(\ref{eq:vvfth}), pictured in the lower right panel of Fig.~\ref{fig:variation}.
Here the $\BAR$ distribution at threshold depends only on the mass spectrum but not on 
the chirality parameter $\gamma$. The spin correlation effects are encoded in the two functions
$\mathrm{J}_\mathrm{V1}$ and $\mathrm{J}_\mathrm{V2}$, represented in Fig.~\ref{fig:factorJ} 
by the green dotted and magenta dot-dashed lines, respectively.
$\mathrm{J}_\mathrm{V1}$ has local maxima at $\BAR=0$ and $\BAR= \pm 0.881$ 
and local minima at $\BAR=\pm 0.456$, while $\mathrm{J}_\mathrm{V2}$
has local maxima at $\BAR=\pm 0.7$ and a local minimum at  $\BAR=0$.
The lower right panel in Fig.~\ref{fig:variation} shows the resulting variation
in the $\BAR$ shape as we vary $y$ from $y=1$ (the pure phase space limit)
to $y=0$ (when the $\mathrm{J}_\mathrm{V1}$ term is maximal). 
We see only very minor variations, due to the fact that the coefficients of
both the $\mathrm{J}_\mathrm{V1}$ and $\mathrm{J}_\mathrm{V2}$
terms in (\ref{eq:vvfth}) are numerically very small.

This concludes our discussion of eqs.~(\ref{eq:ssfth}-\ref{eq:vvfth}), which 
manifestly describe the spin effects at threshold. As Fig.~\ref{fig:shatstudy} showed, 
although most events are produced {\em near} threshold, virtually none are produced exactly 
at threshold, thus eqs.~(\ref{eq:ssfth}-\ref{eq:vvfth}) are perhaps of limited practical interest. 
Their true value is in developing some intuition about
the shape of the $\BAR$ distribution, which could prove useful for the interpretation 
of the finite $\sqrt{s}$ results below.

\subsection{Spin effects in the large energy limit}
\label{sec:infiniteS}

We now turn our attention to the other $\sqrt{s}$ extreme, namely $\sqrt{s}\to \infty$.
In that case, as discussed in Sec.~\ref{sec:BarrPhase} and shown in (\ref{dBAReqdTheta}),
the $\BAR$ distribution reduces to the $\cos{\Theta^\ast}$ distribution 
for $2\rightarrow 2$ processes, whose dependence on the spins of the particles is well known,
so we have
\bea
\frac{\ud   N}{\ud \BAR } \Bigr\rvert^{\mathbf{SXY}}_{\mathbf{\infty}} && \propto 1\, , \quad \textrm{for  } \,  \{X\, ,Y\}  \subset \{\textrm{S, F, V}\} \,  , 
\label{flat}\\[2mm]
\frac{\ud   N}{\ud \BAR } \Bigr\rvert^{\mathbf{VFY}}_{\mathbf{\infty}}&& \propto 1+\BAR^2\, ,  \quad \textrm{for  } \,  Y \in \{\textrm{S, V}\} \, ,
\label{oneplus} \\[2mm]
\frac{\ud   N}{\ud \BAR } \Bigr\rvert^{\mathbf{VYF}}_{\mathbf{\infty}} && \propto1-\BAR^2\, ,  \quad \textrm{for  } \,  Y \in \{\textrm{S, V}\} \, .
\label{oneminus}
\eea
These equations predict three generic shapes for the $\BAR$ distribution at large energies:
\begin{itemize}
\item Flat as in (\ref{flat}). This is the case whenever $A$ is a scalar particle.
\item $1+\BAR^2$ as in (\ref{oneplus}). This case occurs when $A$ has spin 1 and $B$ is a fermion,
e.g. the UED example from Section~\ref{sec:variables} falls into this category.
The distribution (\ref{oneplus}) peaks at $\BAR=\pm 1$ and has a minimum at $\BAR=0$.
\item $1-\BAR^2$ as in (\ref{oneminus}). This is the case when $A$ has spin 1 and $B$ is a boson,
as in the SUSY example considered earlier.
The $\BAR$ distribution then has a peak at $\BAR=0$ and vanishes at the endpoints $\BAR=\pm 1$.
\end{itemize}
As shown in Sec.~\ref{sec:BarrPhase}, the nice correlation (\ref{dBAReqdTheta}) between $\BAR$ and $\cos\Theta^\ast$ 
emerges only at sufficiently large energies, which might not be realistically achieved at a hadron collider.
Therefore, the practical value of the limits (\ref{flat}-\ref{oneminus}) is also debatable,
just like the formulas (\ref{eq:ssfth}-\ref{eq:vvfth}) from Sec.~\ref{sec:dNdBARth}.
However, with the use of (\ref{eq:ssfth}-\ref{eq:vvfth}) and (\ref{flat}-\ref{oneminus}),
one can begin to understand the observed $\BAR$ shapes in Fig.~\ref{fig:var},
where we noticed that the MUED case gives a flatter $\BAR$ distribution than the MSSM.
In the case of the MSSM, the $\BAR$ distribution at threshold (\ref{eq:vsfth}) peaks at $\BAR=0$
and vanishes at $\BAR=\pm 1$, while the large energy limit (\ref{oneminus}) {\em also}
peaks at $\BAR=0$ and vanishes at $\BAR=\pm 1$. 
The $\BAR$ distribution at finite $\sqrt{s}$ would then interpolate between those two {\em very similar} limiting cases
and inherit their common properties --- which explains why the actual MSSM distribution in Fig.~\ref{fig:var}
has a relatively sharp peak at $\BAR=0$ and vanishes at $\BAR=\pm 1$.

In the case of MUED, we have a different story --- here at intermediate $\sqrt{s}$ we have to interpolate between 
two very different shapes: the threshold distribution (\ref{eq:vsfth}) which peaks at $\BAR=0$
and vanishes at $\BAR=\pm 1$, and the asymptotic distribution (\ref{oneplus}) which peaks 
at $\BAR=\pm 1$ and has a minimum at $\BAR=0$. At the end of the day, the resulting $\BAR$ shape 
will depend on the exact form of the underlying $\sqrt{s}$ distribution, but Fig.~\ref{fig:shatstudy} suggests that
the threshold behavior will dominate, since a typical event is produced closer to threshold.
This is indeed what we observe in Fig.~\ref{fig:var}, where the $\BAR$ distribution in the MUED case
{\em also} peaks at $\BAR=0$, in spite of the asymptotic expectation (\ref{oneplus}).

\subsection{Spin effects near threshold}
\label{sec:offshell}

When particle $A$ is off-shell (i.e.~$M_A<2M_B$), the onset of the $pp\to B_1B_1$ production cross-section
as a function of $\sqrt{s}$ is sensitive to the spins of $A$ and $B$. 
The partonic cross section at a given $\sqrt{s}$ can be expressed as
\beq
\hat \sigma{\left(\sqrt{s}\right)}=\frac{1}{2 s} \int\ud \Pi_2^* \left\lvert M{\left(\sqrt{s}\right)}\right\rvert^2
=\left(\frac{1}{8\pi}\right)^2 \frac{\beta_{\textrm{CM}} }{s} \int\ud\Omega^*\left\lvert M{\left(\sqrt{s}\right)}\right\rvert^2 \, ,
\eeq
where $\beta_{\textrm{CM}}$ is the $B_i$ boost factor in the CMBB frame
\beq
\beta_{\textrm{CM}} \equiv \tanh \eta^\ast. 
\eeq 
The cross-section near threshold is suppressed by a certain power of $\beta_{\textrm{CM}}$,
which depends on the spin scenario: 
\bea
\left\lvert M{\left(\sqrt{s}\right)}\right\rvert^2_{(\textrm{SS})} && \propto 1+\mathcal{O}{\left( \beta_{\textrm{CM}}^2\right)}\, ,\\
\left\lvert M{\left(\sqrt{s}\right)}\right\rvert^2_{(\textrm{SF})} && \propto \left(1-\cos{\delta}\right)+\left(1+\cos{\delta}\right) \beta_{\textrm{CM}}^2+\mathcal{O}{\left( \beta_{\textrm{CM}}^4\right)}\, ,\\
\left\lvert M{\left(\sqrt{s}\right)}\right\rvert^2_{(\textrm{SV})} && \propto 1+\mathcal{O}{\left( \beta_{\textrm{CM}}^2\right)}\, ,\\
\left\lvert M{\left(\sqrt{s}\right)}\right\rvert^2_{(\textrm{VS})} && \propto  \beta_{\textrm{CM}}^2 
\left(1-\cos^2{\Theta^\ast}\right)+\mathcal{O}{\left( \beta_{\textrm{CM}}^4\right)}\, ,\\
\left\lvert M{\left(\sqrt{s}\right)}\right\rvert^2_{(\textrm{VF})} && \propto 1+\mathcal{O}{\left( \beta_{\textrm{CM}}\right)}\, ,\\
\left\lvert M{\left(\sqrt{s}\right)}\right\rvert^2_{(\textrm{VV})} && \propto  \beta_{\textrm{CM}}^2
\left(1-\frac{3}{19} \cos^2{\Theta^\ast}\right)
+\mathcal{O}{\left( \beta_{\textrm{CM}}^4\right)}\, .
\eea
Interestingly, the $SF$ case can be subdivided into two categories: a real scalar 
\beq
\left\lvert M{\left(\sqrt{s}\right)}\right\rvert^2_{(\textrm{S}^{{\scriptscriptstyle ++}}\textrm{F}) } \propto  \beta_{\textrm{CM}}^2+\mathcal{O}{\left( \beta_{\textrm{CM}}^4\right)}\, ,
\eeq
or a pseudo-scalar
\beq
\left\lvert M{\left(\sqrt{s}\right)}\right\rvert^2_{(\textrm{S}^{{\scriptscriptstyle -+}}\textrm{F}) } \propto 1+\mathcal{O}{\left( \beta_{\textrm{CM}}^4\right)}\, .
\eeq
Thus, the $ \beta_{\textrm{CM}}$ suppression factor near threshold is as follows: 
\beq
\frac{\ud \hat\sigma(\sqrt{s}) }{\ud\Omega^\ast }  \propto \left\{
    \begin{array}{rl}
       \beta_{\textrm{CM}}, & \quad \mathrm{for\ }  (\textrm{SS})\, ,(\textrm{S}^{{\scriptscriptstyle PC}}\textrm{F})=(\textrm{S}^{{\scriptscriptstyle -+}}\textrm{F}) \, ,(\textrm{SV})\, ,(\textrm{VF})\, ,\\[3mm]
       \beta_{\textrm{CM}}^3, & \quad\mathrm{for\ }  (\textrm{S}^{{\scriptscriptstyle PC}}\textrm{F})=(\textrm{S}^{{\scriptscriptstyle ++}}\textrm{F})\, ,(\textrm{VS})\, ,(\textrm{VV}).
    \end{array} \right.
\label{eq:boost}
\eeq
 
\section{Comparison of the different spin scenarios}
\label{sec:discrimination}

We are now ready to contrast the different spin scenarios and discuss the prospects for spin discrimination.
In principle, one could pose two questions: 
\begin{enumerate}
\item How well can two different spin scenarios be discriminated experimentally, 
given the uncertainty in the coupling chiralities and instrumental effects like
SM backgrounds and the finite detector resolution. Clearly, the answer to this question
will depend on many quantitative factors - the chosen study point, the size of the signal, etc.,
and it is difficult if not impossible to give a``one size fits all" answer.
\item Which spin scenarios (and under what circumstances) are in danger of being 
confused with each other? This question is easier to tackle theoretically, 
because if we can identify the cases where two spin scenarios look identical
at the parton level, the conclusions will remain unchanged when we add
all the usual experimental complications. This is why in this section we shall
use the intuition developed in previous sections to pinpoint the difficult 
cases for spin discrimination. 
\end{enumerate}

 \TABULAR[t]{|c||c|c|c||c|}{\hline
Configuration  &  $\sqrt{s} \to 2 M_{B}  $ &  $\sqrt{s} \rightarrow \infty$  & $|M|^2 \propto \beta_{CM}^n$  & evil twin \\  \hline \hline
SSF   &  $\mathrm{J}_{\mathrm{PS}}  $  &  1 &  1      & SVF\\  \hline
SVF   &  $\mathrm{J}_{\mathrm{PS}} + \epsilon_{2}\, \mathrm{J}_{\mathrm{V}}  - \frac{\gamma^2}{4} \, \epsilon_{3}\, \mathrm{J}_{\mathrm{F}} $     &1 	 &    1     & SSF\\  \hline\hline
SFS   &  $\mathrm{J}_{\mathrm{PS}} +\frac{\gamma^2}{4} \,\mathrm{J}_{\mathrm{F}} $ & 1&
\begin{minipage} {0.19\linewidth}
\centering
\begin{tabular}{c|c}
$\scriptstyle{\mathrm{S}^{{\scriptscriptstyle -+}}\mathrm{FS}}$ & $\scriptstyle{\mathrm{S}^{{\scriptscriptstyle ++}}\mathrm{FS}}$\\ 
1 & 2
\end{tabular}
\end{minipage}
 &  SFV   \\  \hline
SFV  &  $\mathrm{J}_{\mathrm{PS}} +\frac{\gamma^2}{4} \, \epsilon_1\,\mathrm{J}_{\mathrm{F}} $   &  1 &  
\begin{minipage} {0.19\linewidth}
\centering
\begin{tabular}{c|c}
$\scriptstyle{\mathrm{S}^{{\scriptscriptstyle -+}}\mathrm{FV}}$ & $\scriptstyle{\mathrm{S}^{{\scriptscriptstyle ++}}\mathrm{FV}}$\\ 
1 & 2
\end{tabular}
\end{minipage}
 & SFS      \\  \hline\hline
VSF   &   $\mathrm{J}_{\mathrm{PS}}  $  &  $1-\BAR^2$   &  2       & VVF \\  \hline
VVF   &  $\mathrm{J}_{\mathrm{PS}} - \epsilon_{4}\, \mathrm{J}_{\mathrm{V1}}  -  \,\epsilon_{5}\, \mathrm{J}_{\mathrm{V2}} $      &  $1-\BAR^2$    &  2  & VSF \\  \hline\hline
VFS   &  $\mathrm{J}_{\mathrm{PS}} -\frac{\gamma^2}{4} \,\mathrm{J}_{\mathrm{F}} $    &  $1+\BAR^2$   &   1  & VFV  \\  \hline
VFV   & $\mathrm{J}_{\mathrm{PS}} -\frac{\gamma^2}{4} \, \epsilon_1 \,\mathrm{J}_{\mathrm{F}} $      & $1+\BAR^2$    &   1 & VFS   \\  \hline\hline
}{\label{table:results} The asymptotic behavior of the $\BAR$ distribution at threshold (second column) and 
at $\sqrt{s}\to\infty$ (third column) for different spin configurations in the antler topology. The third column 
lists the power of any additional $\beta_{\textrm{CM}}$ threshold suppression coming from the matrix element.
The last column shows the alternative spin scenario with similar properties.
}

Let us focus on the $\BAR$ distribution. Table \ref{table:results} summarizes 
its salient features identified in Sec.~\ref{sec:Chiralities}: the behavior at
threshold (second column) or asymptotically at $\sqrt{s}\to \infty$ (third column),
and the power of any additional $\beta_{\textrm{CM}}$ threshold suppression arising from
the matrix element. For convenience, formulas (\ref{eq:ssfth}-\ref{eq:vvfth})
are rewritten in terms of the shorthand notation
\begin{align}
\epsilon_1=\left(\frac{1-2y}{1+2y}\right)^2&& \epsilon_2=\frac{(1-y)^2}{4(1+2y)}&&\epsilon_3=\frac{2}{(1+2y)}\\
\epsilon_4=\frac{(1-y)^2}{4(1+y)}&&\epsilon_5=\frac{y(1-y)}{8(1+y)}.
\end{align}
In Table~\ref{table:results}, the 8 spin configurations from Table~\ref{tab:models} 
are arranged in pairs. The two spin scenarios within each pair can exhibit identical behavior both at threshold 
(for a suitably chosen mass spectrum) and at $\sqrt{s}\to\infty$, {\em and} have the same $\beta_{\textrm{CM}}$ threshold suppression. 
Those pairs are therefore the problematic cases where spin discrimination based on $\BAR$ might be difficult.
Of course, this is simply a conjecture based on the asymptotic behavior, so it is worth 
checking if it holds for the $\BAR$ distribution at any $\sqrt{s}$. Then
if it turns out that the $\BAR$ distribution is indeed very similar for a given pair 
of spin configurations, we will then check whether any of the remaining variables discussed 
in Sec.~\ref{sec:variables} can offer an alternative tool for discrimination.

\FIGURE[t]{
\includegraphics[width=\textwidth]{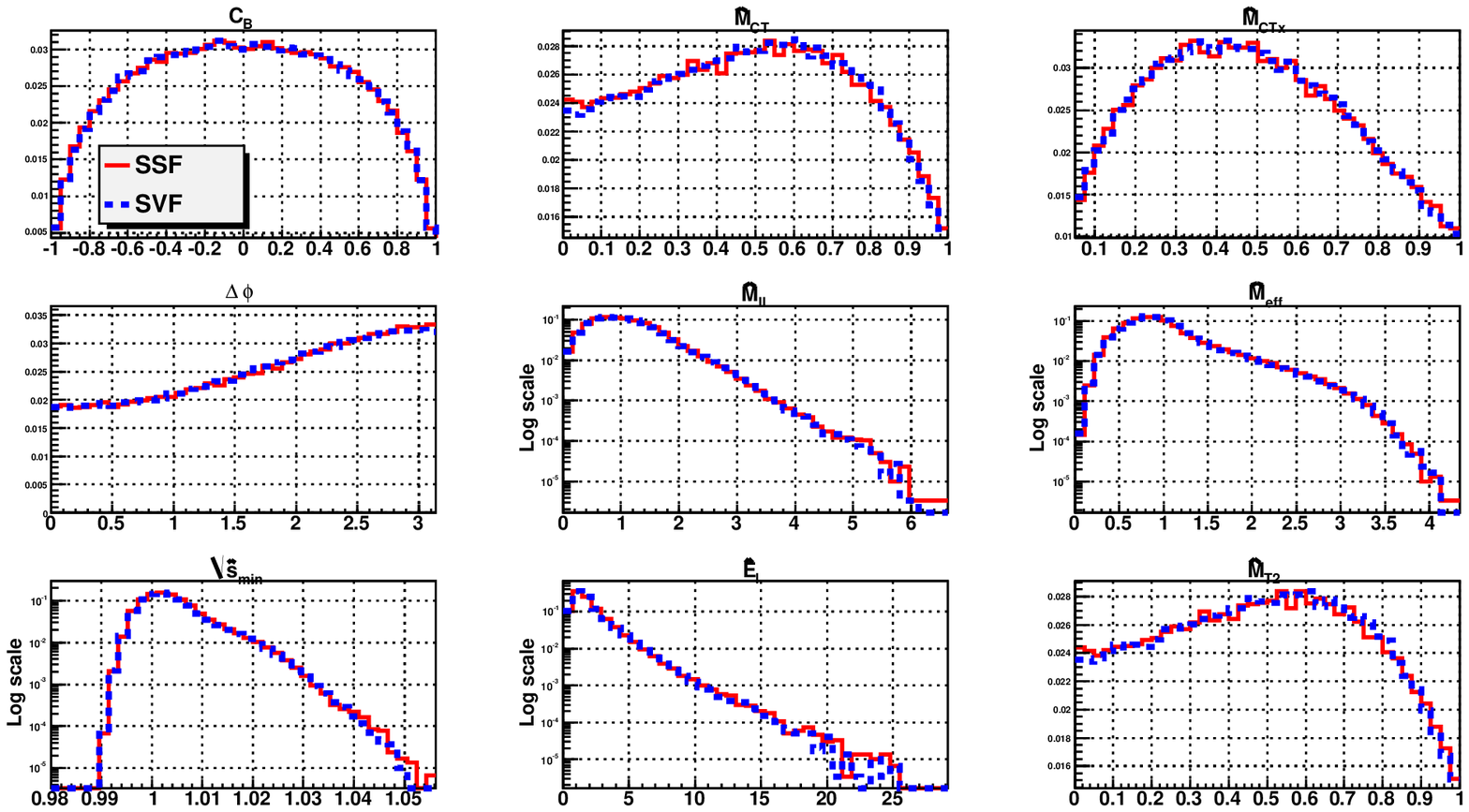} 
\caption{The same as Fig.~\ref{fig:var}, but this time comparing kinematic distributions 
for the SSF and SVF spin scenarios at an 8 TeV LHC. 
The couplings and masses in each case are the same, and were chosen to match the 
asymptotic $\BAR$ distribution at threshold ($y \sim 1$ and $\gamma=0$):
$M_A=1200$ GeV, $M_B=500$ GeV and $M_C=495$ GeV. 
For this case, $\beta$ is not defined, while the choice of $\alpha$ is inconsequential.
}
\label{fig:ssf_svf}
}

Figs.~\ref{fig:ssf_svf}-\ref{fig:vfs_vfv} show the distributions of the nine variables from Sec.~\ref{sec:variables},
for each of the twin spin scenarios in Table~\ref{table:results}. All simulations were done with MadGraph v4\cite{Alwall:2007st}, in the 
narrow width approximation, for an 8 TeV LHC. In all figures, we consider the case of a heavy resonance $A$ with $M_A=1.2$ TeV,
and intermediate particles $B_i$ with mass $M_B=500$ GeV. We then adjust the mass of $M_C$
appropriately in order to obtain an exact match in the $\BAR$ distributions at threshold. 
For example, Fig.~\ref{fig:ssf_svf}
compares the case of SSF to SVF. The former has a $\BAR$ distribution which is given by the pure phase space formula
at threshold, while the $\BAR$ distribution for the latter involves terms proportional to $\epsilon_2$
and $\epsilon_3\gamma^2/4$ (see Table~\ref{table:results}). Then the worst case scenario would have a degenerate spectrum with $y\to 1$
($M_B\sim M_C$), which would set $\epsilon_2=0$, and purely vector-like couplings with $\gamma=0$.
This is precisely the case shown in Fig.~\ref{fig:ssf_svf}, where for definiteness we choose $M_C=495$ GeV. 
As expected, the two $\BAR$ distributions are almost identical
and can hardly be used to measure the spins. The remaining plots in Fig.~\ref{fig:ssf_svf} then check
whether any of the other eight distributions show observable differences. 
As it turns out, they are all pretty well matched as well, and the spin discrimination
between SSF and SVF is indeed very problematic under these circumstances.

\FIGURE[t]{
\includegraphics[width=\textwidth]{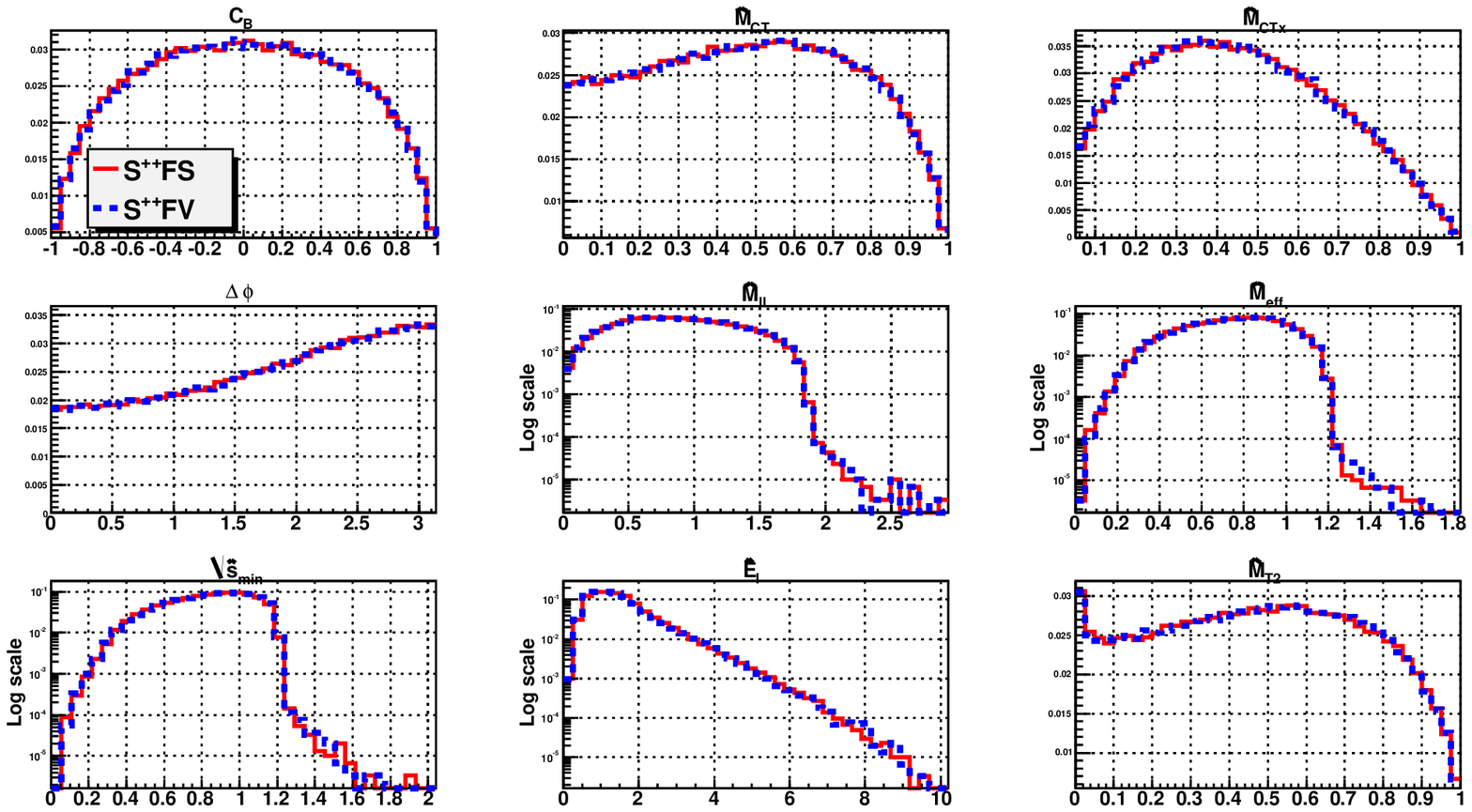} 
\caption{The same as Fig.~\ref{fig:ssf_svf}, but for the SFS and SFV spin scenarios with a scalar particle $A$. 
Here the threshold behavior for $\BAR$ is obtained with $y \sim 0$, so unlike Fig.~\ref{fig:ssf_svf}, here
we choose $M_C=5$ GeV.}
\label{fig:sfs_sfv_scalar}
}
 
In the next two figures we show a similar comparison between the SFS and SFV
spin scenarios: in Fig.~\ref{fig:sfs_sfv_scalar} for the case when particle $A$ is a scalar ($S^{{\scriptscriptstyle ++}}$) 
and in Fig.~\ref{fig:sfs_sfv_pseudoscalar} for the case when particle $A$ is a pseudoscalar ($S^{{\scriptscriptstyle -+}}$)
Here, to match the threshold $\BAR$ distributions, it is sufficient to set $y=0$
(for definiteness, we choose $M_C=5$ GeV). Figs.~\ref{fig:sfs_sfv_scalar}  
and \ref{fig:sfs_sfv_pseudoscalar} reveal a similarly grim situation --- the distributions 
of all 9 variables match almost perfectly in the two cases, 
and any spin discrimination thus appears to be virtually impossible.

\FIGURE[t]{
\includegraphics[width=\textwidth]{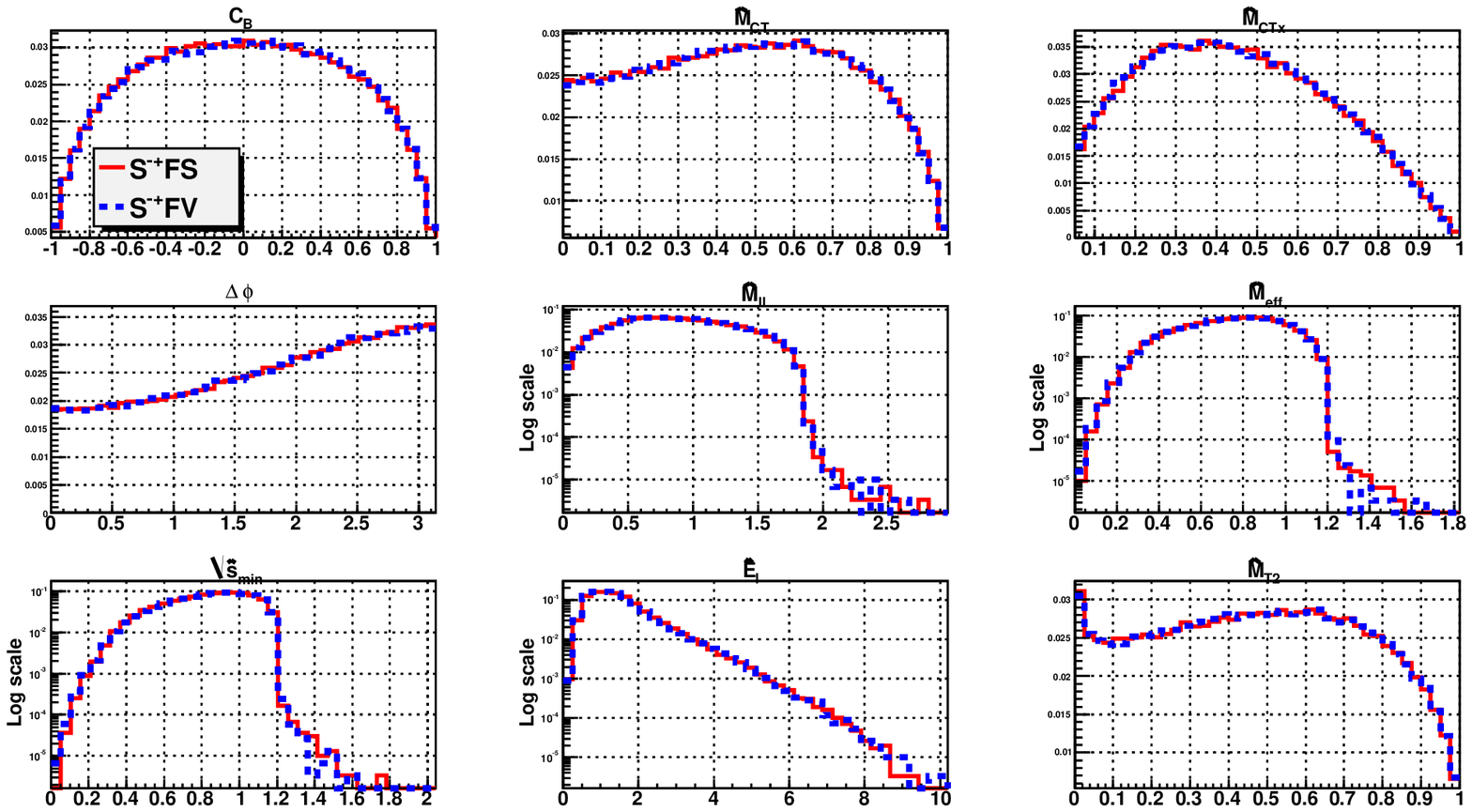} 
\caption{The same as Fig.~\ref{fig:sfs_sfv_scalar}, but for a pseudoscalar particle $A$.  }
\label{fig:sfs_sfv_pseudoscalar}
}

The next pair of similar spin configurations in Table~\ref{table:results} is
VSF and VVF, which are compared in Fig.~\ref{fig:vsf_vvf}. The threshold
$\BAR$ distribution for VSF is given by the pure phase space result (\ref{eq:PHTH})
while for VVF, it reduces to the same formula if $y=1$.
In order to mimic the case of $y=1$, we again choose $M_C=495$ GeV.
Once again, we find that the distributions of all 9 variables are pretty similar.

\FIGURE[t]{
\includegraphics[width=\textwidth]{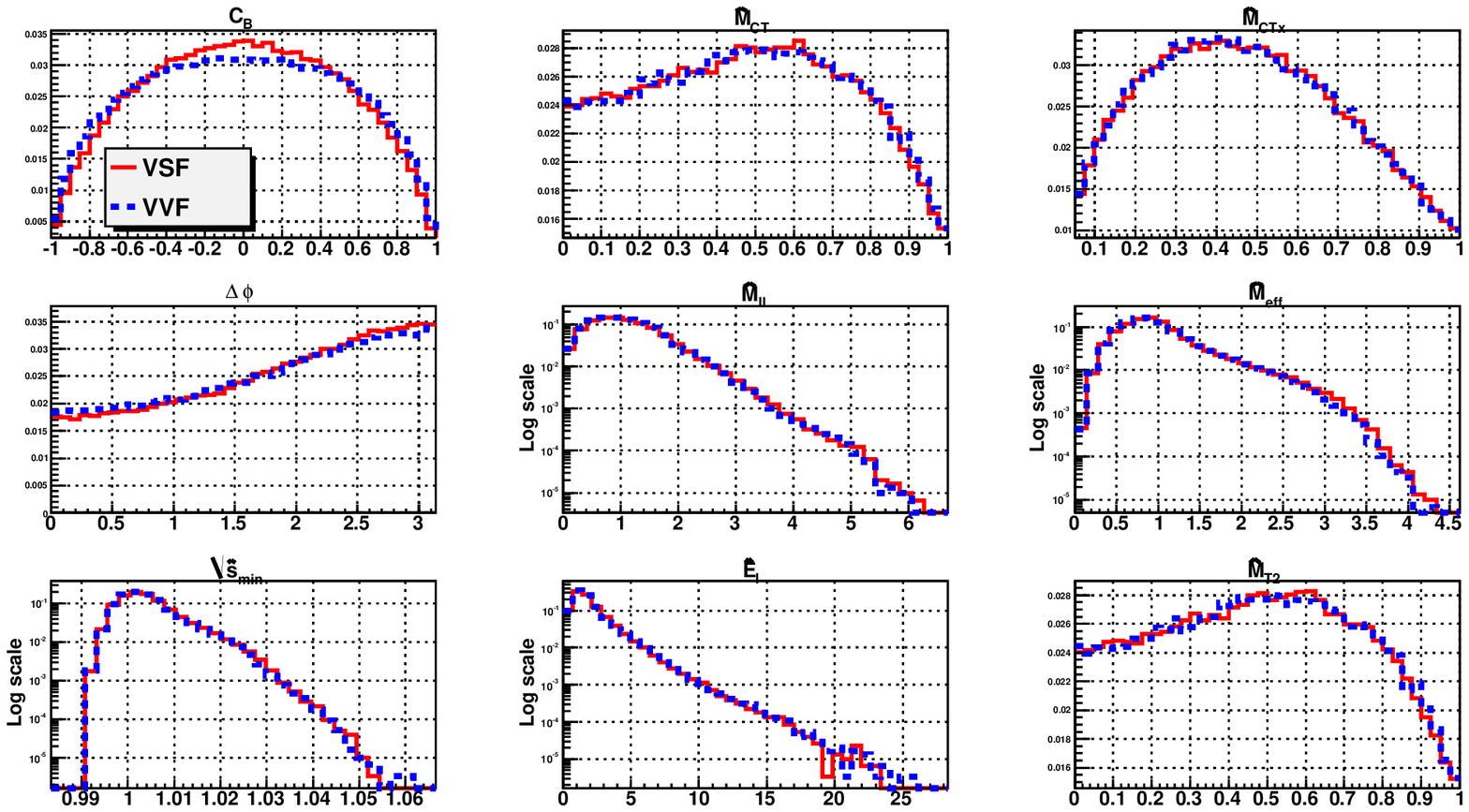} 
\caption{The same as Fig.~\ref{fig:ssf_svf}, but for the VSF and VVF spin scenarios with $y \sim 1$ ($M_C=495$ GeV).
}
\label{fig:vsf_vvf}
}

The last problematic pair of spin configurations in Table~\ref{table:results} is VFS and VFV. 
Their $\BAR$ distributions at threshold become identical if $\gamma=0$ or
if $y \sim 0$. Fig.~\ref{fig:vfs_vfv} shows a case with $\gamma=0$ and $y \sim 0$.
Once again we observe that all 9 kinematic variables behave very 
similarly and spin discrimination is very difficult.

\FIGURE[!ht]{
\includegraphics[width=\textwidth]{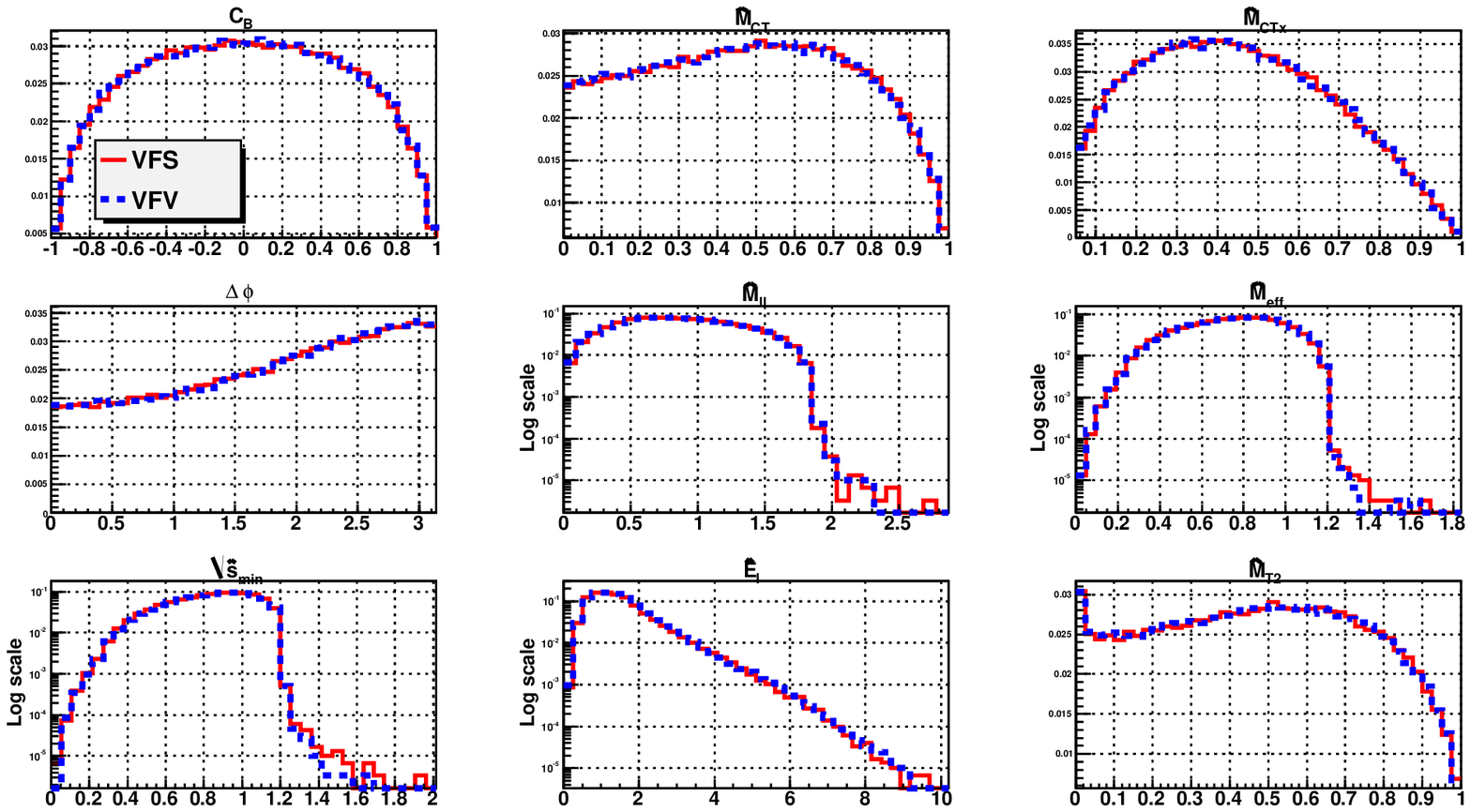} 
\caption{The same as Fig.~\ref{fig:ssf_svf}, but for the VFS and VFV spin scenarios with $\alpha=\beta=\gamma=0$ and $y \sim 0$ ($M_C=5$ GeV).}
\label{fig:vfs_vfv}
}

In conclusion of this section, we can summarize its main two lessons as follows:
\begin{itemize}
\item For the spin scenarios where particle $B$ is a fermion, the two situations in which
it will be quite difficult to determine the spin of particle $C$ are:
\begin{enumerate}
\item Vectorlike couplings between $B$ and $C$ with $\gamma=0$. The presence of chiral couplings is
a necessary condition to observe spin correlations \cite{Wang:2006hk,Wang:2008sw,Burns:2008cp,Cheng:2010yy}.
\item Models with $y\sim 0$, where particle $C$ is highly boosted. When $C$ is a vector boson, it is 
longitudinally polarized and effectively behaves as a scalar.
\end{enumerate}
\item For the spin scenarios where particle $B$ is a boson, it may be difficult to determine whether that
boson is a spin 0 or spin 1 particle. Models with $\gamma=0$ and/or $y \sim 1$ are again the most problematic.
\end{itemize}

\section{Discussion}\label{sec:conclusion}

In this paper we investigated the antler event topology of Fig.~\ref{fig:top} with regards to spin
effects in the kinematic distributions of the two visible particles.
Given the simplicity of Fig.~\ref{fig:top}, one might have hoped to be able to describe 
analytically the observable kinematic distributions, including explicitly the effects of spin correlations.
Unfortunately, this is not the case, and spin studies must be done numerically by template methods.
Nevertheless, useful physics intuition can be developed in the two $\sqrt{s}$ extremes, namely, 
at threshold $\sqrt{s} \sim 2M_B$ and at $\sqrt{s}\to \infty$. In Section~\ref{sec:Chiralities}
we analyzed the asymptotic analytical expressions for the $\BAR$ distribution,
which helped us identify in Sec.~\ref{sec:discrimination} the difficult cases for spin discrimination.
In Table~\ref{table:results} we identified pairs of spin configurations which can easily mimic each other, 
and the exact circumstances when this is most likely to occur. Typically, the chirality $\gamma$ of the 
$BC\ell$ coupling and the mass splitting between $B$ and $C$ play an important role --- 
the spin correlations tend to be diminished when $\gamma=0$ and depending on case, 
\begin{itemize}
\item For the spin scenarios where particle $B$ is a fermion: $M_C\sim M_B$.
\item For the spin scenarios where particle $B$ is a boson: $M_C << M_B$.
\end{itemize}
 
While our discussion centered mostly on the $\BAR$ variable, which has been most often utilized for spin studies,
we also considered 8 other kinematic variables, which were previously developed mainly in the context 
of mass determination. Using the popular examples of SUSY and MUED, we investigated their
power for spin discrimination. We generally found that some variables are better suited for spin studies 
than others. Among the better performers were $\BAR$, $\Delta\varphi$, $M_{\ellp\ellm}$, $M_{\mathrm{eff}}$ and $\smin$.
It is interesting to note the complementarity between the $\BAR$ variable and the others in this group.
As shown in Fig.~\ref{fig:corr}, the $\BAR$ distribution becomes most sensitive to spins at high energies,
where the correlation (\ref{dBAReqdTheta}) is manifest. In contrast, the distributions of the 
other variables are probing the different $\sqrt{s}$ behavior {\em near threshold} (see, e.g.~Fig.~\ref{fig:shatstudy}).
Therefore it appears that a multifaceted approach, utilizing a number of different and complementary variables, 
would be most beneficial for spin determinations.

\acknowledgments
We thank A.~Barr for insightful comments.
LE has been supported by the German Ministry of Education and Research (BMBF) under contract no. 05H09WWE and 05H09PAE, 
and thanks the theory group at the University of Florida for hospitality during the completion of this project. 
The work of KM is supported by a US Department of Energy grant DE-FG02-97ER41029.  
MP is supported by a CERN-Korea fellowship.

\appendix

\section{Derivation of the $\BAR$ angular distribution at threshold}
\label{sec:derivation}

In this Appendix, we show the derivation of the $\BAR$ angular distributions in the threshold limit
$\sqrt{s} \to 2M_B$. According to (\ref{BAReqtanh}), in that case the $\BAR$ variable can be expressed 
in terms of the pseudorapidities of the visible particles in the corresponding CMB1 and CMB2 frames as
\beq
\BAR=\tanh\left(\frac{\eta_1'-\eta_2'}{2}\right),
\eeq
with $\eta_1'$ and $\eta_2'$ defined in (\ref{eq:etaprime}). Since the initial state at the LHC is symmetric
(both beams are proton beams), one should symmetrize the $\BAR$ distribution with respect to $\BAR\leftrightarrow -\BAR$, 
since we do not know which proton beam the initial state quark came from.

\subsection{$\BAR$ distribution with no spin correlation}

In the pure phase space limit,
\bea
\frac{\ud N}{\ud \BAR } &\propto& \iiint \ud \Omega^* \ud \Omega_1' \ud \Omega_2' ~|\mathcal{M}|^2
~\delta \left(\BAR- \BAR\left[\Theta^\ast,\varphi^\ast,\eta^*,\phi_i',\eta_i'\right]\right) \nonumber \\
&\propto& \iint_{-\infty}^{\infty}\ud \eta_1' \ud\eta_2' \sech^2{\eta_1'} \sech^2{\eta_2'}
~\delta \left( \BAR- \tanh\left(\frac{\eta_1'-\eta_2'}{2}\right)\right) \nonumber \\
&\propto&\int_{-\infty}^{\infty} \ud \eta_2' \left\{\cosh\left({\eta_1'}^{(0)}-\eta_2'\right)
+1\right\} \sech^2 {\eta_1'}^{(0)} \sech^2\eta_2' \, ,
\eea
where ${\eta_1'}^{(0)}$ is the zero of the argument of the delta function
\beq
 \BAR- \tanh\left(\frac{\eta_1'-\eta_2'}{2}\right) =0 \Longleftrightarrow 
 {\eta_1'}^{(0)} = \eta_2'+ \ln\left[\frac{1+\BAR}{1-\BAR}\right].
 \eeq
 After integrating over $\eta_2'$, we get the following unit-normalized distribution
\beq
\mathrm{J}_{\mathrm{PS}} =\frac{\ud N}{\ud\BAR } =
\frac{1- \BAR^2}{4\BAR^3}
\left\{ -2 \BAR+\left(1+\BAR^2\right)
 \ln\left[\frac{1+ \BAR}{1- \BAR}\right]\right\} ,
\label{eq:disTHA}
\eeq
which is eq.~(\ref{dNdCBth}). This result is already symmetrized with respect to
$ \BAR \leftrightarrow  -\BAR $.

\subsection{$\BAR$ distribution when particle $B$ is a fermion}
\label{secA:phasespace}

When particle $B$ is a fermion, the matrix element has the generic form 
\beq
|\mathcal{M}|^2 \propto 1+A \left(\cos{\theta_1'}-\cos{\theta_2'}\right)
+\left(B_1\cos{ \Delta\phi'}+B_2 \sin{\Delta\phi'}\right) \sin{\theta_1'} \sin{\theta_2'}
+C \cos{\theta_1'} \cos{\theta_2'} \, .
\label{mevfv}
\eeq
The $B_1$ and $B_2$ terms integrate out to zero and we get
\bea
\frac{\ud  N}{\ud \BAR } 
&\propto& \iiint \ud \Omega^* \ud \Omega_1' \ud \Omega_2' ~|\mathcal{M}|^2
~\delta \left(\BAR- \BAR\left[\Theta^\ast,\varphi^\ast,\eta^*,\phi_i',\eta_i'\right]\right) \nonumber \\
&=& \frac{1}{N_{tot}} 
\left(\frac{\ud  N_{\mathrm{PS}}}{\ud \BAR}+A \frac{\ud  N_A}{\ud \BAR }
+C \frac{\ud  N_C}{\ud \BAR } 
\right),
\eea
where 
\beq
\frac{\ud  N_{\mathrm{PS}}}{\ud \BAR} \propto  \iint \ud \Omega_1' \ud \Omega_2' 
~\delta \left( \BAR-\BAR\left[\eta_1',\eta_2'\right]\right)
\eeq
is the pure phase space contribution already derived in Sec.~\ref{secA:phasespace},
while the remaining two contributions are
\bea
\frac{\ud  N_A}{\ud \BAR }
\propto &&  \iint  \ud \Omega_1' \ud \Omega_2' ~\left(\cos{\theta_1'}-\cos{\theta_2'}\right)
~\delta \left( \BAR-\BAR\left[\eta_1',\eta_2'\right]\right),    \label{Acontribution}\\ [2mm]
\frac{\ud  N_C}{\ud \BAR}
\propto &&   \iint  \ud \Omega_1' \ud \Omega_2' ~\cos{\theta_1'}\cos{\theta_2'}
~\delta \left( \BAR-\BAR\left[\eta_1',\eta_2'\right]\right) .
\eea
The contribution from (\ref{Acontribution}) vanishes after the symmetrization 
$ \BAR \leftrightarrow  -\BAR $ and we are left to evaluate
%
\bea
\frac{\ud  N_C}{\ud \BAR}
 \propto &&   \iint  \ud \Omega_1' \ud \Omega_2' ~\cos{\theta_1'}\cos{\theta_2'}
~\delta \left( \BAR- \BAR\left[\eta_1',\eta_2'\right]\right)  \nonumber \\
\propto&& \iint_{-\infty}^{\infty}\ud \eta_1' \ud\eta_2' \sech^2{\eta_1'} \sech^2{\eta_2'} \tanh{\eta_1'} \tanh{\eta_2'}
~\delta \left(\BAR- \tanh\left(\frac{\eta_1'-\eta_2'}{2}\right)\right) \nonumber \\
=&&\int_{-\infty}^{\infty} \ud \eta_2' \left\{\cosh\left({\eta_1'}^{(0)}-\eta_2'\right)
+1\right\} \sech^2 {\eta_1'}^{(0)}  \tanh{{\eta_1'}^{(0)}}  \sech^2\eta_2' \tanh{\eta_2'} \nonumber \\
=&& \frac{2 }{1- \BAR^2}\int_{-\infty}^{\infty} \ud \eta_2' \left\{
\frac{\sech^2{\eta_2'} \tanh{\eta_2'}  \tanh{\left(\eta_2'+\ln\left[\frac{1+\BAR}{1-\BAR}\right]\right)} }
{\cosh^2{\left(\eta_2'+\ln\left[\frac{1+\BAR}{1-\BAR}\right]\right)}} 
\right\} 
\, .
\label{Ccontribution}
\eea
Upon integrating over $\eta_2'$, this gives the contribution which is due to spin correlations.
we use a subscript $\mathrm{F}$ for it as a reminder that it corresponds to the case of a fermion particle $B$:
\beq
\frac{\ud  N}{\ud\BAR } =\frac{1}{N_{tot}}\left( \frac{\ud  N_{\mathrm{PS}}}{\ud \BAR }+C \frac{\ud N_\mathrm{F}}{\ud \BAR } \right)
=\mathrm{J}_{\mathrm{PS}} +\frac{C}{4}\mathrm{J}_{\mathrm{F}} \, ,
\label{dNdCBapp2}
\eeq
where (\ref{Ccontribution}) evaluates to
\beq
\mathrm{J}_{\mathrm{F}} =\frac{\ud N_\mathrm{F}}{\ud \BAR }  = \frac{1-\BAR^2}{4 \BAR^5}
\left\{6\BAR+4 \BAR^3+6 \BAR^5-(3+\BAR^2+\BAR^4+3 \BAR^6)\ln\left(\frac{1+\BAR}{1-\BAR}\right) \right\}\, ,
\label{eq:intF}
\eeq
which is the result in (\ref{eq:JF}). The prefactor $C$ in (\ref{dNdCBapp2}) depends on the 
particular spin configuration, as seen in eqs.~(\ref{eq:sfsth},\ref{eq:sfvth},\ref{eq:vfsth},\ref{eq:vfvth}).

\subsection{$\BAR$ distribution in the SVF  scenario}

In the SVF spin scenario, the matrix element has the form
\bea
|\mathcal{M}|^2 \propto &&1+\frac{(1-y)^2}{1+2 y}\left(\cos{\Delta\phi'} \sin{\theta_1'} \sin{\theta_2'}-
\cos{\theta_1'} \cos{\theta_2'}\right)^2 \nonumber \\
&&+\frac{2 \gamma^2}{1+2 y}\left(\cos{\Delta\phi'} \sin{\theta_1'} \sin{\theta_2'}-
\cos{\theta_1'} \cos{\theta_2'}\right).
\eea
The $\BAR$ distribution is
\bea
\frac{\ud  N}{\ud \BAR } 
&\propto& \iiint \ud \Omega^* \ud \Omega_1' \ud \Omega_2' ~|\mathcal{M}|^2
~\delta \left(\BAR- \BAR\left[\Theta^\ast,\varphi^\ast,\eta^*,\phi_i',\eta_i' \right]\right) \nonumber \\
&=& \frac{1}{N_{tot}} 
\left(
\frac{\ud  N_{\mathrm{PS}}}{\ud \BAR}+\frac{(1-y)^2}{1+2 y} \frac{\ud  N_\mathrm{V}}{\ud \BAR }
-\frac{2 \gamma^2}{1+2 y} \frac{\ud  N_\mathrm{F}}{\ud \BAR } 
\right),
\eea
where $\frac{\ud  N_{\mathrm{PS}}}{\ud \BAR}$ is the pure phase space contribution
(\ref{eq:disTHA}), while $\frac{\ud  N_\mathrm{F}}{\ud \BAR} $ is the contribution  (\ref{eq:intF})
from the helicity structure of the $B\ell C$ interaction. The new term here 
$\frac{\ud  N_\mathrm{V}}{\ud \BAR}$ is given by 
\beq
\frac{\ud  N_\mathrm{V}}{\ud \BAR }
\propto  \iint  \ud \Omega_1' \ud \Omega_2' ~\left(\cos{\Delta\phi'} \sin{\theta_1'} \sin{\theta_2'}-
\cos{\theta_1'} \cos{\theta_2'}\right)^2
~\delta \left( \BAR-\BAR\left[\eta_1',\eta_2'\right]\right)\, .
\eeq
After integrating out $\phi_i'$, we get
\bea
\frac{\ud  N_\mathrm{V}}{\ud \BAR}
 \propto &&   \iint  \ud\cos{\theta_1'} \ud\cos{\theta_2'} ~\left(\frac{1}{2}\sin^2{\theta_1'} \sin^2{\theta_2'}+
\cos^2{\theta_1'} \cos^2{\theta_2'}\right)
~\delta \left( \BAR-\BAR\left[\eta_1',\eta_2'\right]\right) \nonumber \\
\propto&& \iint_{-\infty}^{\infty}\ud \eta_1' \ud\eta_2' \sech^4{\eta_1'} \sech^4{\eta_2'}\left(\frac{1}{2}
 +\sinh^2{\eta_1'} \sinh^2{\eta_2'} \right)
~\delta \left(\BAR- \tanh\left(\frac{\eta_1'-\eta_2'}{2}\right)\right) \nonumber \\
=&&\int_{-\infty}^{\infty} \ud \eta_2' \left\{\cosh\left({\eta_1'}^{(0)}-\eta_2'\right)
+1\right\} 
 \sech^4{{\eta_1'}^{(0)}} \sech^4{\eta_2'}\left(\frac{1}{2}
 +\sinh^2{{\eta_1'}^{(0)}} \sinh^2{\eta_2'} \right)  \nonumber \\
=&& \frac{2 }{1- \BAR^2}\int_{-\infty}^{\infty} \ud \eta_2'
\left\{
\frac{\sech^4{\eta_2'} \left(\frac{1}{2} +  \sinh^2{\eta_2'} \sinh^2{\left(\eta_2'+\ln\left[\frac{1+\BAR}{1-\BAR}\right]\right)}
\right)}
{\cosh^4{\left(\eta_2'+\ln\left[\frac{1+\BAR}{1-\BAR}\right]\right)}} 
\right\} \, .
\eea
Finally, after integrating over $\eta_2'$, we get (\ref{eq:svfth})
\bea
\frac{\ud  N}{\ud\BAR } &=&\frac{1}{N_{tot}} 
\left(
\frac{\ud  N_{\mathrm{PS}}}{\ud \BAR}+\frac{(1-y)^2}{1+2 y} \frac{\ud  N_\mathrm{V}}{\ud \BAR }
-\frac{2 \gamma^2}{1+2 y} \frac{\ud  N_\mathrm{F}}{\ud \BAR } 
\right) \nonumber \\
&=&\frac{3(1+2 y)}{\left(2+y\right)^2} 
\left(
\mathrm{J}_{\mathrm{PS}}+\frac{(1-y)^2}{4(1+2 y)}\mathrm{J}_{\mathrm{V}}
-\frac{\gamma^2}{2(1+2 y)}\mathrm{J}_{\mathrm{F}}
\right)\, ,
\eea
where the new contribution is given by (\ref{JVdef})
\bea
\mathrm{J}_{\mathrm{V}}=\frac{\ud N_V}{\ud \BAR }  &=&\frac{1-\BAR^2}{48 \BAR^7} \cdot
\bigg\{ 
-2 \BAR\left(45+22 \BAR^4+45 \BAR^8\right) \nonumber \\
&+& 3\left(15-5\BAR^2+6\BAR^4+6\BAR^6-5 \BAR^8+15\BAR^{10}\right)\ln\left(\frac{1+\BAR}{1-\BAR}\right)\bigg \}\, .
\eea

\subsection{$\BAR$ distribution in the VVF scenario}

After integrating out $\phi_i'$ from the matrix element, we get
\bea
\iint\ud \phi_1' \ud \phi_2' |\mathcal{M}|^2 \propto && 1
- \frac{(1-y)^2}{1+y} \cos^2{\theta_1'} \cos^2{\theta_2'}
-\frac{y (1-y)}{2 (1+y)} \left(\cos^2{\theta_1'}+\cos^2{\theta_2'}\right) \nonumber \\
&&+\frac{\alpha \gamma }{1+y}\left\{1-(1-y) \cos{\theta_1'} \cos{\theta_2'}\right\} \left(\cos{\theta_1'}+\cos{\theta_2'}\right)\, .
\eea
The $\BAR$ distribution is given by
\bea
\frac{\ud  N}{\ud \BAR } 
\propto&& \iiint \ud \Omega^* \ud \Omega_1' \ud \Omega_2' ~|\mathcal{M}|^2
~\delta \left(\BAR- \BAR\left[\Theta^\ast,\varphi^\ast,\eta^*,\phi_i',\eta_i'\right]\right) \nonumber \\
=&& \frac{1}{N_{tot}} 
\bigg\{
\frac{\ud  N_{\mathrm{PS}}}{\ud \BAR}
- \frac{(1-y)^2}{1+y} \frac{\ud  N_\mathrm{V1}}{\ud \BAR } 
-\frac{y (1-y)}{2 (1+y)} \frac{\ud  N_\mathrm{V2}}{\ud \BAR } 
+\frac{\alpha \gamma }{1+y}  \frac{\ud  N_\mathrm{CV}}{\ud \BAR } 
\bigg\},
\label{dNdCBapp4}
\eea
but the last term integrates out to zero:
\bea
\frac{\ud  N_\mathrm{CV}}{\ud \BAR}
&\propto& \iint  \ud\cos{\theta_1'} \ud\cos{\theta_2'} ~ \left(\cos{\theta_1'}+\cos{\theta_2'}\right)
\left\{1-(1-y) \cos{\theta_1'} \cos{\theta_2'}\right\}
~\delta \left( \BAR-\BAR\left[\eta_1',\eta_2'\right]\right) \nonumber \\
&\propto& \iint_{-\infty}^{\infty}\ud \eta_1' \ud\eta_2' \sech^3{\eta_1'} \sech^3{\eta_2'} \sinh{(\eta_1'+\eta_2')}
\left\{1-(1-y) \tanh{\eta_1'} \tanh{\eta_2'}\right\} \times \nonumber\\
&\times& \delta \left(\BAR- \tanh\left(\frac{\eta_1'-\eta_2'}{2}\right)\right) \nonumber \\
&=&0 \, .
\eea
Thus, in the VVF case, the chiralities $\alpha$ and $\gamma$ do not have an effect on the $\BAR$ distribution
at threshold. The second term in (\ref{dNdCBapp4}) is
\bea
\frac{\ud  N_\mathrm{V1}}{\ud \BAR}
&\propto& \iint  \ud\cos{\theta_1'} \ud\cos{\theta_2'} ~\cos^2{\theta_1'}\cos^2{\theta_2'} 
~\delta \left( \BAR-\BAR\left[\eta_1',\eta_2'\right]\right) \nonumber \\
&\propto& \iint_{-\infty}^{\infty}\ud \eta_1' \ud\eta_2' \sech^2{\eta_1'} \sech^2{\eta_2'}
\tanh^2{\eta_1'} \tanh^2{\eta_2'}
~\delta \left(\BAR- \tanh\left(\frac{\eta_1'-\eta_2'}{2}\right)\right) \nonumber \\
&=&\int_{-\infty}^{\infty} \ud \eta_2' \left\{\cosh\left({\eta_1'}^{(0)}-\eta_2'\right)
+1\right\} 
 \sech^2{{\eta_1'}^{(0)}} \sech^2{\eta_2'}
\tanh^2{{\eta_1'}^{(0)}} \tanh^2{\eta_2'}
  \nonumber \\
&=& \frac{2 }{1- \BAR}\int_{-\infty}^{\infty} \ud \eta_2'
\left\{
\frac{\sech^2{\eta_2'}\tanh^2{\eta_2'}
\tanh^2{\left(\eta_2'+\ln\left[\frac{1+\BAR}{1-\BAR}\right]\right)}}
{\cosh^2{\left(\eta_2'+\ln\left[\frac{1+\BAR}{1-\BAR}\right]\right)}} 
\right\} \, .
\eea
After integrating out $\eta_2'$, we get (\ref{JV1def})
\bea
\mathrm{J}_{\mathrm{V1}}=\frac{\ud N_\mathrm{V1}}{\ud \BAR }  &=& \frac{1-\BAR^2}{24 \BAR^7} \cdot
\bigg\{ 
-2 \BAR\left(15+8 \BAR^2+10\BAR^4+8\BAR^6+15\BAR^8 \right) 
\nonumber \\
&+& 3\left(5+\BAR^2+2\BAR^4+2\BAR^6+\BAR^8+5\BAR^{10}\right)\ln\left(\frac{1+\BAR}{1-\BAR}\right)\bigg \}\, .
\eea
By a similar calculation, we obtain (\ref{eq:JV2})
\beq
\mathrm{J}_{\mathrm{V2}}=\frac{\ud N_\mathrm{V2}}{\ud \BAR }  = \frac{1}{3 \BAR^5} 
\left\{
 -2 \BAR\left(3+\BAR^2-\BAR^4-3\BAR^6 \right) 
 + 3\left(1-\BAR^8\right)\ln\left(\frac{1+\BAR}{1-\BAR}\right)\right \}\, .
\eeq
In total, we get the answer in (\ref{eq:vvfth})
\bea
\frac{\ud  N}{\ud \BAR } 
=&& \frac{1}{N_{tot}} 
\bigg\{
\frac{\ud  N_{\mathrm{PS}}}{\ud \BAR}
- \frac{(1-y)^2}{1+y} \frac{\ud  N_\mathrm{V1}}{\ud \BAR } 
-\frac{y (1-y)}{2 (1+y)} \frac{\ud  N_\mathrm{V2}}{\ud \BAR } 
\bigg\} \nonumber \\[3mm]
=&& \frac{9(1+y)}{2(2+y)^2} 
\left\{ 
\mathrm{J}_{\mathrm{PS}}
-\frac{(1-y)^2}{4 (1+ y)}  \mathrm{J}_{\mathrm{V1}}
-\frac{y(1-y)}{8 (1+y)}
   \mathrm{J}_{\mathrm{V2}}
 \right\}  \, .
\eea
 
\subsection{$\BAR$ distribution in SUSY and MUED}

For SUSY (MSSM), the threshold behavior of $\BAR$ is given by (\ref{eq:PHTH}),
since $B$ is a scalar particle (slepton):
\bea
\frac{\ud N}{\ud \BAR} =\mathrm{J}_{\mathrm{PS}}\, .
\eea
For the minimal UED model (MUED), the matrix element is
\beq
|\mathcal{M}|^2 \propto 1-A_1 (\cos{\theta_1'}-\cos{\theta_2'})-A_2 \cos{\theta_1'}\cos{\theta_2'}\, ,
\label{meued}
\eeq 
with
\bea
A_1&=&2\left(\frac{1-2 y}{1+2 y}\right) \frac{\left(3-\frac{M_Z^2}{M_B^2} C_W^2\right)}
{1+\left(3-\frac{M_Z^2}{M_B^2} C_W^2\right)^2}\, , \nonumber \\
A_2&=&\left(\frac{1-2 y}{1+2 y}\right)^2 \, .
\eea
$M_Z$ is the mass of the $Z$ boson and $C_W\equiv \cos\theta_W$ is the cosine of the Weinberg angle $\theta_W$.
Comparing (\ref{meued}) to (\ref{mevfv}), we find $C=-A_2$ and from (\ref{dNdCBapp2}) we get
\beq
\frac{\ud N}{\ud \BAR}  = \mathrm{J}_{\mathrm{PS}}-\frac{1}{4}\left(\frac{1-2 y}{1+2 y}\right)^2 \mathrm{J}_{\mathrm{F}} \, .
\eeq


\end{document}